\documentclass[preprint,prc,
tightenlines,
nofootinbib
,superscriptaddress]{revtex4}
\usepackage{amssymb}
\usepackage{amsmath}
\usepackage{graphicx}
\usepackage{bm} % for bold greek letters e.g. \bm{\tau}
\usepackage[loose]{subfigure}
\usepackage{xcolor}
\usepackage[squaren]{SIunits}
\usepackage{multirow}
\usepackage{braket}
\usepackage{tensor}
\usepackage{slashed}
\usepackage{array}

\newcommand{\di}{\mathrm{d}}

\newcommand{\punkt}{\;\text{.}}
\renewcommand{\vec}[1]{\boldsymbol{#1}}
\newcommand{\komma}{\;\text{,}}
\renewcommand{\i}{\mathrm{i}}

\begin{document}
\title{Elastic pion-nucleon scattering in chiral perturbation theory:\\
  A fresh look }

\author{D.~Siemens}
\email[]{dmitrij.siemens@rub.de}
\affiliation{Institut f\"ur Theoretische Physik II, Ruhr-Universit\"at Bochum, D-44780 Bochum, Germany}

\author{V.~Bernard}
\email[]{bernard@ipno.in2p3.fr}
\affiliation{Groupe de Physique Th\'eorique, Institut de Physique
Nucl\'eaire, UMR 8608, CNRS, Univ. Paris-Sud, Universit\'e Paris Saclay,
F-91406 Orsay Cedex, France}

\author{E.~Epelbaum} 
\email[]{evgeny.epelbaum@rub.de}
\affiliation{Institut f\"ur Theoretische Physik II, Ruhr-Universit\"at Bochum, D-44780 Bochum, Germany}

\author{A.~Gasparyan}
\email[]{ashot.gasparyan@rub.de}
\affiliation{Institut f\"ur Theoretische Physik II, Ruhr-Universit\"at Bochum, D-44780 Bochum, Germany}
\affiliation{SSC RF ITEP, Bolshaya Cheremushkinskaya 25, 117218 Moscow, Russia}

\author{H.~Krebs}
\email[]{hermann.krebs@rub.de}
\affiliation{Institut f\"ur Theoretische Physik II, Ruhr-Universit\"at Bochum, D-44780 Bochum, Germany}

\author{Ulf-G.~Mei{\ss}ner}
\email[]{meissner@hiskp.uni-bonn.de}
\affiliation{Institute~for~Advanced~Simulation, Institut~f\"{u}r~Kernphysik and
J\"{u}lich~Center~for~Hadron~Physics, ~Forschungszentrum~J\"{u}lich,
D-52425~J\"{u}lich, Germany}
\affiliation{Helmholtz-Institut f\"ur Strahlen- und
             Kernphysik and Bethe Center for Theoretical Physics, \\
             Universit\"at Bonn,  D--53115 Bonn, Germany}
\affiliation{JARA~-~High~Performance~Computing, Forschungszentrum~J\"{u}lich, 
D-52425 J\"{u}lich,~Germany}

\begin{abstract}
Elastic pion-nucleon scattering is analyzed in the framework of
chiral perturbation theory  up to fourth order within the heavy-baryon
expansion and a covariant
approach based on an extended on-mass-shell renormalization
scheme. We discuss in detail the renormalization of the various
low-energy constants and provide explicit expressions for the relevant
$\beta$-functions and the finite subtractions of the power-counting
breaking terms within the covariant formulation. To estimate
the theoretical uncertainty from the truncation of the chiral
expansion, we employ an approach which has been successfully
applied in the most recent analysis of the nuclear forces. This 
allows us to reliably extract the relevant low-energy constants
from the available scattering data at low energy. The obtained results
provide a clear evidence that the breakdown scale of the chiral
expansion for this reaction is related to the $\Delta$-resonance. 
The explicit inclusion of the leading contributions of the
$\Delta$-isobar is demonstrated to substantially increase the 
range of applicability of the effective field theory. The resulting predictions for the
phase shifts are in an excellent agreement with the ones from the recent 
Roy-Steiner-equation analysis of pion-nucleon scattering. 
\end{abstract}

\maketitle

\section{Introduction}
Chiral perturbation theory ($\chi$PT) provides a systematically
improvable theoretical framework to analyze
low-energy hadronic reactions. It relies on the chiral symmetry of
QCD and its breaking patterns, in particular the spontaneous chiral symmetry
breaking with the appearance of a triplet of Goldstone bosons, the pions.
$\chi$PT corresponds to an expansion of the scattering
amplitude around the chiral and zero-energy limits. Here, we 
consider the two-flavor chiral limit with vanishing up and
down quark masses and the strange quark mass fixed at its physical value.
%with the expansion parameter 
%given by $Q \in \{ M_\pi /\Lambda_\chi, \; p/\Lambda_\chi \}$. Here, $M_\pi$ denotes the
%pion mass, $p$ refers to generic three-momenta of external particles
%and $\Lambda_\chi$ is the breakdown scale which is
%expected to be of the order of $\Lambda_\chi \sim M_\rho$. 
Characteristic for any effective
field theory (EFT), effects of higher energy physics are accounted for via 
low-energy constants (LECs) accompanying the interaction terms in the
effective Lagrangian.

Pioneered in the meson sector 
\cite{Weinberg:1978kz,Gasser:1983yg,Gasser:1984gg}
and extended to the single-baryon 
\cite{Gasser:1987rb,Bernard:1995dp,Bernard:2006gx,Bernard:2007zu} as well as to
few-baryon sectors \cite{Weinberg:1990rz,Ordonez:1995rz,Epelbaum:2008ga,Machleidt:2011zz},
numerous applications and extensions of $\chi$PT have been 
performed over the
last decades.  Historically, most of the studies in the baryon
sector have been  carried out utilizing the so-called heavy-baryon (HB)
approach \cite{Jenkins:1990jv,Bernard:1992qa}. In this formulation, the effective chiral
Lagrangian is expanded in inverse powers of the nucleon mass 
treated on the same footing as the breakdown scale of the chiral
expansion $\Lambda_b$, also referred to as the chiral symmetry
breaking scale $\Lambda_\chi$.  With only negative powers of the
nucleon mass appearing in the HB Lagrangian, this
formulation offers the simplest way to maintain the power counting for
dimensionally regularized loop integrals which enter the scattering amplitude. On the other hand, the
strict HB approach does not correctly reproduce certain analytic
properties of the scattering amplitude
\cite{Bernard:1993ry,Bernard:1996cc,Becher:1999he}. 
Using manifestly covariant versions of $\chi$PT
does lead to a correct representation of the analytic properties of the
scattering amplitude but requires special care in order to maintain the chiral
power counting for loop contributions due to the appearance of
positive powers of the nucleon mass $m_N$. In the so-called
infrared renormalization (IR) scheme proposed by Becher and Leutwyler
\cite{Becher:1999he}, see also Ref.~\cite{Ellis:1997kc} for a related earlier work, only the
infrared-singular (in the limit of vanishing pion masses) pieces of
the loop integrals are kept, which are responsible for 
non-integer powers of the soft scales in the scattering amplitude. 
On the other hand, the IR scheme of Ref.~\cite{Becher:1999he,Bernard:2002pw} 
induces unphysical singularities in the amplitude at high momenta. Alternatively, one may
employ  the so-called extended on-mass-shell scheme (EOMS)
\cite{Gegelia:1999gf,Fuchs:2003qc} which makes use of the freedom in
the choice of renormalization conditions to maintain the chiral power
counting.  For a detailed discussion and comparison of the various formulations
of $\chi$PT the reader is referred to Ref.~\cite{Bernard:2007zu}.

In this paper we analyze in detail the reaction $\pi N\to\pi N$
at low energies within the HB$\chi$PT and EOMS formulations at the full
one-loop order. 
Pion-nucleon scattering certainly belongs to the most extensively studied
processes in $\chi$PT, see Refs.~\cite{Bernard:1992qa,Mojzis:1997tu,Fettes:1998ud,
Buettiker:1999ap,Fettes:2000gb,Fettes:2000xg,Becher:2001hv,Hoferichter:2009gn,
Gasparyan:2010xz,Alarcon:2012kn,Chen:2012nx} for the analyses of the
elastic channel and Refs.~\cite{Beringer:1992ic,Bernard:1994wf,Bernard:1995gx,
Jensen:1997em,Fettes:1999wp,Bernard:1997tq,Mobed:2005av,Siemens:2014pma}
for studies of the single-pion production $\pi N\to\pi\pi N$. It also has attracted renewed
interest in recent years in light of its importance for understanding
the long-range behavior of the nuclear forces \cite{Krebs:2012yv,Wendt:2014lja,Entem:2014msa,Entem:2015xwa}. In particular,
the state-of-the-art nucleon-nucleon potentials of Ref.~\cite{Epelbaum:2014sza} include
the two-pion exchange contribution derived from the fourth-order
approximation of the pion-nucleon scattering amplitude \cite{Krebs:2012yv}. 
It was demonstrated in Ref.~\cite{Epelbaum:2014sza} that nucleon-nucleon
scattering data show clear evidence of the resulting two-pion exchange potential,
see also Refs.~\cite{Rentmeester:1999vw,Birse:2003nz} for similar findings at lower chiral orders. 
Given the ongoing efforts towards pushing the precision frontier in
nuclear chiral EFT \cite{Epelbaum:2015pfa}, a reliable determination
of pion-nucleon LECs entering the two-pion exchange contributions to
the two- and three-nucleon forces with quantified uncertainties becomes an
important task. This is a non-trivial issue given that most
of the $\chi$PT studies  of pion-nucleon scattering in $\chi$PT rely
on the  Karlsruhe-Helsinki \cite{Koch:1980ay} and GWU-SAID \cite{Arndt:2006bf} partial-wave
analyses (PWA) which do not provide information about systematic
uncertainties. An important step towards resolving this issue was
made recently in Refs.~\cite{Hoferichter:2015dsa,Hoferichter:2015tha}, where pion-nucleon scattering was
analyzed in the framework of Roy-Steiner equations (RS) and  detailed
error estimates of all input quantities, the solution procedure and
truncations were performed, see  Ref.~\cite{Hoferichter:2015hva} for a review. 
The resulting phase shifts with quantified uncertainties provide a
solid basis for a reliable determination of the LECs. In this paper
we, however, follow a different path and analyze directly the available
pion-nucleon scattering data at low energies, see also Ref.~\cite{Wendt:2014lja}
for a related study. To quantify the theoretical uncertainty from the
truncation of the chiral expansion, we employ the approach suggested
in Ref.~\cite{Epelbaum:2014efa}  which has also been employed in recent few-nucleon
studies \cite{Epelbaum:2014sza,Binder:2015mbz}. The resulting phase shifts are compared with the ones
of Ref.~\cite{Hoferichter:2015dsa} obtained from the Roy-Steiner analysis. 
We also 
%provide detailed expressions for the renormalized
%pion-nucleon scattering amplitude at fourth order in the EOMS scheme
%and 
discuss the role of the $\Delta$(1232) resonance in this reaction. 

Our paper is organized as follows. In section~\ref{sec:basicdef}, the necessary
definitions for a study of $\pi N\to\pi N$ in the HB and covariant
approach are given. The renormalization procedures in both chiral approaches are
discussed in section~\ref{sec:pow-count-renorm}, whereas the details
of the fitting procedure can be found in section~\ref{sec:fitting-procedure}. Our predictions for observables not used
in the fitting procedure are collected in section~\ref{sec:predictions} which also provides a discussion of the obtained
results.  Next, the explicit inclusion of the lowest-order
$\Delta$(1232) contributions is presented in section~\ref{sec:delta}.
Finally, the main results
of our study are summarized in section \ref{sec:sum}. The appendix
contains explicit expressions for the renormalized LECs.

\section{Basic definitions}
\label{sec:basicdef}
In this section, we provide some basic definitions which
are necessary for the description of the reaction $\pi N\to \pi N$. The
reader familiar with this is invited to skip this section.
Throughout this work, the kinematical variables are defined as follows:
\begin{equation}
  \label{eq:6}
  \pi^a(q) \, N(p=m_N v +k) \; \to \; \pi^b(q') \, N^\prime(p^\prime=m_N v +k^\prime)\komma
\end{equation}
where $N$ denotes a nucleon and $\pi^a$ a
pion with the isospin quantum number $a$. Note that the
decomposition of the nucleon four-momenta in terms of the
four-velocity $v_\mu$ and the residual small momentum $k_\mu$ is
only relevant for the heavy baryon approach.
To relate  the $T$-matrix to phase shifts, we follow the procedure of
Ref.~\cite{Hohler:1984ux} (Ref.~\cite{Fettes:1998ud}) for the
covariant (HB) approach as described below. 

\subsection{Covariant chiral perturbation theory}
\label{sec:covariant-chiral}
In the covariant approach, the $T$-matrix can be decomposed in the
following way
\begin{equation}
  \label{eq:1}
  T^{ba}=\chi^\dagger_{N^\prime}\left(\delta^{ab} T^+ +\i
    \epsilon^{bac} \tau_c T^-\right)\chi_N \komma
\end{equation}
where
\begin{equation}
  \label{eq:3}
  T^\pm=\bar{u}^{(s^\prime)}\left(A^\pm+\slashed{q}B^\pm\right)u^{(s)}
\end{equation}
and the amplitudes $A^\pm$ and $B^\pm$ depend on the
Mandelstam variables
\begin{equation}
  \label{eq:17}
     s=(p+q)^2\komma\quad t=(q-q^\prime)^2\komma\quad
  u=(p^\prime-q)^2\komma\quad s+t+u=2m_N^2+2M_\pi^2\punkt
\end{equation}
The partial wave amplitudes can be expressed in terms of $A^\pm$ and
$B^\pm$  as follows:
\begin{equation}
  \label{eq:12}
  \begin{aligned}
    f^I_{l\pm}(s) =& \frac{1}{16\pi\sqrt{s}}\left( (E+m_N)\left(  
        A^I_l(s)+(\sqrt{s}-m_N)B^I_l(s)\right)\right.\\
        &\left.+(E-m_N)\left( -A^I_{l\pm}(s)+(\sqrt{s}+m_N)B^I_{l\pm}(s) \right)\right) \komma
  \end{aligned}
\end{equation}
where for $X\in\{A,B\}$ 
\begin{equation}
  \label{eq:13}
  \begin{aligned}
    X^I_l(s)&= \int_{-1}^{+1} \di z\, X^I(s,t) P_l(z)\komma\\
  \end{aligned}
\end{equation}
with $t=-2 \vec{q}^2(1-z)$, $E=\sqrt{m_N^2+\vec{q}^2}$ and the relations to the isospin basis read
\begin{equation}
  \label{eq:15}
  \begin{aligned}
        X^{I=1/2}&=X^++2X^-\komma\quad
    X^{I=3/2}&=X^+-X^-\punkt
  \end{aligned}
\end{equation}
The phase shifts are obtained by using the unitarization prescription
\begin{equation}
  \label{eq:16}
  \delta^I_{l \pm}(s) =\arctan(|\vec{q}| \Re\, f^I_{l\pm}(s))\punkt
\end{equation}

\subsection{Heavy-baryon chiral perturbation theory}
\label{sec:heavy-baryon-chiral}
In the HB approach, the decomposition reads
\begin{equation}
  \label{eq:1}
  T^{ba}=\chi^\dagger_{N^\prime}\left(\delta^{ab} T^+ +\i
    \epsilon^{bac} \tau_c T^-\right)\chi_N \komma
\end{equation}
where
\begin{equation}
  \label{eq:4}
  T^\pm=\bar{u}_v^{(s^\prime)}\left(g^\pm+\,2\i\,\vec S\cdot \vec
    q\times \vec q^\prime h^\pm\right)u_v^{(s)}\punkt
\end{equation}
The amplitudes $g^\pm$ and $h^\pm$ depend on the four momenta $k$,
$k^\prime$, $q$, $q^\prime$ and are related to the partial wave
amplitudes via
\begin{equation}
  \label{eq:18}
  f_{l\pm}^I(s)= \frac{E+m_N}{16\pi\sqrt{s}}\int_{-1}^{+1}\di z\,
\left( g^I P_l(z) +\vec{q}^2 h^I(P_{l\pm}(z)-zP_l(z)) \right)\punkt
\end{equation}
The relation to the isospin basis is the same as in Eq.~\eqref{eq:15}
with $X\in\{g,h\}$.

\subsection{Observables}
\label{sec:observables}
The observables of interest are differential cross sections $\di\sigma
/ \di\Omega$ and
polarizations $P$ for the three channels $\pi^+p\to\pi^+p$,
$\pi^-p\to\pi^-p$ and $\pi^-p\to\pi^0n$. At low energy and/or forward
angles, these observables are strongly affected by electromagnetic
interactions which are taken into account following the procedure
described in Ref.~\cite{Tromborg:1976bi}.
This paper also provides all the necessary formula to relate the strong phase shifts in
Eq.~\eqref{eq:16} to the observables we are interested in. 
%(XXX must add some caution here XXX)
Still, it should be understood that the treatment of the
electromagnetic effects in that paper is approximative.

\section{Power Counting and Renormalization}
\label{sec:pow-count-renorm}
In $\chi$PT, the invariant amplitudes are calculated in the chiral
expansion with the expansion parameter
\begin{equation}
  \label{eq:12}
  Q=\left\{\frac{q}{\Lambda_b},\frac{M_\pi}{\Lambda_b}\right\}\komma
\end{equation}
where $M_\pi$ is the pion mass, $q$ denotes generic three- (four-) momenta of external nucleons
(pions) and $\Lambda_b$ is the breakdown scale of the chiral
expansion whose value will be specified below. 
%which is expected to be of the order of $\Lambda_\chi \sim M_\rho$. 
Since the nucleon mass $m_N$ does not vanish in the chiral limit,
the power counting employed in the Goldstone boson sector breaks
down for dimensionally regularized loop integrals in the presence of baryons.
The traditional way of curing this problem is the HB approach
\cite{Jenkins:1990jv,Bernard:1992qa}, where
the nucleon mass is treated as an additional large scale, $m_N \sim \Lambda_b$, and a
$1/m_N$ expansion is performed at the level of the effective
Lagrangian. 
%This additional expansion breaks Lorentz invariance. 
For certain observables such as some of the
nucleon 
%and electroweak and scalar 
form factors,
the HB expansion exhibits a very limited rage of convergence
\cite{Bernard:1996cc,Becher:1999he}~\footnote{It should, however, be noted that
these deformations of the analytic structure of the underlying amplitudes  can be overcome easily by including 
the first $1/m_N$ correction into the heavy fermion propagator, $i/(v\cdot k) \to i/(v\cdot k + k^2/(2m_N))$.}.
It is, therefore, advantageous to employ the Lorentz covariant
formulations of baryon $\chi$PT  using either the IR \cite{Becher:1999he} or 
the EOMS scheme \cite{Gegelia:1999gf,Fuchs:2003qc} in order to maintain the power counting. 
In this work, we will employ the HB and covariant EOMS approaches.
In both schemes, the effective Lagrangian needed to describe pion-nucleon
dynamics at one-loop level consists 
of the following pieces (see Ref.~\cite{Fettes:2000gb} for a full list of terms):
\begin{equation}
  \label{eq:2}
  \mathcal{L}_\mathrm{eff}=\mathcal{L}_{\pi\pi}^{(2)}+\mathcal{L}_{\pi\pi}^{(4)}
    +\mathcal{L}_{\pi N}^{(1)}+\mathcal{L}_{\pi N}^{(2)}+\mathcal{L}_{\pi N}^{(3)}
    +\mathcal{L}_{\pi N}^{(4)}\komma
\end{equation}
where the superscripts refer to the chiral dimension. 
Further, for the HB approach, we will also show results corresponding
to the power counting assignment $m_N \sim \Lambda_b^2/M_\pi$, which is commonly
used in the studies of the nuclear forces \cite{Epelbaum:2008ga}  and will be referred to as
HB-NN. The above assignment results in the relativistic corrections being
pushed to higher orders in the EFT expansion as compared to the
standard HB approach used in the single-baryon sector, which will be
referred to as HB-$\pi$N.

Before discussing the renormalization of the $\pi N\to\pi N$ amplitudes, we need to express the bare quantities in
the leading-order Lagrangian in terms of physical ones. The
expressions for $m_N$ and the nucleon axial vector coupling $g_A$
for both chiral approaches are given in Appendix \ref{sec:renorm-rules}.
Throughout this work, we express all results in terms of the effective axial vector coupling
constant $g_A$ which takes into account the Goldberger-Treiman
discrepancy and is related to the physical axial vector coupling $g_{A,ph}$ via 
\begin{equation}
  \label{eq:7}
  g_A=g_{A,ph}-2M_\pi^2 d_{18} +\mathcal{O}(Q^5)\,.
\end{equation}
The value of $g_A$ is fixed by the Goldberger-Treiman relation 
\begin{equation}
  \label{eq:9}
  g_A= \frac{g_{\pi NN}F_\pi}{m_N}\,.
\end{equation}
For the pion-nucleon coupling constant $g_{\pi NN}$, we adopt the
value from Ref.~\cite{Baru:2010xn}, $g_{\pi NN}^2/4\pi=13.7(2)$
leading to $g_A=1.289(1)$. Note
that we do not study the effects of the uncertainty of $g_A$ in this
work and only employ the mean value.
In addition to removing the redundant (for the considered reaction)
LEC $d_{18}$, using $g_A$ ensures a correct reproduction of the 
analytic structure of the  $\pi N\to\pi N$ scattering amplitude.

\begin{figure}[ht]
  \centering
\includegraphics[width=0.8\textwidth]{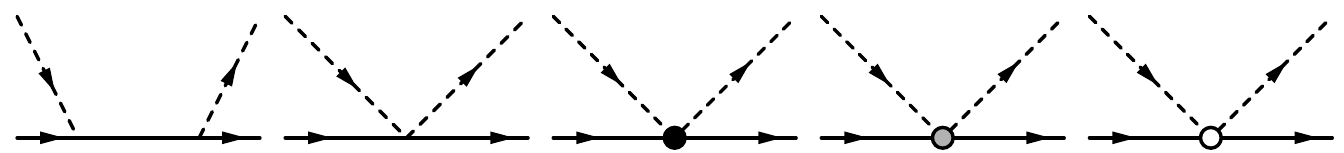}
  \caption{Tree graphs for the reaction $\pi N\to\pi N$. The black/gray/white
    blob denotes an insertion of the $c_i$/$d_i$/$e_i$-
    vertices. Dashed and solid lines refer to pions and nucleons,
    respectively. Crossed
    diagrams are not shown.}
  \label{fig:TreeGraphs}
\end{figure}

\begin{figure}[ht]
  \centering
\includegraphics[width=0.8\textwidth]{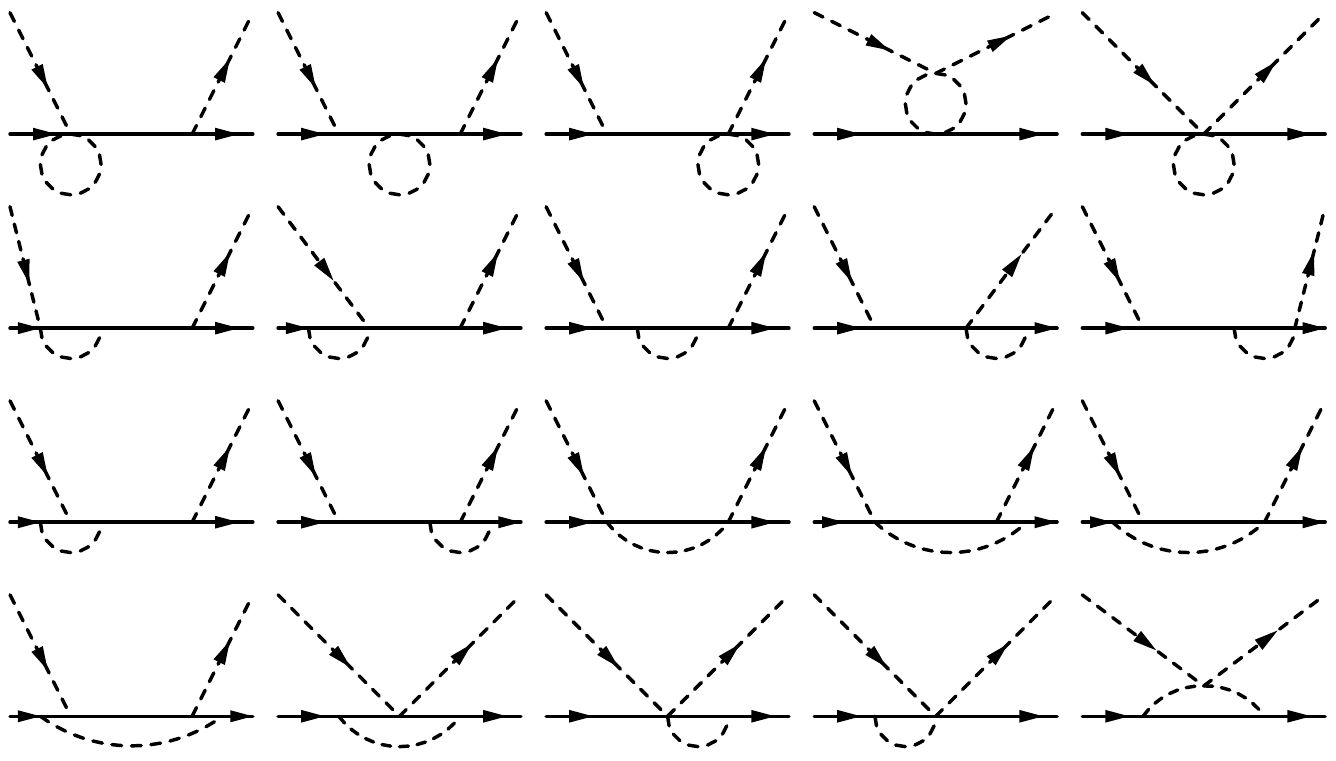}
  \caption{One-loop graphs for the reaction $\pi N\to\pi N$. For notation see Fig.~\ref{fig:TreeGraphs}.}
  \label{fig:LoopGraphsSelfEnergy}
\end{figure}

\begin{figure}[ht]
  \centering
\includegraphics[width=0.65\textwidth]{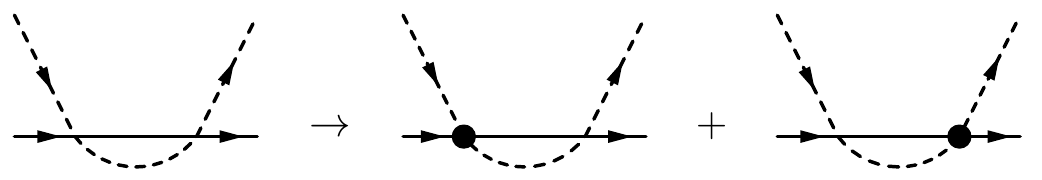}
  \caption{Transition from leading to next-to-leading order loop
    graphs. For notation see Fig.~\ref{fig:TreeGraphs}.}
  \label{fig:LoopRule}
\end{figure}

The relevant tree-level diagrams for $\pi N\to\pi N$ to order $Q^4$
are visualized in
Fig.~\ref{fig:TreeGraphs} while the leading-order loop diagrams are shown
in Fig.~\ref{fig:LoopGraphsSelfEnergy}.
The next-to-leading order loop diagrams are not shown explicitly but
can be easily generated by replacing one of the lowest-order $\pi N$-vertices with an even number of pions
in the shown loop diagrams by a subleading one from $\mathcal{L}_{\pi
  N}^{(2)}$ as visualized in
Fig.~\ref{fig:LoopRule}. Notice that there are no $\pi N $-vertices with an odd number of
pions in $\mathcal{L}_{\pi N}^{(2)}$.

The leading-order tree-level diagrams are constructed solely from
the lowest-order vertices and thus depend only on the well-known LECs
$F_\pi$ and $g_A$. The higher-order tree-level  graphs involve insertions of
vertices with the LECs $c_i$ from $\mathcal{L}_{\pi N}^{(2)}$, $d_i$ from
$\mathcal{L}_{\pi N}^{(3)}$, $e_i$ from $\mathcal{L}_{\pi N}^{(4)}$
and the purely mesonic LECs $l_i$ from $\mathcal{L}_{\pi\pi}^{(4)}$.
Some of the LECs $e_i$  enter the $\pi N$ scattering amplitude only
within 
linear combinations with the LECs $c_i$. In order to get rid of the
redundant LECs, we 
make the following redefinitions on the level of
the renormalized LECs discussed below \cite{Fettes:2000xg}
\begin{equation}
  \label{eq:46}
  \begin{aligned}
    \bar c_1&\to \bar c_1 + 2M_\pi^2(\bar e_{22}-4 \bar e_{38}+ \bar c_1 \beta_{l_3}\bar l_3/(32\pi^2F_\pi^2))\komma\\
    \bar c_2&\to \bar c_2 - 8M_\pi^2(\bar e_{20}+\bar e_{35})\komma\\
    \bar c_3&\to \bar c_3 - 4M_\pi^2(2 \bar e_{19}-\bar e_{22}-\bar e_{36})\komma\\
    \bar c_4&\to \bar c_4 -4M_\pi^2 (2\bar e_{21}-\bar e_{37})\punkt
  \end{aligned}
\end{equation}
This is a general phenomenon in $\chi$PT, namely that working an sufficiently high 
orders, one encounters quark mass renormalizations of certain lower order LECs that
can not be resolved for the physical values of the quark masses.
Finally, the $\pi N$-scattering amplitudes depend on the
LECs $c_{1,2,3,4}$, $d_{1+2,3,5,14-15}$ and
$e_{14,15,16,17,18}$.  This number is consistent with the most general 
polynomial representation of the $\pi N$ scattering amplitude to fourth order,
see e.g. Ref.~\cite{Meissner:1998nh}.

The renormalization of the LECs in the HB formalism can be
performed order-by-order in a complete analogy with the  mesonic
sector, where one has (using dimensional regularization)
\begin{equation}
\begin{aligned}
l_i =   \frac{\beta_{l_i}}{32\pi^2}\bar l_i +
 \beta_{l_i}\left(\bar\lambda+\frac{1}{32\pi^2}\log\left(\frac{M_\pi^2}{\mu^2}\right)\right)
\end{aligned}
\label{eq:14}
\end{equation}
with 
\begin{equation}
  \label{eq:50}
  \begin{aligned}
    \bar \lambda=\frac{\mu^{d-4}}{16\pi^2}\left(
      \frac{1}{d-4}+\frac{1}{2}(\gamma_E-1-\ln 4\pi) \right)\punkt
  \end{aligned}
\end{equation}
The ultraviolet (UV) divergent pieces in the HB scattering amplitude
up to order $Q^4$ are canceled by the counter terms upon expressing the
bare LECs $d_i$ and $e_i$ in terms of the renormalized ones  $\bar
d_i$ and $\bar e_i$ via
\begin{equation}
  \label{eq:10}
  \begin{aligned}
    d_i&= \bar d_i + \frac{\beta_{d_i}}{F_\pi^2}\left(\bar
      \lambda+\frac{1}{32\pi^2}\log\left(\frac{M_\pi^2}{\mu^2}\right)\right)\\
    e_i&= \bar e_i + \frac{\beta_{e_i}}{F_\pi^2}\left(\bar
      \lambda+\frac{1}{32\pi^2}\log\left(\frac{M_\pi^2}{\mu^2}\right)\right)
  \end{aligned}
\end{equation}
where the relevant $\beta$-functions are listed in 
Appendix \ref{sec:renormalization-lecs}. For the LECs $d_i$, the
$\beta$-functions are identical to those of Refs.~\cite{Gasser:2002am},
see also \cite{Fettes:1998ud}. For the LECs $e_i$, we have verified
that the obtained $\beta$-functions are identical to the ones listed
in Ref.~\cite{Becher:1999he} after changing their operator basis to
ours. Note that $c_i=\bar c_i$ in the HB framework.

In the covariant approach, the renormalization of the LECs
is more complicated. After performing dimensional
regularization with the $\overline{\rm MS}$ scheme, 
loop diagrams still contribute at every chiral order which violates
the power counting. The main idea to resolve this issue
is based on the observation that a loop function can be split into an
IR regular and IR singular parts. All power counting breaking terms
(PCBTs) stemming from loop graphs are included in the IR regular part,
which is analytic in the quark mass and momenta in $d$ dimension and thus can be absorbed into LECs of the most general
Lagrangian \cite{Gegelia:1999gf,Fuchs:2003qc}. For our purpose we need to consider the
IR regular parts from the 
loop graphs of order $Q^3$ and $Q^4$ in the naive counting which,
after renormalization of the leading-order couplings $m_N$ and $g_A$,
start to appear at order $Q^2$. Therefore, we perform an
additional finite renormalization of 
the LECs as follows
\begin{equation}
  \label{eq:23}
  \begin{aligned}
    c_i &= \bar c_i +\delta c^{(3)}_i+\delta c^{(4)}_i\\
    d_i &=  \bar d_i +\delta d^{(3)}_i+\delta d^{(4)}_i\\
    e_i &=  \bar e_i +\delta e^{(4)}_i
  \end{aligned}
\end{equation}
where for $x\in\{c,d,e\}$
\begin{equation}
  \begin{aligned}
    \delta x_i^{(n)} &= \frac{\delta\bar x^{(n)}_{i,f}}{F_\pi^2}+
    \frac{\beta^{(n)}_{x_i,B}}{F_\pi^2}\left(\bar
      \lambda+\frac{1}{32\pi^2}\log\left(\frac{m_N^2}{\mu^2}\right)\right)
    + \frac{\beta^{(n)}_{x_i,M}}{F_\pi^2}\left(\bar
      \lambda+\frac{1}{32\pi^2}\log\left(\frac{M_\pi^2}{\mu^2}\right)\right)\\
    &= \frac{\delta\bar x^{(n)}_{i,f}}{F_\pi^2}+
    \frac{\beta^{(n)}_{x_i}}{F_\pi^2}\left(\bar
      \lambda+\frac{1}{32\pi^2}\log\left(\frac{m_N^2}{\mu^2}\right)\right)
    + \frac{\beta_{x_i}}{32 F_\pi^2\pi^2}\log\left(\frac{M_\pi^2}{m_N^2}\right)\punkt
  \end{aligned}
\label{eq:25}
  \end{equation}
Here, $\delta\bar x^{(n)}_{i,f}$ denotes the negative 
of the finite IR regular parts from loops of naive order $n$, while $\beta^{(n)}_{x_i,B}$ and 
$\beta^{(n)}_{x_i,M}$ are the $\beta$-functions which are needed to cancel
the baryonic and mesonic tadpoles, respectively. In order to make the
notation more compact, we made the replacements  
$\beta^{(n)}_{x_i}=\beta^{(n)}_{x_i,B}+\beta^{(n)}_{x_i,M}$
and $\beta_{x_i}=\beta^{(n)}_{x_i,M}$, with $\beta_{x_i}$ from
Eq.~\eqref{eq:10}  in the last line of the above equation.
Note that in Eq.~\eqref{eq:23}, we absorb all IR regular pieces up to
the order we are working at. This procedure
does, strictly speaking, differ from the EOMS approach where only PCBTs are
absorbed into the LECs. In EOMS at order $Q^3$ ($Q^4$), one would only absorb the IR regular
pieces up to the order $Q^2$ ($Q^3$), instead we absorb them up to the
order $Q^3$ ($Q^4$). In addition, we also perform shifts of the LECs proportional
to $\log (M_\pi^2/m_N^2)$, which is not done in EOMS.
This modified version of the  EOMS is employed in
this work to guarantee the equivalence between the results in the HB and covariant approaches up to
the order we are working  with the difference being of higher orders
only. Thus, an expansion of our renormalized covariant amplitudes at
orders $Q^3$ and $Q^4$
in inverse powers of the nucleon mass $m_N$ would give our renormalized HB
amplitudes up to order $Q^3$ and $Q^4$, respectively. Note that our
renormalized amplitudes are equivalent to $\pi N\to\pi N$ amplitudes renormalized in
EOMS. There is no loss of information, just a reshifting of terms from the amplitude to the LECs.
%
%In a deltafull theory, the modified EOMS me employ is also required to
%maintain decoupling of the $\Delta$ degrees of freedom by absorbing all non-negative
%powers and logarithms of the
%$\Delta$ mass into the LECs.

We have determined the finite and UV divergent pieces in the following way.
First, we have changed the  basis for the scattering amplitude such that
every spin structure fullfills the power counting by itself leading to
\cite{Becher:2001hv}
%In $\pi N\to\pi N$ the better suited basis is
\begin{equation}
  \label{eq:26}
    T^\pm=\bar{u}^{(s^\prime)}\left(D^\pm-\frac{1}{4m_N}[\slashed{q}^\prime,\slashed{q}]B^\pm\right)u^{(s)}\punkt
\end{equation}
where $D=A+\nu B$ with $\nu=(s-u)/(4m_N)$. Next, $D$ and $B$ are
expanded in small parameters
\begin{equation}
  \label{eq:27}
  M_\pi\sim \mathcal{O}(Q^1)\komma\quad s-m_N^2\sim \mathcal{O}(Q^1)\komma\quad 
  u-m_N^2\sim \mathcal{O}(Q^1)\komma\quad t\sim \mathcal{O}(Q^2)\punkt
\end{equation}
Note that while the linear combination $s+u-2m_N^2$ counts according
to the above estimations as order-$Q^1$, it actually starts
contributing only at order $Q^2$ due to the cancellation of the
order-$Q^1$ terms, see e.g. Eq.~\eqref{eq:17}. Therefore, for practical reasons, it is advantageous to 
express $D$ and $B$ either in $(s,t)$ or $(u,t)$. Also note that $m_N$
in Eq.~\eqref{eq:27} denotes the physical nucleon mass, whereas the
expansion in the EOMS scheme is, strictly speaking, 
around the nucleon mass in the chiral limit $\mathring{m}_N$.
The difference is of the order of $m_N-\mathring{m}_N\sim \mathcal{O}(Q^2)$ and
is thus affecting the shifts at chiral order $Q^4$. However, due to
our choice to work with the shifted LECs $c_i$, see Eq.~\eqref{eq:46}, 
this amounts merely to a
reshuffling of the terms between the $c_i$ and $e_i$ and does not
affect the final results.

The pertinent $\beta$-functions can be calculated by substituting every
loop function by its UV divergent part and expanding the result in small 
parameters. The determination of the finite IR regular pieces is more demanding. 
It requires the substitution of the loop functions by their IR regular parts.  
This has been achieved by interchanging the loop integration with a Taylor
series in powers of the small parameters.

Several checks on the $\pi N\to\pi N$ amplitudes have been  performed. 
The renormalization of $m_N$, $Z_N$ and $g_A$ was checked by setting
the internal nucleon line in the covariant (heavy baryon) $\pi N$ amplitudes on-shell.
An expansion around $s=m_N^2$ or $u=m_N^2$ corresponding to vanishing
pion energy in the center-of-mass system (CMS), $\omega=0$, showed that
only the leading order diagrams exhibit poles and thus
giving the right analytic structure of the amplitudes. Using the
redefined LECs from Appendix \ref{sec:renormalization-lecs}, the
$\pi N\to\pi N$ amplitudes fullfil power-counting and are UV-finite up to order $Q^3$
and $Q^4$, respectively. Another consistency check was done by using
the same renormalization shifts in the amplitudes of the reaction $\pi
N\to\pi\pi N$, whose analysis will be published elsewhere, 
and verifying the power counting and UV-finiteness by redefining
only the new LECs appearing in $\pi N\to\pi\pi N$. In Appendix \ref{sec:renormalization-lecs}
we list all LECs appearing in both reactions. The pion field was
defined in the most general form given by unitarity
\begin{equation}
  \label{eq:4}
U = 1 + \i \frac{\vec{\tau} \cdot \vec{\pi}}{F_\pi} - \frac{\vec{\pi}^2}{2 F_\pi^2} - \i \alpha  \frac{
\vec{  \pi}^2 \vec{\tau} \cdot \vec{\pi} }{F_\pi^3} 
+ \frac{(8 \alpha - 1)}{8 F_\pi^4} \vec{\pi}^4 + \ldots \,,
\end{equation}
and it was checked that the final renormalized amplitudes are
independent of  the parameter $\alpha$. We checked our amplitudes by comparing them 
with the results of Ref.~\cite{Chen:2012nx}. Notice that the
expressions published in that reference contain some typos. We,
however, were able to reproduce their results by comparing the explicit expressions 
in a Mathematica notebook with the ones provided by one of the authors of Ref.~\cite{Chen:2012nx}. 
To avoid the same problems with typing rather lengthy expressions, we prefer
to provide the amplitudes in a Mathematica notebook upon request.

Finally, we emphasize that we take the isospin limit in all our amplitudes,
i.e.~we take $m_p=m_n=m_N$ and $M_{\pi^\pm}=M_{\pi^0}=M_{\pi}$.
The electromagnetic corrections of Ref.~\cite{Tromborg:1976bi}
employed in our analysis do, of course, 
take into account some of the isospin-breaking effects. However, it is also
clear that this procedure does not include all possible isospin violating effects.
For a fully consistent calculation including all such effects for the $\pi N$ scattering lengths,
see e.g. Ref.~\cite{Hoferichter:2009ez}.

\section{Fitting Procedure}
\label{sec:fitting-procedure}
The amplitudes for the reaction $\pi N\to\pi N$ depend on several LECs as explained in 
section~\ref{sec:pow-count-renorm}. Throughout this work, we use the following
values for the various LECs and masses entering the leading order
effective Lagrangian: $M_\pi = 139.57$ MeV, $F_\pi =92.2$ MeV, 
$m_N= 938.27$ MeV \cite{Agashe:2014kda}. All LECs should be
understood as renormalized quantities as discussed in the previous section.
For convenience, we will suppress in the following the bars on the
renormalized LECs $\bar c_i$, $\bar d_i$ and $\bar e_i$, which values are always given in
units of GeV$^{-1}$, GeV$^{-2}$ and GeV$^{-3}$, respectively.

All fits described below are performed to $\pi N\to\pi N$ scattering data
$\di\sigma / \di\Omega$, $P$ in all three channels simultaneously. In
this least squares fit we minimize the quantity 
\begin{equation}
  \label{eq:31}
  \begin{aligned}
    \chi^2 =\sum_i\left( \frac{\mathcal{O}^{exp}_{i}-N_i
      \mathcal{O}^{(n)}_{i}}{\delta \mathcal{O}_i}\right)^2\qquad
  \mathrm{with} \qquad \delta\mathcal{O}_i=\sqrt{(\delta\mathcal{O}^{exp}_i)^2+
(\delta\mathcal{O}^{(n)}_i)^2}\komma
  \end{aligned}
\end{equation}
where $\mathcal{O}^{exp}_i$, $\delta\mathcal{O}^{exp}_i$ and $N_i$ are
taken from the GWU-SAID data base 
\cite{Workman:2012hx} and  $\mathcal{O}^{(n)}_i$ denotes the
observable calculated in $\chi$PT up to order $n$.
The theoretical error takes into account the uncertainty from the
truncation of the chiral expansion at a given order and is estimated in the way proposed in
Ref.~\cite{Epelbaum:2014efa}, namely
\begin{equation}
  \label{eq:5}
  \delta\mathcal{O}^{(n)}_i =\max( |\mathcal{O}^{(\mathrm{LO})}_i| Q^{n-\mathrm{LO}+1} ,
\{|\mathcal{O}^{(k)}_i-\mathcal{O}^{(j)}_i| Q^{n-j} \})\qquad
\mathrm{with}\qquad j<k\leq n
\end{equation}
and $Q=\omega_{CMS}/\Lambda_b$, where $\omega_{CMS}$ denotes the energy
of the incoming pion in the CMS frame. Further, LO refers to the
chiral order, at which  the
observable $\mathcal{O}_i$ appears receives its first nonvanishing contribution. 
In the Goldstone boson and single-baryon sectors, the
breakdown scale of the chiral expansion is often assummed to be of the order of $\Lambda_b
\sim \Lambda_\chi \sim M_\rho \sim 4 \pi F_\pi \sim 1$~GeV. On the
other hand, a somewhat more conservative estimation of $\Lambda_b \sim
600$~MeV was obtained and employed in a recent study of nucleon-nucleon
scattering in Ref.~\cite{Epelbaum:2014efa}. It was also verified in an analysis
of Ref.~\cite{Furnstahl:2015rha} utilizing the Bayesian approach. Here
and in what follows, we adopt the more conservative estimate of $\Lambda_b \sim
600$~MeV which seems to be justified given the \emph{implicit} inclusion
of  the Roper resonance in our calculations. 
In addition to Eq.~(\ref{eq:5}), the theoretical errors is required to be at least of the size of actual higher-order
contribution
\begin{equation}
  \label{eq:19}
    \delta\mathcal{O}^{(n)}_i \geq\max( 
\{|\mathcal{O}^{(k)}_i-\mathcal{O}^{(j)}_i|\})\qquad
\mathrm{with}\qquad n\leq j<k\punkt
\end{equation}
Both Eqs.~\eqref{eq:5} and \eqref{eq:19} are implemented in the fits
using an iterative procedure.\footnote{As a starting point in this
  iterative procedure, we performed fits without theoretical errors.} 

To give a meaningful uncertainty quantification for other observables
we define the correlation and covariance matrices as follows
\begin{equation}
  \label{eq:8}
  \begin{aligned}
    \mathrm{Cov}(c_i c_j) &= \mathbf{H}^{-1}_{ij}\quad \mathrm{with}
    \quad \mathbf{H}_{ij}=\left.\frac{1}{2}\frac{\partial^2\chi^2}{\partial
    c_i\partial c_j}\right|_{\vec{c}=\vec{c}_*}\komma\\
    \mathrm{Corr}(c_i c_j) &= \mathrm{Cov} (c_i c_j)/\sqrt{\mathrm{Cov} (c_i c_i) \mathrm{Cov}(c_j c_j)}\komma
  \end{aligned}
\end{equation}
where $\vec{c}$ is a set of LECs and $\vec{c}_*$ is the set
which minimizes $\chi^2$. The correlation and covariance matrices
for the fits discussed above are given in Tables~\ref{tab:Q3piNCorrCov} and \ref{tab:Q4piNCorrCov}.
Note the correlations at order $Q^4$ between $c_1$ and $c_2$ and the additional
correlations in the HB countings between $c_2$ and $e_{16}$ and
between $c_4$ and $d_{1+2}$.

\section{Fit results, predictions and discussion}
\label{sec:predictions}

\begin{figure}[ht]
  \centering
\includegraphics[width=0.85\textwidth]{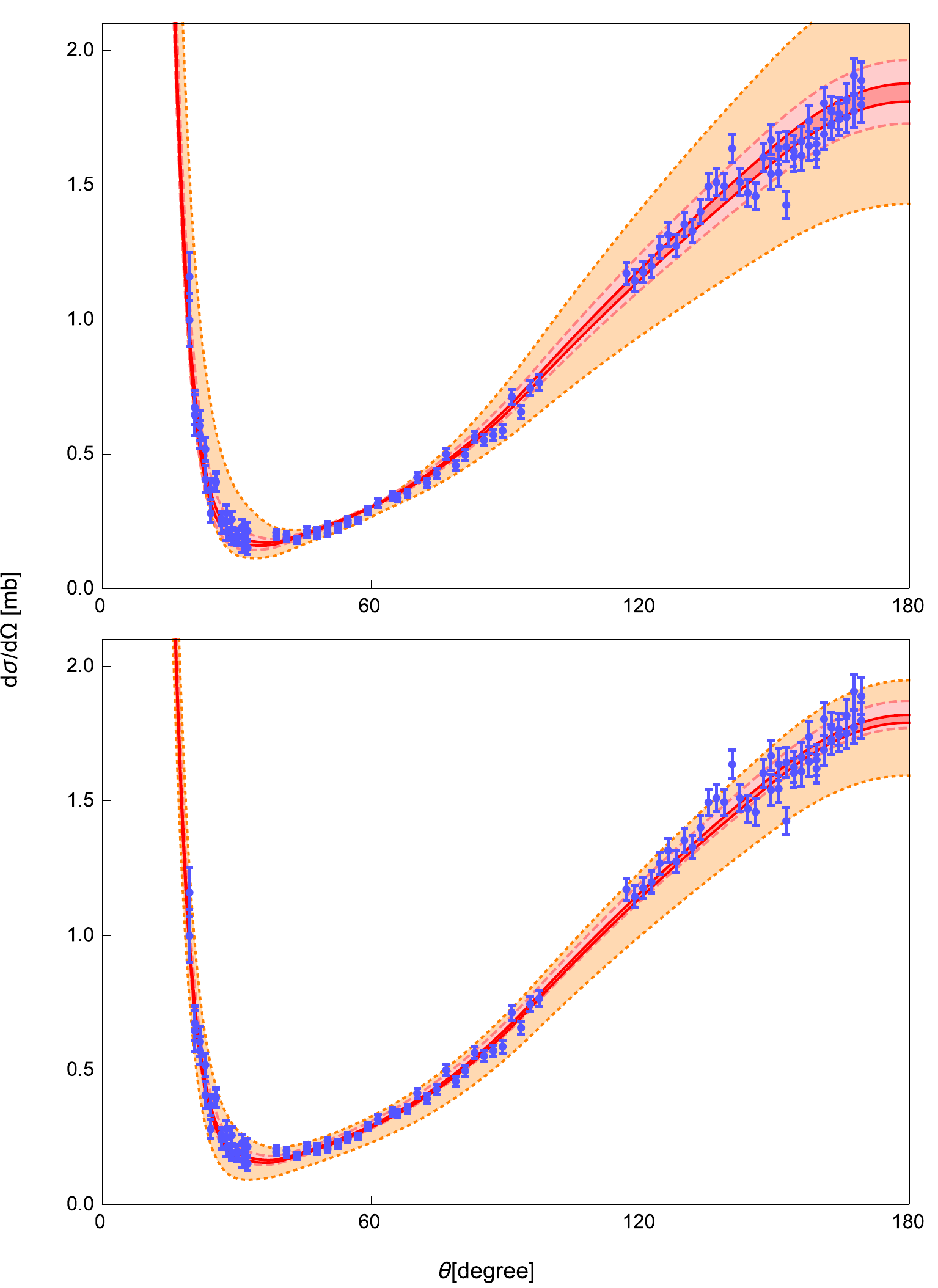}
  \caption{$\pi^+
    p\to\pi^+ p$ differential cross section at $T_\pi=43.3$ MeV as a representative example of the quality of our
    fits (carried out  to all available data for $T_\pi<100$ MeV).   In the upper panel, the orange, pink and red
    (dotted, dashed  and solid) bands refer to $Q^2$, $Q^3$ and $Q^4$
    results in the covariant approach
  including theoretical uncertainties,
  respectively. In the lower panel the orange, pink and red (dotted,
  dashed  and solid) bands refer to $Q^2+\delta^1$, $Q^3+\delta^1$ and
  $Q^4+\delta^1$ results in the covariant approach
  including theoretical uncertainties,
  respectively. Experimental data of Ref.~\cite{Denz:2005jq} are taken from the GWU-SAID data base 
\cite{Workman:2012hx}. }
  \label{fig:SigmaDiff}
\end{figure}

\begin{figure}[ht]
  \centering
\includegraphics[width=\textwidth]{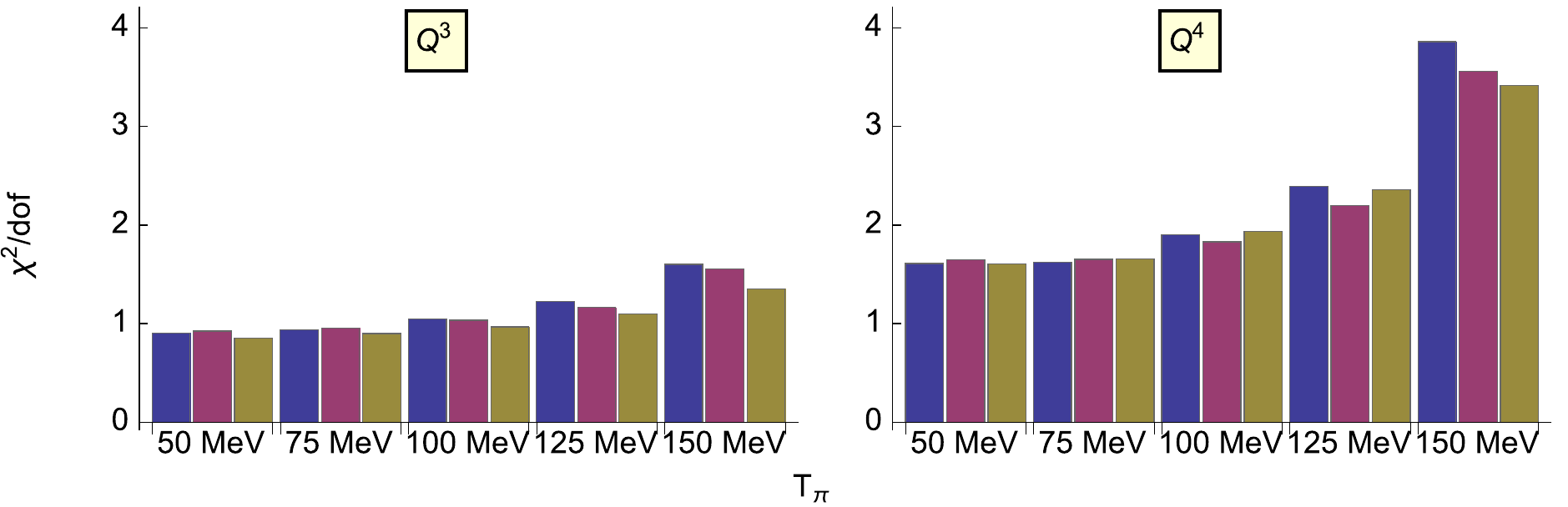}
\includegraphics[width=\textwidth]{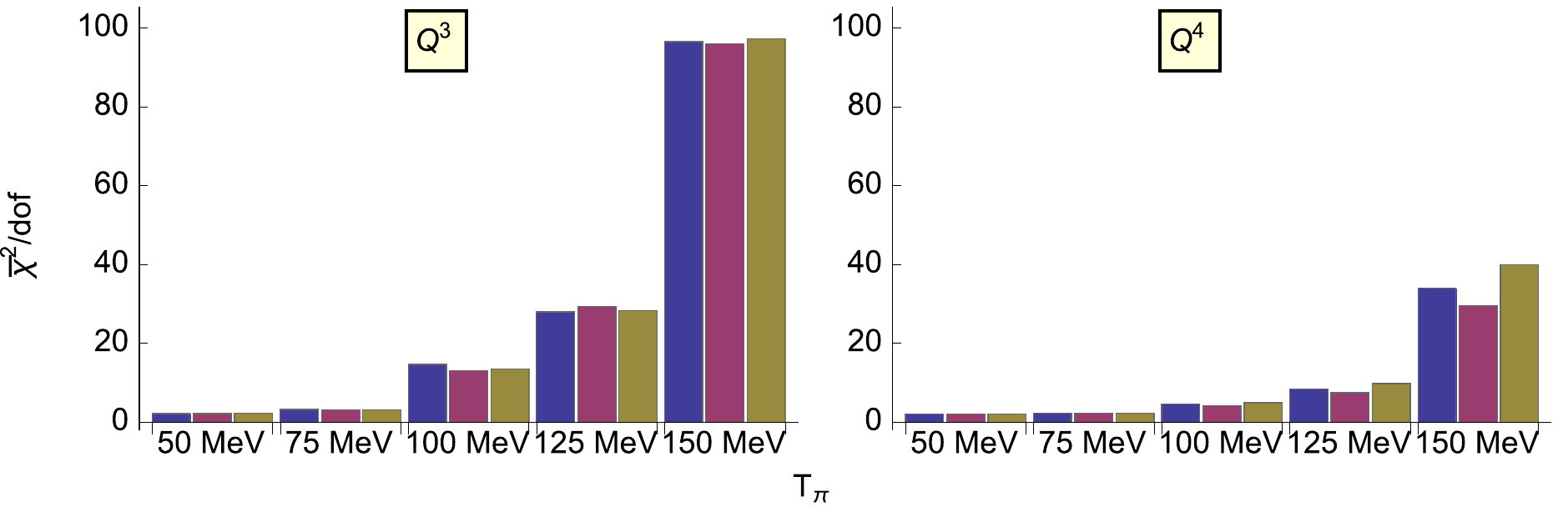}
\includegraphics[width=\textwidth]{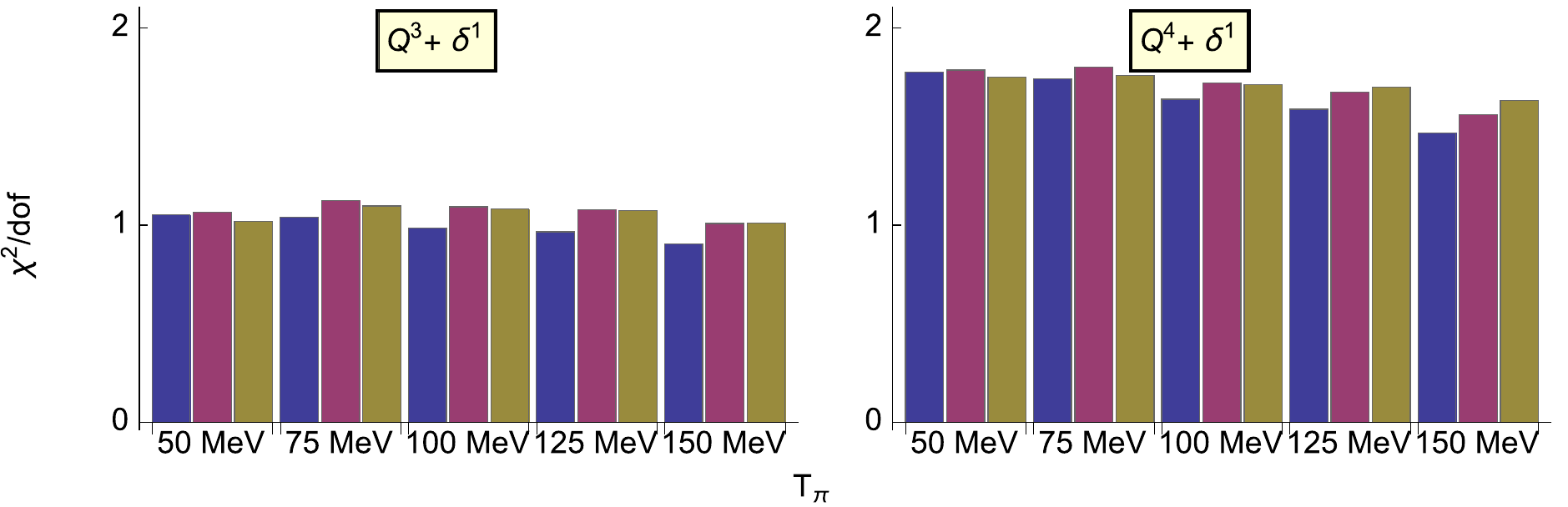}
\includegraphics[width=\textwidth]{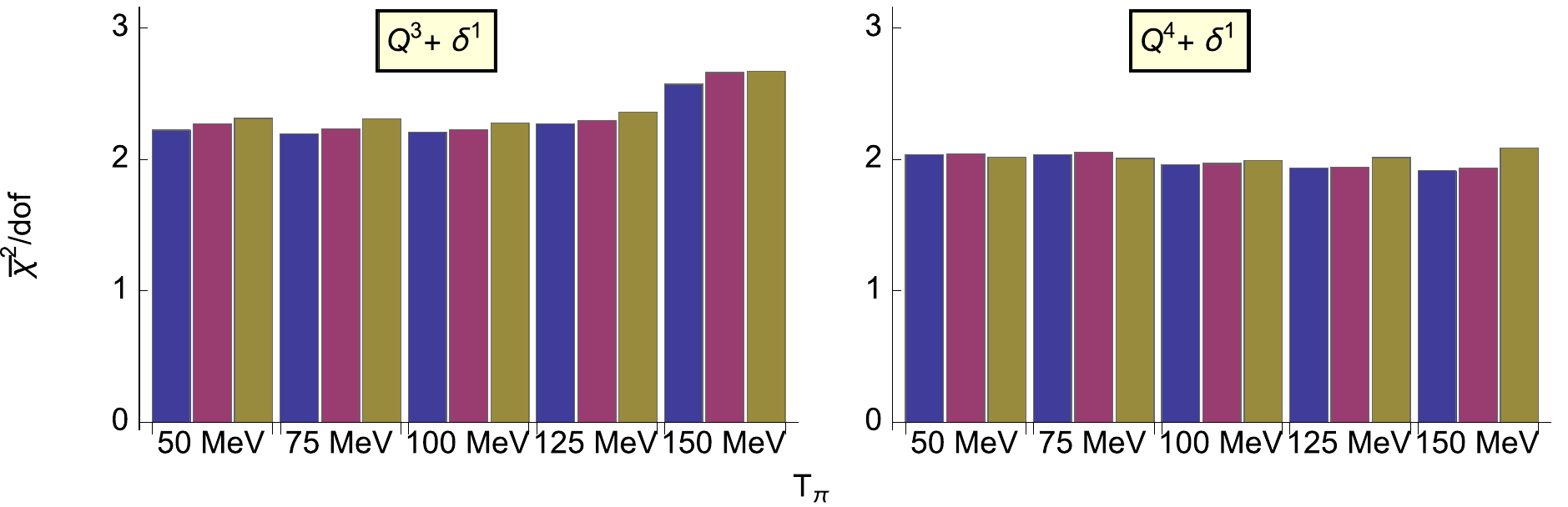}
  \caption{Reduced $\chi^2$ (with theoretical error) and $\bar\chi^2$
    (without theoretical error) for fits up to various maximum energy
    $T_\pi$. The blue/red/green bars denote the results for the HB-NN/HB-$\pi$N/Cov counting.}
  \label{fig:RedChiSq}
\end{figure}

\begin{figure}[ht]
  \centering
\includegraphics[width=0.7\textwidth]{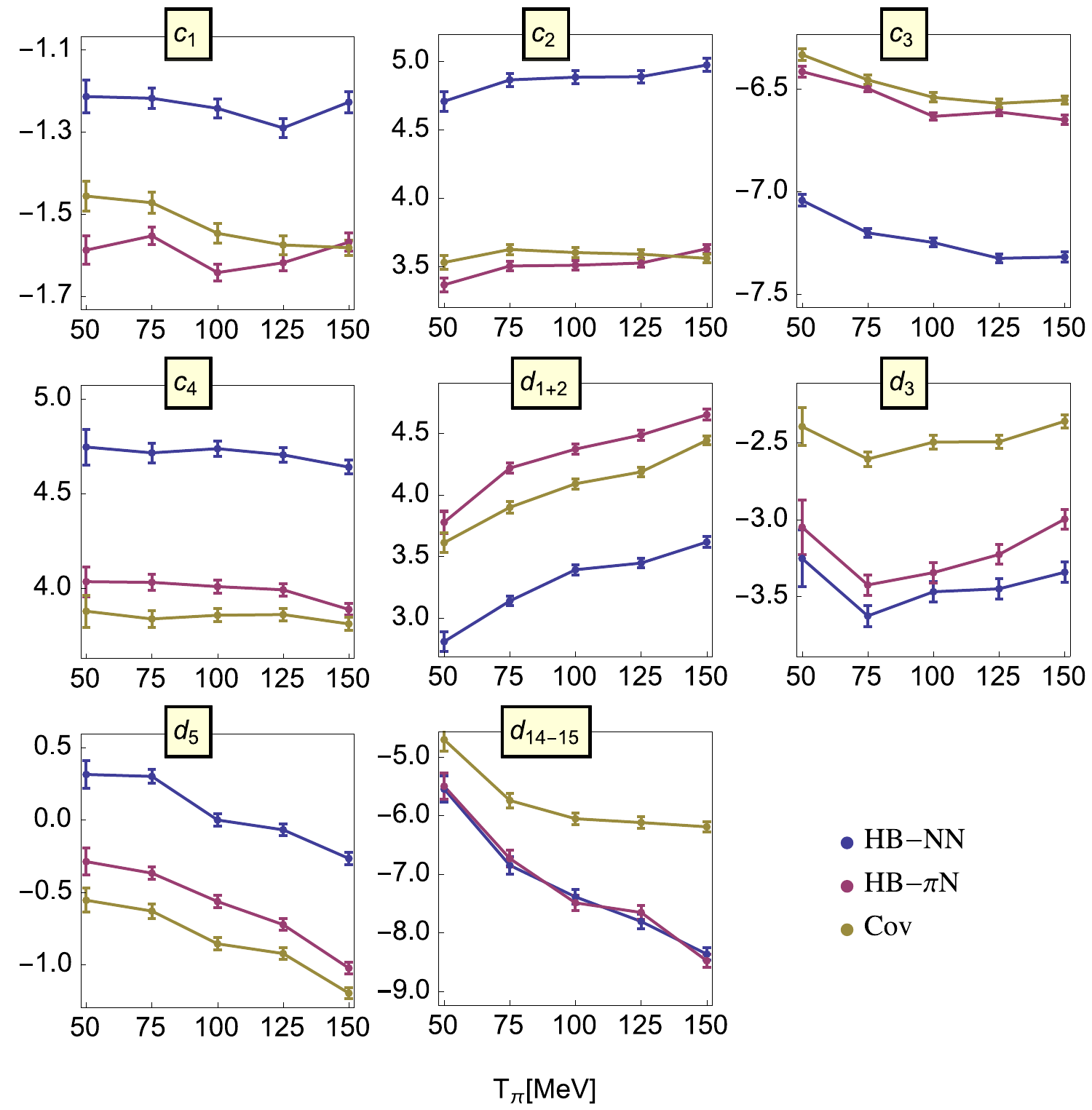}
  \caption{Change of LECs at $Q^3$ over maximum fit energy $T_\pi$.}
  \label{fig:LECsQ3}
\end{figure}

\begin{figure}[ht]
  \centering
\includegraphics[width=0.7\textwidth]{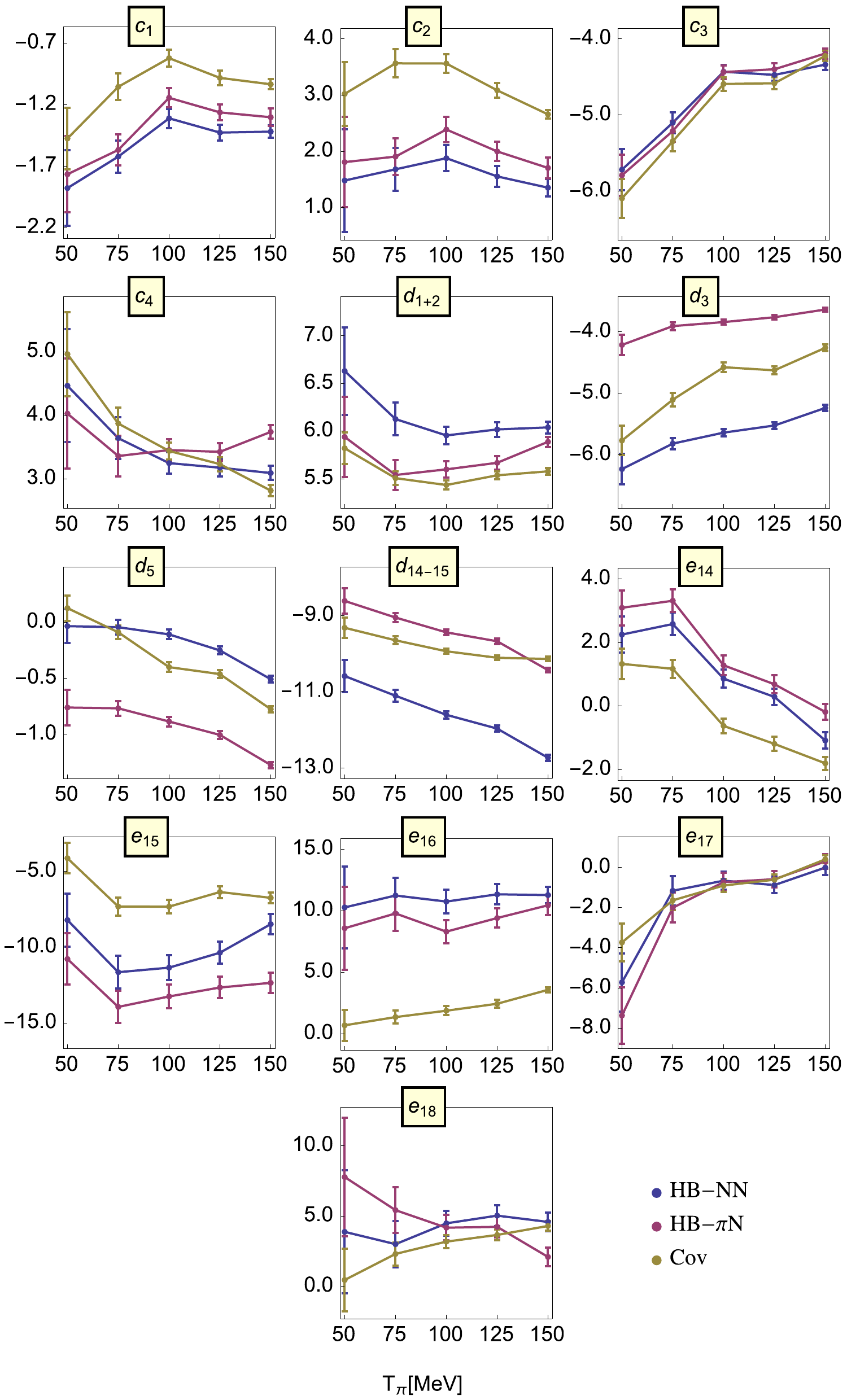}
  \caption{Change of LECs at $Q^4$ over maximum fit energy $T_\pi$.}
  \label{fig:LECsQ4}
\end{figure}

We performed fits to all available data for all scattering angles
and an incoming pion kinetic energy
$T_\pi<\{50, 75, 100, 125, 150\}$~MeV, which corresponds to $\{ 1035,
1368 ,1704, 1854 , 2176 \}$ data points, respectively. 
In the upper panel of Fig.~\ref{fig:SigmaDiff}, we show a representative fit to $\di\sigma/\di\Omega$ for the channel
$\pi^+p\to\pi^+p$ at $T_\pi=43.3$ MeV. A precise definition of the
uncertainty bands will be given below. 
The fitted LECs as a function
of the maximal fitting energy $T_\pi$ are shown in Figs.~\ref{fig:LECsQ3}
and \ref{fig:LECsQ4} while the reduced $\chi^2$ ($\bar\chi^2$) with (without)
theoretical errors as a function of $T_\pi$ is plotted in Fig.~\ref{fig:RedChiSq}.
As can be seen in the figures, most of the fitted LECs exhibit a
plateau-like behavior for the maximal fitting energy in the range between $75$~MeV
and $125$~MeV yielding, at the same 
time, a reasonable reduced $\chi^2$ close to $1$. On the other hand, the 
$\chi^2/$dof starts increasing when experimental data at higher
energies are included in contradiction with an expected flat
behavior. This feature is also reflected in the deviation of the 
LECs viewed as functions of $T_\pi$ from a plateau-like behavior when
higher-energy data are included in the fit as
visualized in Figs.~\ref{fig:LECsQ3} and \ref{fig:LECsQ4}.  The
observed instability of the fits at higher energies provide a clear
indication that the actual theoretical uncertainty is larger than the
one estimated as described in the previous section. As will be shown
below, the slow convergence pattern of the chiral expansion is
caused by the $\Delta$(1232) resonance which is not explicitly
included in the considered formulations of $\chi$PT. 

The extracted values of the LECs at orders 
$Q^2$, $Q^3$, $Q^4$ are listed in Table~\ref{tab:Fit} for all
considered approaches along with the corresponding values of the
reduced $\chi^2$ and $\bar\chi^2$.
For the sake of compactness, we restrict ourselves here and in what
follows to the fits with $T_\pi<100$~MeV which can be regarded as
representative examples.  As expected, the value of 
$\bar\chi^2/$dof decreases with an increasing chiral order showing
the improved description of the data. 
%On the other hand, $\chi^2/$dof
%shows the opposite behavior indicating again the underestimated
%theoretical uncertainty. 
Notice further that all considered approaches
lead to a similar quality of the fits. The extracted values of the
LECs do not show a strong dependence on the counting scheme except
for some of the $e_i$'s at order $Q^4$ and are generally in a reasonably 
good agreement with the values reported in the literature. 
Specifically, except for the value of $c_2$, the LECs $c_i$ and $d_i$
extracted at order $Q^4$ in the HB-NN approach
are compatible with the ones determined in Ref.~\cite{Krebs:2012yv} from  the KH
and GW-SAID PWA if the spread between the results based  on the two
different PWA is interpreted as the uncertainty. The large differences
in the values of the LECs $c_2$ and $e_{16}$ are naturally explained
by the very strong correlation between these LECs, see Table~\ref{tab:Q4piNCorrCov}. On the other hand, it is comforting to see
that the LECs $e_{14}$ and $e_{17}$ which enter the order-$Q^5$
contribution to the three-nucleon force \cite{Krebs:2012yv} are rather stable. 
Similar conclusions apply to a comparison with the recent
determination of the LECs from the subthreshold coefficients obtained
in the RS analysis \cite{Hoferichter:2015tha}, although the differences between the LECs
generally appear to be somewhat larger.  In any case, the sizable (large) shifts in
the LECs $c_i$ ($d_i$) extracted at different orders in the chiral expansion
indicate that the uncertainties in their values are presently
dominated by the truncation of the chiral expansion. 

We are now in the position to discuss predictions of other observables
not used in the fits. 
%It is important to employ exactly the same fitting protocol for the various
%approaches in order to allow for meaningful comparison. 
Here and in what follows, we will use the values of the 
LECs collected in Table~\ref{tab:Fit}. All predictions are
supplemented with an estimated uncertainty which includes both the statistical and theoretical errors.
Here and in what follows, the error associated with the uncertainty in the values of the LECs
determined by the fitting procedure specified in the previous section  
will be referred to as statistical. 
%error for the fitting procedure specified in the previous section 
It is calculated via 
\begin{equation}
  \label{eq:11}
  (\delta\mathcal{O}^{stat}_i)^2=\vec{J}^T
  \mathbf{H}^{-1}\vec{J}\quad\mathrm{with}\quad J_j=\left.\frac{\partial
    \mathcal{O}_i}{\partial c_j}\right|_{c=c_*}\komma
\end{equation}
whereas the theoretical uncertainty from the
truncation of the chiral expansion is estimated using
Eqs.~(\ref{eq:5}) and (\ref{eq:19}) using the central values of the LECs
determined in a corresponding fit.

\begin{figure}[ht]
\vspace{-1cm}
  \centering
  \includegraphics[width=0.65\textwidth]{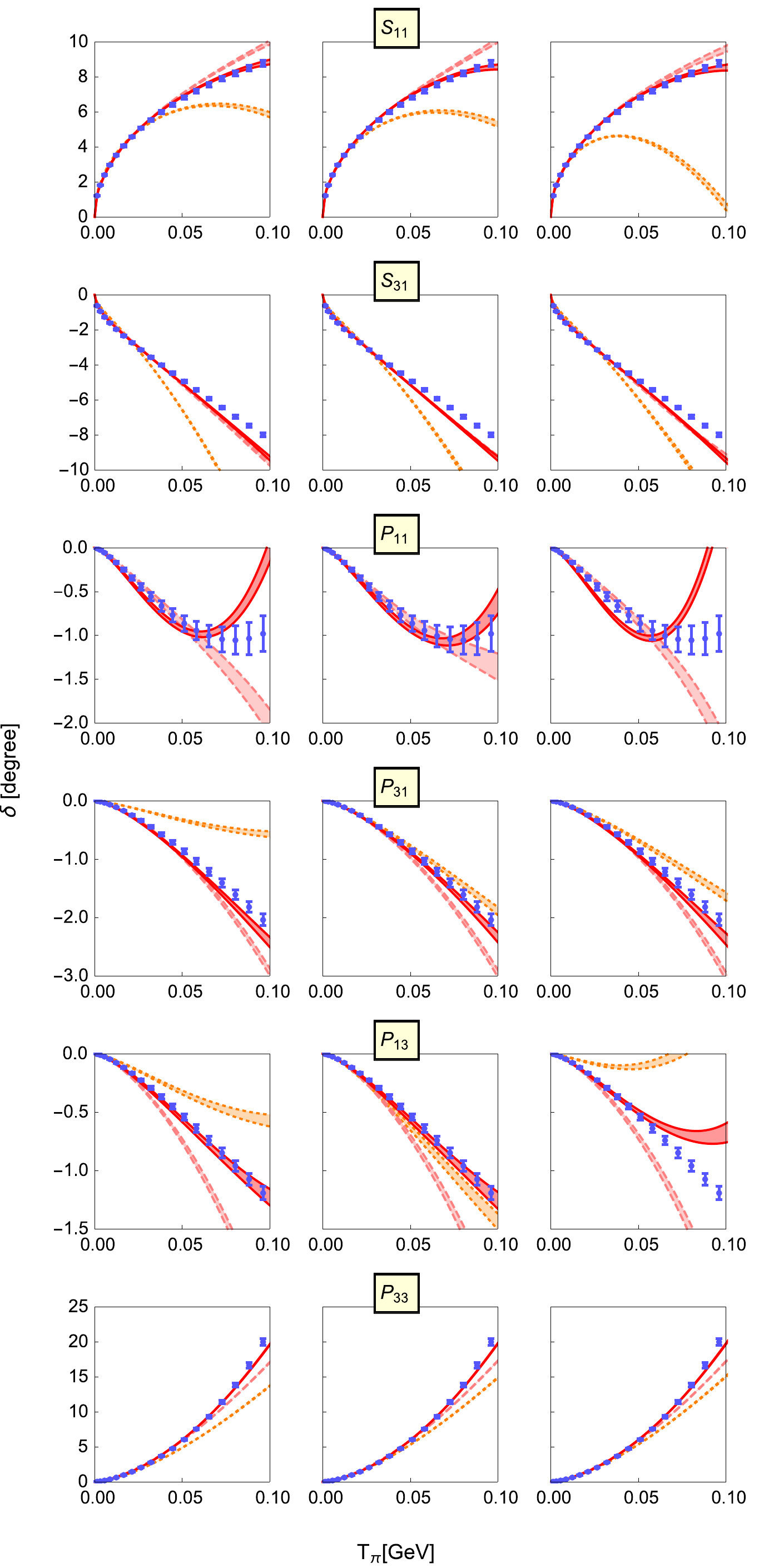}
\caption{(Color online) Predictions for $S$ waves up to
  $T_\pi=100$ MeV. Columns from left to right corresponds to the
  to the predictions in the HB-NN, HB-$\pi$N and Covariant counting, respectively.
  The orange, pink and red (dotted, dashed  and solid) bands refer to $Q^2$, $Q^3$ and $Q^4$ results
  including statistical uncertainties,
  respectively. }
\label{fig:SnPwavesStat}
\end{figure}

\begin{figure}[ht]
\vspace{-1cm}
  \centering
  \includegraphics[width=0.65\textwidth]{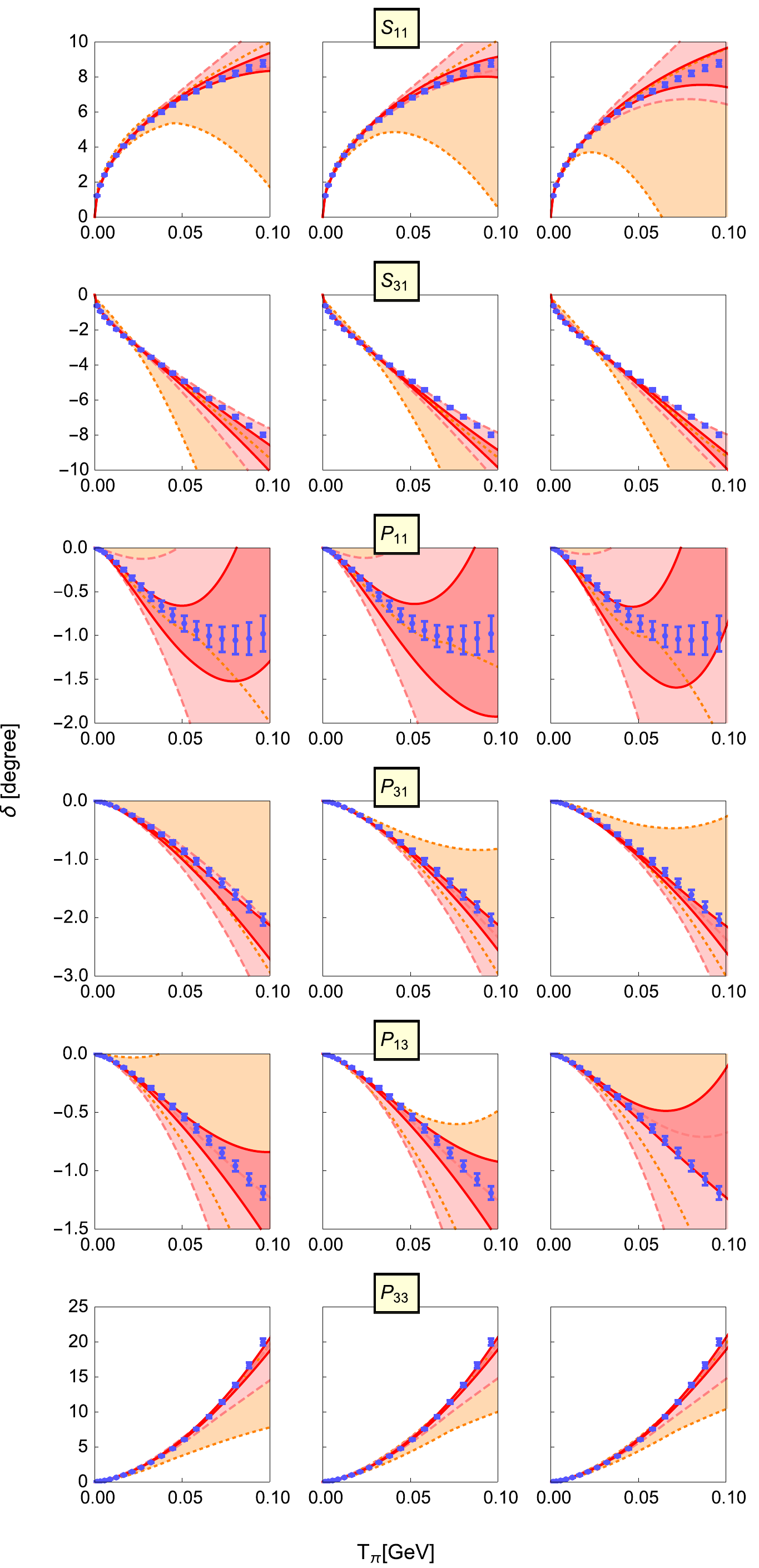}
\caption{(Color online) Predictions for $S$ waves up to
  $T_\pi=100$ MeV. Columns from left to right corresponds to the
  to the predictions in the HB-NN, HB-$\pi$N and Covariant counting,
  respectively.
  The orange, pink and red (dotted, dashed and solid) bands refer to $Q^2$, $Q^3$ and $Q^4$ results
  including theoretical uncertainties,
  respectively. }
\label{fig:SnPwaves}
\end{figure}

\begin{figure}[ht]
\vspace{-1cm}
  \centering
  \includegraphics[width=0.65\textwidth]{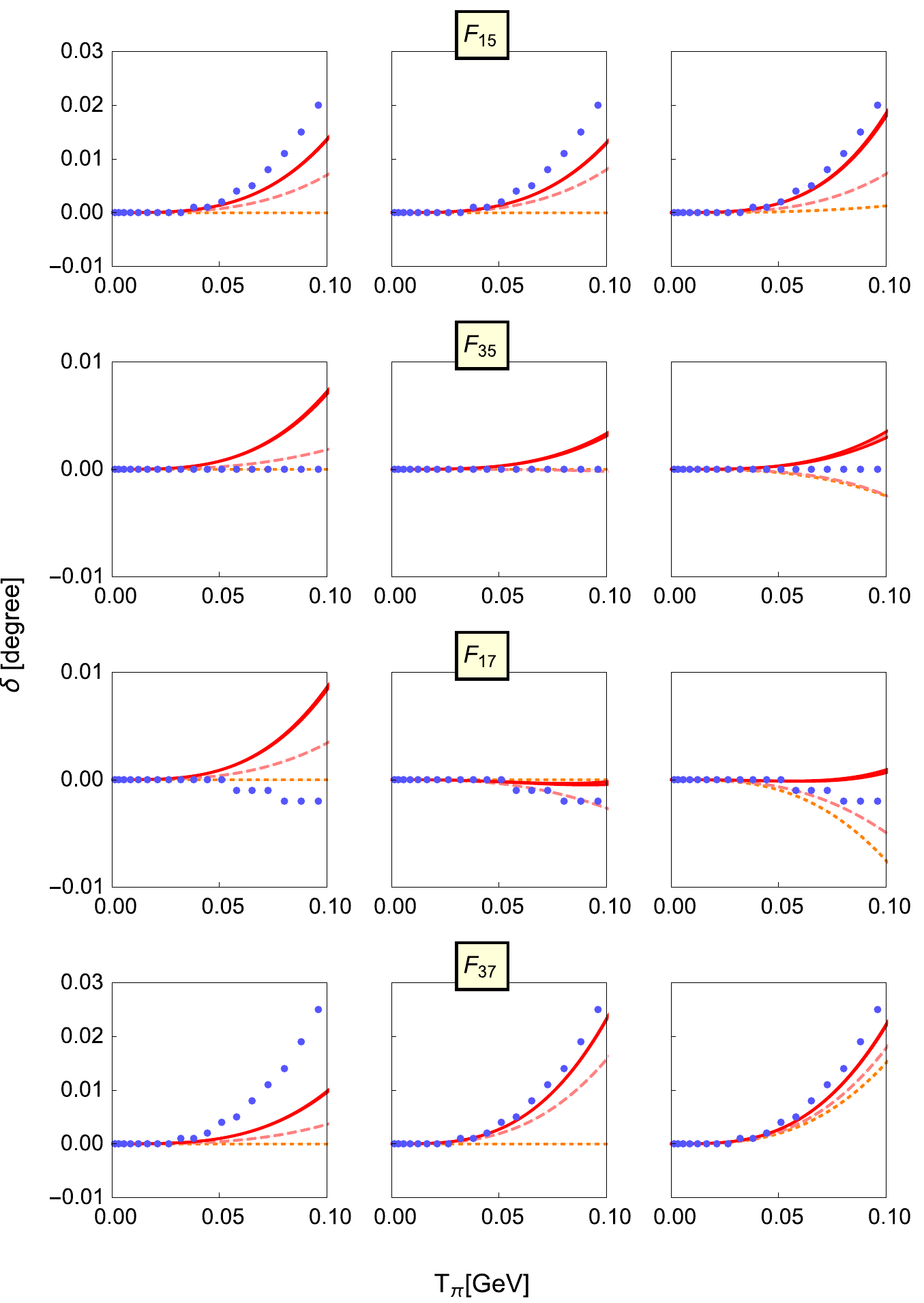}
\caption{(Color online) Predictions including statistical uncertainties for $F$ waves up to
  $T_\pi=100$~MeV. For remaining notation see Fig.~\ref{fig:SnPwavesStat}.}
\label{fig:FwavesStat}
\end{figure}

The predicted phase shifts in the $S$, $P$, $D$ and $F$ partial waves with
pion energies up to $100\,$MeV at orders $Q^2$, $Q^3$ and $Q^4$ are shown 
in Figs.~\ref{fig:SnPwavesStat} - \ref{fig:Fwaves} for all three considered
formulations of $\chi$PT in comparison with the phase shifts from the
RS results of Ref.~\cite{Hoferichter:2015dsa} for  $S$- and $P$-waves
and with the GWU-SAID solution \cite{Arndt:2006bf,Igor}  for $D$- and $F$-waves. 
% for the deltaless
%theory and in Figs.~\ref{fig:SnPwavesStatD} - \ref{fig:FwavesD} for
%the deltafull theory. In the deltaless case, 
Given that for predictions we use the same definition of the
theoretical error as employed in the fits, the statistical and theoretical
uncertainties for a predicted quantity are not
really independent from each other and it is not clear to us how to
combine them in a meaningful way. For this reason, we will show in the
following both kinds of uncertainties separately. The extracted phase
shifts in the $S$
and $P$ waves shown in Figs.~\ref{fig:SnPwavesStat} and \ref{fig:SnPwaves}
agree with the RS results for energies up to $T_\pi<70$ MeV. 
For energies above $70$~MeV, the difference between the $Q^3$ and $Q^4$
predictions increases which results in rather large theoretical
uncertainties. This applies especially to the $P_{11}$ partial wave
which is not surprising given the smallness of the corresponding phase
shift. On the other hand, the statistical uncertainties appear to be
negligibly small for the $S$ and  $P$ waves. One also observes that 
all considered formulations lead to nearly identical results for these
phase shifts which is consistent with the similar values of $\bar
\chi^2_{\pi N}/$dof, see Table~\ref{tab:Fit}.

The situation is rather different for the $D$ waves which are shown in Figs.~\ref{fig:DwavesStat} and
\ref{fig:Dwaves} in comparison with the results of the GWU-SAID
partial wave analysis. Note that the GWU-SAID PWA does not provide an
uncertainty for their phase shifts so that a comparison with our
predictions should be taken with care. Similarly to the $S$- and
$P$-waves, one observes large shifts between the order-$Q^3$ and $Q^4$
predictions which result in a very large theoretical uncertainty at
order $Q^3$. Statistical errors appear to be completely negligible at
this order. At order $Q^3$, our predictions are consistent with the 
GWU-SAID PWA (within the very large theoretical
uncertainties). However, at the highest considered order  $Q^4$, our results
do show significant disagreements with the GWU-SAID PWA
especially in the $D_{35}$ partial wave for the HB-NN counting and 
 $D_{33}$ and $D_{15}$ partial waves for the covariant approach, see
 Fig.~\ref{fig:Dwaves}. We, however, emphasize that the statistical uncertainty
 is not negligible anymore at this order. It stems mainly 
from correlations, see Table~\ref{tab:Q4piNCorrCov}, as well as from
the relatively large uncertainty in the determined values of the LECs $e_i$. 
Our predictions for $F$-waves are visualized in Figs.~\ref{fig:FwavesStat} and \ref{fig:Fwaves} and show a better agreement
with the GWU-SAID PWA except for the HB-NN scheme.

We also show in Table~\ref{tab:SubThrPara} the predictions 
for the threshold and subtreshold parameters in comparison with the values from the RS-analysis
\cite{Hoferichter:2015tha}, see also 
Refs.~\cite{Baru:2010xn,Baru:2011bw}. For the subthreshold and threshold region we
used $Q=M_\pi/\Lambda_b$ as expansion parameter in the theoretical
error in Eq. \eqref{eq:5}. We calculated the subthreshold
parameters and scattering lengths in all three counting schemes. We
reproduced the analytic expressions in Ref.~\cite{Becher:2001hv} for the HB formulation. 
The covariant expressions are lengthy and can be provided upon request.
%The deltapredictions in the HB framework are troublesome. 
While the predictions at $Q^2$, $Q^3$  are mostly in 
agreement with the empirical values within uncertainties, the
results at order $Q^4$ do exhibit significant discrepancies in many
cases. Furthermore, the $Q^4$ results show often no improvement
compared with the $Q^3$
ones.  
%Both HB approaches fail to reproduce the RS
%values, whereas the $Q^4$ results often look worse than the $Q^3$
%ones.   

The above findings within the HB-NN, HB-$\pi$N and the covariant
approaches appear to be not quite satisfactory in the following respects: 
%To summarize the above findings, the observed pattern in our
%results such as especially: 
\begin{itemize}
\item
The resulting $\chi^2/$dof is found to increase if scattering data at energies
above $T_\pi \sim 100$~MeV are included in the fits in contradiction
with the
expected nearly constant behavior.
\item
There are deviations from a plateau-behavior for the extracted LECs as a
function of the maximal fitting energy which indicates that the fits
become unstable if the energy is increased.
\item
One observes large disagreements between the predicted $D$-wave phase shifts and
the results of the GWU-SAID PWA at order $Q^4$.
\item
Large deviations are observed for some of the predicted subthreshold
coefficients at order $Q^4$. 
\end{itemize}
These inconsistencies indicate that the actual breakdown scale of the
chiral expansion in our calculations is smaller than the assumed $\Lambda_\chi \simeq
600$~MeV and, as a result, that  the theoretical uncertainty has been
underestimated. Given that the results are similar for all
considered approaches, there is no indication that the slow convergence of the chiral expansion
is to be attributed to the treatment of relativistic
corrections.  Clearly, the most natural explanation of the observed
pattern is provided by the $\Delta$(1232) resonance, which has a
low excitation energy with $m_\Delta - m_N \simeq 2 M_\pi$ and 
couples strongly to the $\pi N$ system \cite{Jenkins:1991es}. To validate this hypothesis, 
we redo our analysis in the next section with the leading-order contributions of the 
$\Delta$-resonance being included \emph{explicitly}. 

\section{The explicit inclusion of the lowest-order $\Delta$(1232) contributions}
\label{sec:delta}

\begin{figure}[ht]
  \centering
\includegraphics[width=0.17\textwidth]{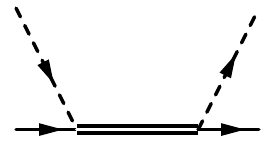}
  \caption{Leading-order $\Delta$ pole diagram. The double solid line
    refers to $\Delta$. For notation see Fig.~\ref{fig:TreeGraphs}.}
  \label{fig:DeltaPole}
\end{figure}

To quantify the importance of the $\Delta$(1232) resonance for the description
of $\pi N$ scattering at low energy,  we include the leading-order
$\Delta$ pole diagrams ($\delta^1$) shown in Fig.~\ref{fig:DeltaPole} and repeat the fitting
procedure described above. Note that the standard treatment of the
$\Delta$ in the HB framework breaks down in the vicinity of the
$\Delta$ pole. Therefore, we use the $\delta^1$ amplitudes calculated in
the covariant framework based on the Lagrangian in
\cite{Siemens:2014pma} for all three counting schemes. It has to be emphasized that the inclusion
of the  $\Delta$ in such a way is a phenomenological procedure which
is not based on a consistent power-counting such as the ones
formulated in Refs.~\cite{Hemmert:1997ye,Pascalutsa:2002pi}. 
A consistent inclusion of the $\Delta$ including
loop contributions is deferred to a future publication. We use the
same unitarization as in the previously discussed delta-less case (see
Eq. \eqref{eq:16}) and do not include explicitly the width of the $\Delta$
in our amplitudes. The only two new parameters which appear in the
$\delta^1$-amplitudes are the mass of the $\Delta$, which is fixed to
its Breit-Wigner value $m_\Delta=1.232$~GeV, and the pion-nucleon-$\Delta$
coupling constant, which is fixed to its large $N_C$ value $g_{\pi
  N\Delta}=3/(2\sqrt{2}) g_{A,ph}=1.35$, where we have used $g_{A,ph}=1.27$.
Notice that this value of $g_{\pi  N\Delta}$ is close to the one
extracted  from the
$\Delta$ width at leading order in the EFT expansion, see e.g.~\cite{Bernard:2012hb}.
%  (XXX should we not use the GTR modified value?XXX)

As in the case without $\Delta$, a representative fit to $\di\sigma/\di\Omega$ for the channel
$\pi^+p\to\pi^+p$ at $T_\pi=43.3$~MeV is shown in Fig.~\ref{fig:SigmaDiff}. 
In Fig.~\ref{fig:RedChiSq}, we show the reduced $\chi^2$ and $\bar
\chi^2$ as a function of the maximal energy used in the fits.  As
expected and differently to the delta-less calculations, one observes 
in the all three counting schemes a fairly flat behavior of  $\chi^2/$dof as function
of $T_\pi$ indicating that our estimation of the theoretical
uncertainty is reasonable. Actually,  $\chi^2/$dof even tend to
decrease with energy which may be viewed as an indication, that the
actual breakdown scale $\Lambda_b$ of the resulting approach is
somewhat higher than $600$~MeV.  
%The strong
%increase of  $\chi^2/$dof with energy in the HB approaches for $T_\pi \ge 100$~MeV is to be
%expected and can be traced back to the fact, that the employed
%unitarization procedure is valid in the delta region only for the
%covariant formulation and breaks down when the HB approach is employed 
Comparing the values of $\chi^2/$dof in the delta-less
and delta-full formulations, one furthermore realizes a significant
improvement in the quality of the fits upon the explicit inclusion of
the $\Delta$ isobar. A different treatment of the relativistic
corrections does not have a significant impact on the quality of the
fit except for the HB-NN results at order $Q^2 + \delta^1$ which 
are considerably less accurate than those of the HB-$\pi$N and
covariant approaches at the same orders. 

It is also comforting to see that the extracted LECs  
are now indeed rather stable with respect to increasing the energy range
used in the fits contrary to the observed pattern in the delta-less
case.  The resulting values of the LECs at different chiral orders in the
delta-full approach 
are collected  in Table~\ref{tab:FitD} for all three counting schemes, while the corresponding
correlation and covariance matrices are listed in 
Tables~\ref{tab:Q3piNCorrCovD} and \ref{tab:Q4piNCorrCovD}. 
Here, we refrain from comparing the values of the
LECs to the delta-less analyses available in the literature although
such a comparison could, in principle, be done by explicitly taking into account the
contributions of the $\Delta$ in the framework of resonance saturation.
Remarkably, \emph{all} extracted LECs
including $e_i$ from the order-$Q^4$ pion-nucleon Lagrangian come out 
of a natural size for all considered counting schemes which is clearly not the case in the delta-less
approach. Further, the differences between the values of the LECs
extracted based on the different treatments of the relativistic
corrections are much smaller as compared with the delta-less
calculations. Also, the shifts in the LECs $c_i$ and $d_i$ when increasing
the chiral order are now strongly reduced. All these findings provide a
strong evidence that convergence of the EFT
expansion for $\pi N$ scattering is considerably improved upon the explicit treatment of the
$\Delta$-resonance.   

\begin{figure}[ht]
\vspace{-1cm}
  \centering
  \includegraphics[width=0.65\textwidth]{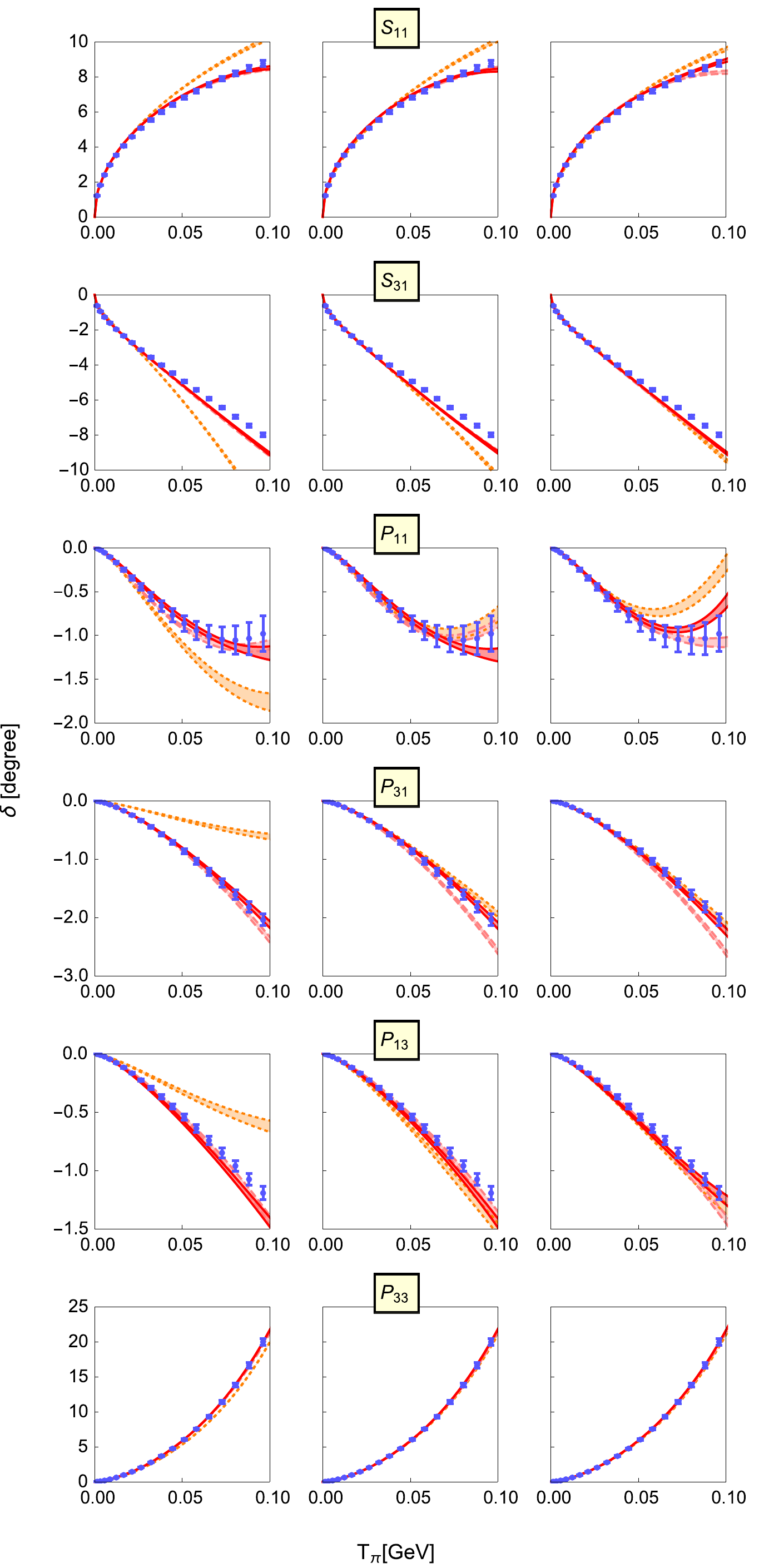}
\caption{(Color online) Predictions for $S$ waves up to
  $T_\pi=100$ MeV. Columns from left to right corresponds to the
  to the predictions in the HB-NN, HB-$\pi$N and Covariant counting, respectively.
  The orange, pink and red (dotted, dashed  and solid) bands refer to $Q^2+\delta^1$, $Q^3+\delta^1$ and $Q^4+\delta^1$ results
  including statistical uncertainties,
  respectively. }
\label{fig:SnPwavesStatD}
\end{figure}

\begin{figure}[ht]
\vspace{-1cm}
  \centering
  \includegraphics[width=0.65\textwidth]{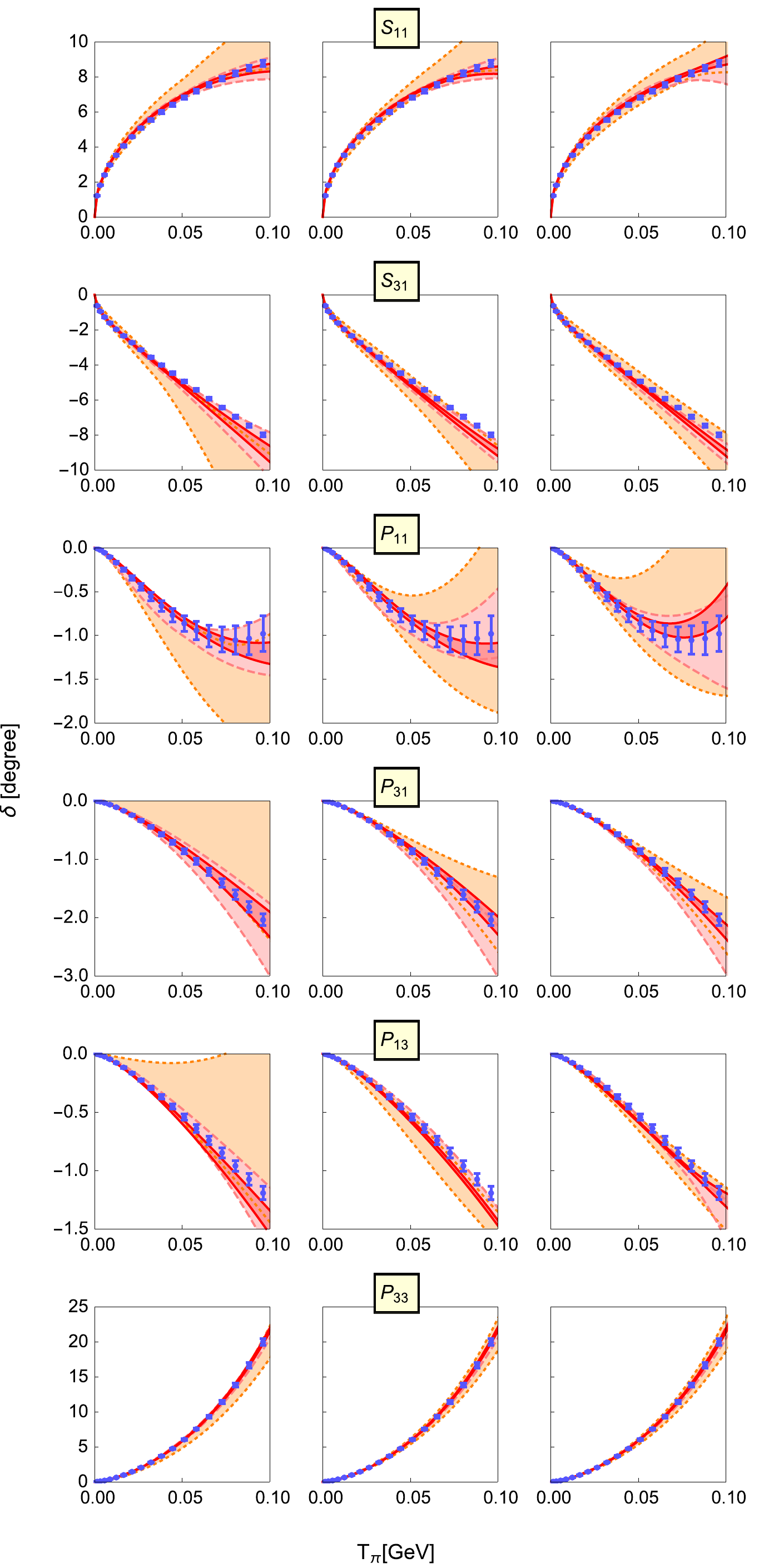}
\caption{(Color online) Predictions for $S$ waves up to
  $T_\pi=100$ MeV. Columns from left to right corresponds to the
  to the predictions in the HB-NN, HB-$\pi$N and Covariant counting, respectively.
  The dotted, dashed  and solid bands refer to $Q^2+\delta^1$, $Q^3+\delta^1$ and $Q^4+\delta^1$ results
  including theoretical uncertainties,
  respectively. }
\label{fig:SnPwavesD}
\end{figure}

\begin{figure}[ht]
\vspace{-1cm}
  \centering
  \includegraphics[width=0.65\textwidth]{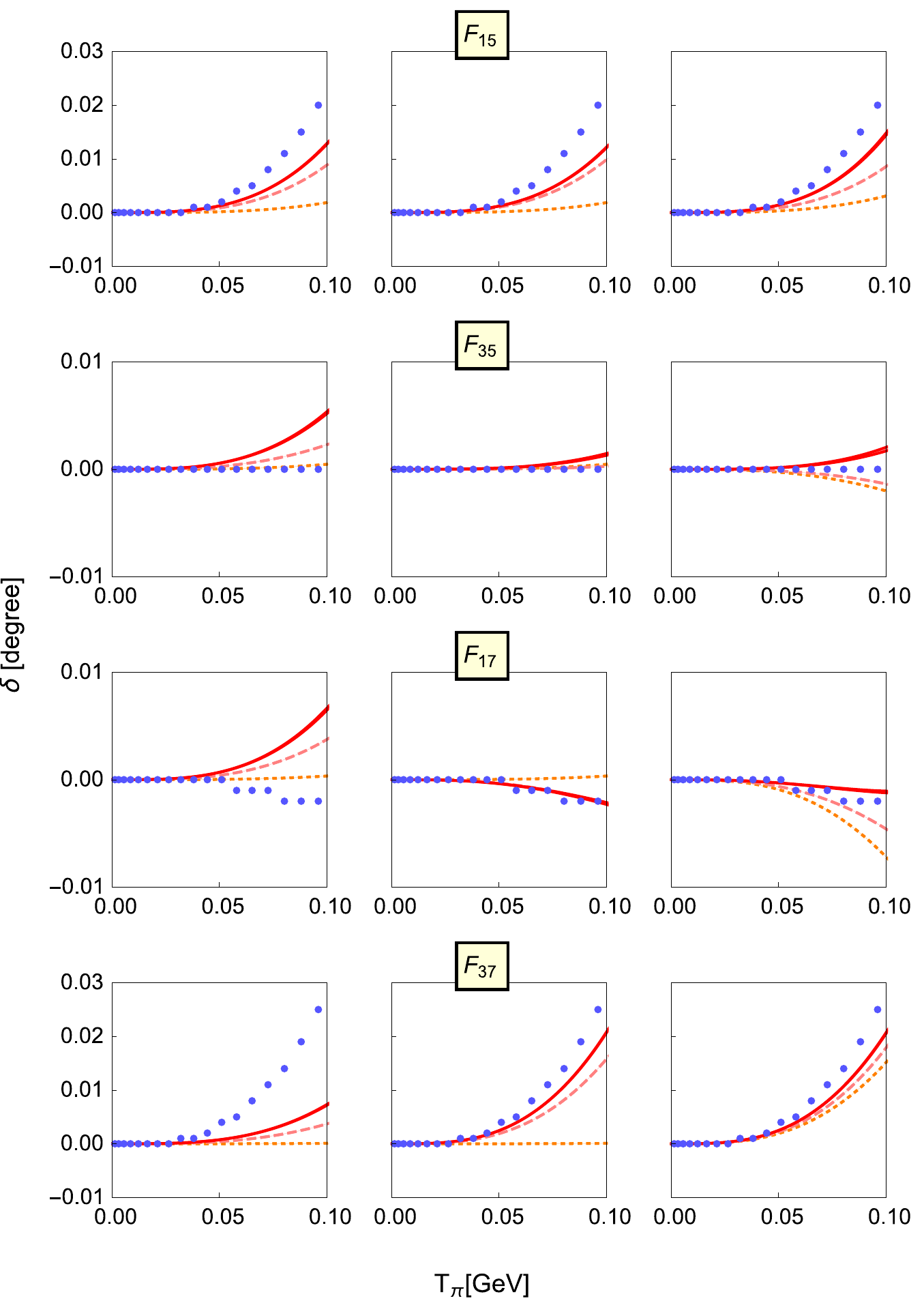}
\caption{(Color online) Predictions including statistical uncertainties for $F$ waves up to
  $T_\pi=100$~MeV. For remaining notation see Fig.~\ref{fig:SnPwavesStatD}.}
\label{fig:FwavesStatD}
\end{figure}

Our predictions for the $S$-, $P$-, $D$- and $F$-wave phase shifts are
summarized in Figs.~\ref{fig:SnPwavesStatD}-\ref{fig:FwavesD}. 
A comparison of the size of the uncertainty bands for  $S$-, $P$- and most
of the $D$-waves with the ones of the delta-less approaches confirms
the improved convergence of the delta-full theory. For the $S$- and
$P$-waves,  one observes
excellent agreement between the predicted phase shifts and the 
ones of Ref.~\cite{Hoferichter:2015dsa} determined
from the RS analyses. Furthermore, our predictions for the $D$-wave
phase shifts agree rather well (within uncertainties) with the ones of
the GWU-SAID PWA. For the $F$-waves, the treatment of the
$1/m_N$-corrections seems to play a more important role. In particular, 
for the $F_{35}$- $F_{17}$ and $F_{37}$-waves, one observes significant differences
between the HB-NN results at order $Q^4 + \epsilon^1$ and the ones based on the HB-$\pi$N and
covariant approaches, which appear to agree  rather well with the
GWU-SAID PWA. The employed approach to uncertainty quantification
clearly underestimates the error for the $F$-waves in the HB-NN
approach. On the other hand, it is comforting to see that a more
complete treatment of the relativistic corrections leads to a better
agreement with the GWU-SAID PWA. It is, however,
difficult to make conclusive statements due to the absence of
uncertainties in the GWU-SAID PWA. 

For the subthreshold coefficients and the scattering length,  
the explicit inclusion of the $\Delta$ resonance does, with very few
exceptions, noticeably
improve both the order-$Q^3$ and order-$Q^4$ results for all counting
schemes. Further, our predictions within the covariant approach show clearly a better
agreement with the values found in the RS analysis as compared to
the predictions within the HB formulations. This holds true for
both orders $Q^3 + \delta^1$ and $Q^4 + \delta^1$. We also observe
that our theoretical uncertainty for the subthreshold coefficients is
underestimated at order $Q^4 + \delta^1$. It
remains to be seen whether a more complete inclusion of the $\Delta$
resonance will allow for a better description of these quantities. It
would also be interesting to study in detail the convergence of the
$1/m_N$ expansion for the subthreshold coefficients and to estimate
the impact of other sources of uncertainties such as e.g. the one in the value of the
pion-nucleon coupling constant and/or isospin-breaking effects 
which are not included in our analysis. Work along these lines is in
progress.

\section{Summary and outlook}
\label{sec:sum}

The pertinent results of our paper can be summarized as follows:
\begin{itemize}
\item
We have calculated the pion-nucleon scattering amplitude in the
covariant formulation of $\chi$PT up to the order $Q^4$ within the
modified EOMS scheme. We discuss in detail the renormalization and finite
shifts of the various parameters in the effective Lagrangian. 
%provide explicit expressions for the scattering amplitude and compare
%our results to those published in Ref.~\cite{Chen:2012nx}. 
\item
We have implemented the novel approach to estimate the theoretical
uncertainty from the truncation of the chiral expansion formulated in
Ref.~\cite{Epelbaum:2014efa} and performed fits to the available
low-energy $\pi N$ scattering data using the HB-NN, HB-$\pi$N and the
covariant versions of  $\chi$PT. The extracted values of the
various LECs are found to be in a reasonably good agreement with the
ones reported in the literature. All three approaches lead to the
description of the experimental data of a similar quality which, however, exhibits
a fairly small breakdown scale of the chiral expansion. 
\item
By \emph{explicitly} including the lowest-order contributions of the
$\Delta$ isobar, we were able to unambiguously demonstrate that the slow
convergence of the chiral expansion for $\pi$N scattering is related
to the implicit treatment of the $\Delta$ resonance  in the considered formulations of $\chi$PT.
After including the lowest-order contributions of the $\Delta$ in the
scattering amplitude,  the breakdown scale of the resulting EFT is found to be
consistent with and probably even slightly larger than $\Lambda_b \sim 600$~MeV.  
All LECs determined from the corresponding fits to the experimental
data are found to be of a natural size at all orders and for all three
counting schemes. Further, the extracted values of the LECs appear to be
remarkably stable against increasing the maximal fitting energy,
changing the order of the calculation and
employing different counting rules for the $1/m_N$-corrections. 
The predicted phase shifts in the $S$- and $P$-waves at
order $Q^4 + \delta^1$ are in excellent agreement with the ones
extracted in  Ref.~\cite{Hoferichter:2015dsa} within the RS analyses
of $\pi N$ scattering; the predictions for the  $D$- and $F$-waves
are found to agree reasonably well with the GWU-SAID PWA. We also compare
our  predictions for the subthreshold coefficients and the scattering
lengths with their empirical values. 
\end{itemize}

The results of our study provide an important step towards performing
a combined analysis of the $\pi N \to \pi N$ and $\pi N \to \pi \pi N$
reactions, which is expected to result in an even more reliable determination
of the various LECs. Given that one has to use experimental data for
the second reaction, see Ref.~\cite{Siemens:2014pma} for a recent study along this
line, it would be inappropriate to employ empirical phase shift
analyses for the first one. The results of our work thus pave the way for a
unified treatment of both reactions with regard to the available
experimental  information.  Moreover, the inclusion of the theoretical
uncertainty when performing the fits as implemented in our work is
shown to stabilize the results against the variation of the maximal fitting energy (provided
the effects of the $\Delta$ isobar are explicitly taken into account)
which is a necessary prerequisite for carrying out a combined analysis
of the $\pi N \to \pi N$ and $\pi N \to \pi \pi N$ processes. 
Apart from extending the  calculations presented here  to the single pion production
reaction, it would also be interesting to directly confront the
$\chi$PT results for the phase shifts with their recent determination 
in the framework of the Roy-Steiner equation \cite{Hoferichter:2015dsa}
 and to
perform a more complete and consistent treatment of the delta
contributions.     Work along these lines is in progress.

\section*{Acknowledgments}
We would like to thank Jambul Gegelia for helpful discussions and
useful comments on the EOMS scheme
and De-Liang Yao for cross checking all amplitudes.
One of the authors (DS) is grateful to the staff at the Institute for
Nuclear Theory at the University of Washington, Seattle, where a part
of this work has been done.
This work was supported by the DFG (SFB/TR 16,
``Subnuclear Structure of Matter''), 
%the European
%Community-Research Infrastructure Integrating Activity ``Study of
%Strongly Interacting Matter'' (acronym HadronPhysics3,
%Grant Agreement n. 283286) under the Seventh Framework Programme of EU,
the ERC project 259218 NUCLEAREFT  the Ruhr University Research
School PLUS, funded by Germany's Excellence Initiative [DFG GSC 98/3] and
by the Chinese Academy of Sciences (CAS)
President's International Fellowship Initiative (PIFI) (Grant No.
2015VMA076).

\clearpage
\appendix

\section{Renormalization Rules}
\label{sec:renorm-rules}
In this appendix, the formulae related to the renormalization of the
amplitudes are given. The notation for the integrals is the following
\begin{align}
  \label{eq:12}
  %\begin{aligned}
    A_0(m_0^2)&=\frac{1}{\i}\int \frac{\di^d l}{(2\pi)^d} \,
    \frac{1}{l^2-m_0^2}\komma\nonumber\\
    B_0(p^2,m_0^2,m_1^2)&=\frac{1}{\i}\int \frac{\di^d
      l}{(2\pi)^d} \, \frac{1}{(l^2-m_0^2)((l+p)^2-m_1^2)}\komma\\
    J_0(\omega)&=\frac{1}{\i}\int \frac{\di^d
      l}{(2\pi)^d} \, \frac{1}{(l^2-M_\pi^2)(\omega+v\cdot l)}\komma\nonumber\\
    C_0(p_1^2,(p_1-p_2)^2,p_2^2,m_0^2,m_1^2,m_2^2)&=\frac{1}{\i}\int \frac{\di^d
      l}{(2\pi)^d} \, \frac{1}{(l^2-m_0^2)((l+p_1)^2-m_1^2)((l+p_2)^2-m_2^2)}\nonumber
 % \end{aligned}
\end{align}
where the $+\i\epsilon$ prescription was suppressed.

\subsection{Mesonic Sector}
\label{subsection}
The renormalization rules for the pion mass, Z-factor and decay constant read
\begin{equation}
\begin{aligned}
  \label{eq:Mpi}
  M^2 &= M^2_\pi + \delta M^{(4)}\komma\\
  \delta M^{(4)} &=-\frac{2 l_3 M_\pi^4}{F_\pi^2}+\frac{M_\pi^2 A_0(M_\pi^2) }{2 F_\pi^2}\komma\\
    Z_\pi &= 1+ \delta Z_\pi^{(4)}\komma\\
    \delta Z_\pi^{(4)} &= -\frac{2 l_4 M_\pi^2}{F_\pi^2}-\frac{(-1+10 \alpha) A_0(M_\pi^2) }{F_\pi^2}\komma\\
 F &= F_\pi+ \delta F_\pi^{(4)}\komma\\
    \delta F_\pi^{(4)} &=-\frac{l_4 M_\pi^2}{F_\pi}-\frac{A_0(M_\pi^2)
    }{F_\pi}\punkt
  \end{aligned}
\end{equation}

\subsection{Baryonic sector}
\label{sec:baryonic-sector}
\begin{figure}[ht]
  \centering
  \includegraphics[width=0.6\textwidth]{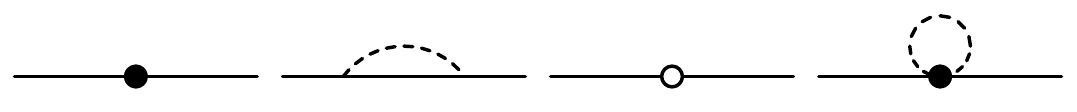}
  \caption{Diagrams contributing to the nucleon self energy. For notation see Fig.~\ref{fig:TreeGraphs}.}
  \label{fig:nuclmass}
\end{figure}

\begin{figure}[ht]
  \centering
  \includegraphics[width=0.6\textwidth]{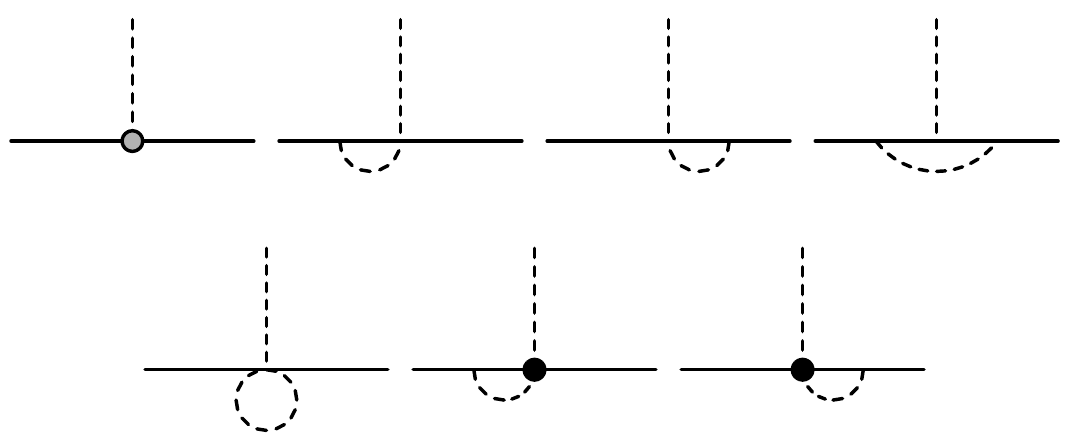}
  \caption{Diagrams contributing to the axial vector coupling of the nucleon. 
For notation see Fig.~\ref{fig:TreeGraphs}.}
  \label{fig:axialcoupl}
\end{figure}

In the baryonic sector one has to differentiate between the covariant
and heavy baryon approaches. The self-energy diagrams necessary for mass
renormalization are shown in Fig.~\ref{fig:nuclmass}. The axial coupling
constant was renormalized at the pion-nucleon vertex and the
contributing diagrams are shown in Fig.~\ref{fig:axialcoupl}.

\subsubsection{Covariant chiral perturbation theory}
\label{sec:cov-baryon}
In covariant $\chi$PT, the renormalization rule for the nucleon mass reads
\begin{equation}
\begin{aligned}
  \label{eq:2}
  m &= m_N + \delta m^{(2)}+ \delta m^{(3)}+ \delta m^{(4)} \komma\\
  \delta m^{(2)} &=4 c_1 M_\pi^2 \komma\\
  \delta m^{(3)} &= -\frac{3 g_A^2 m_N A_0\left(m_N^2\right)}{2 F_\pi^2 }-\frac{3 g_A^2 M_\pi^2 m_N B_0\left(m_N^2,M_\pi^2,m_N^2\right)}{2 F_\pi^2 }\komma\\
  \delta m^{(4)} &= M_\pi^4 \left(2 e_{115}+2 e_{116}+16
    e_{38}-\frac{8 c_1 l_3}{F_\pi^2}-\frac{3 c_2}{128 F_\pi^2 \pi
      ^2}\right)\\
  &+\frac{(32 c_1-3 (c_2+4 c_3)) M_\pi^2 A_0\left(M_\pi^2\right)}{4 F_\pi^2}\komma
\end{aligned}
\end{equation}
whereas the expression  for the Z-Factor is given by
\begin{equation}
  \begin{aligned}
    \label{eq:3}
    Z_N &= 1+ \delta Z_N^{(3)}+\delta Z_N^{(4)} \komma\\
    \delta Z_N^{(3)} &= \frac{3 g_A^2 M_\pi^2 m_N^2}{16 F_\pi^2
      \left(M_\pi^2-4 m_N^2\right) \pi ^2}+\frac{3 g_A^2 \left(5
        M_\pi^2-12 m_N^2\right) A_0\left(M_\pi^2\right)}{4 F_\pi^2
      \left(M_\pi^2-4 m_N^2\right) }\\
    &-\frac{3 g_A^2 M_\pi^2 A_0\left(m_N^2\right)}{F_\pi^2 \left(M_\pi^2-4 m_N^2\right)}-\frac{3 g_A^2 M_\pi^2 \left(M_\pi^2-3 m_N^2\right) B_0\left(m_N^2,M_\pi^2,m_N^2\right)}{F_\pi^2 \left(M_\pi^2-4 m_N^2\right) }\komma\\
    \delta Z_N^{(4)} &= \frac{3 c_2 M_\pi^4}{64 F_\pi^2 m_N \pi ^2}+\frac{3 c_2 M_\pi^2 A_0\left(M_\pi^2\right)}{2 F_\pi^2 m_N}\punkt
  \end{aligned}
\end{equation}

The effective axial coupling constant is renormalized via
\begin{equation}
  \begin{aligned}
    \label{eq:4}
    g &= g_A + \delta g^{(3)}+ \delta g^{(4)} \komma\\
    \delta g^{(3)} &= -\frac{M_\pi^2 \left(3 g_A^3 m_N^2+32 F_\pi^2 (2
        d_{16}-d_{18}) \left(M_\pi^2-4 m_N^2\right) \pi ^2\right)}{16
      F_\pi^2 \left(M_\pi^2-4 m_N^2\right) \pi ^2}\\
    &-\frac{g_A \left(\left(1+4 g_A^2\right) M_\pi^2-2 \left(2+5
          g_A^2\right) m_N^2\right) A_0\left(M_\pi^2\right)}{
      F_\pi^2 \left(M_\pi^2-4 m_N^2\right)}\\
    &+\frac{g_A \left(\left(2+3 g_A^2\right) M_\pi^2-8 m_N^2\right)
      A_0\left(m_N^2\right)}{F_\pi^2 \left(M_\pi^2-4 m_N^2\right)
      }-\frac{g_A^3 m_N^2
      B_0\left(M_\pi^2,m_N^2,m_N^2\right)}{F_\pi^2}\\
    &+\frac{g_A M_\pi^2 \left(\left(2+3 g_A^2\right) M_\pi^2-\left(8+9
          g_A^2\right) m_N^2\right)
      B_0\left(m_N^2,M_\pi^2,m_N^2\right)}{ F_\pi^2 \left(M_\pi^2-4
        m_N^2\right) }\\
    &-\frac{g_A^3 M_\pi^2 m_N^2
      C_0\left(m_N^2,M_\pi^2,m_N^2,M_\pi^2,m_N^2,m_N^2\right)}{F_\pi^2}\komma\\
 \end{aligned}
\end{equation}
\begin{equation*}
  \begin{aligned}
    \delta g^{(4)} &= -\frac{g_A c_3 \left(-3 M_\pi^4+10 M_\pi^2
        m_N^2+8 m_N^4\right)}{144 F_\pi^2 m_N \pi ^2}-\frac{g_A c_4
      \left(-3 M_\pi^4+10 M_\pi^2 m_N^2+8 m_N^4\right)}{144 F_\pi^2
      m_N \pi ^2}\\
&-\frac{g_A c_2 \left(-33 M_\pi^6+224 M_\pi^4 m_N^2+64 M_\pi^2
    m_N^4+36 m_N^6\right)}{2304 F_\pi^2 m_N^3 \pi ^2}\\
&+\left(\frac{2 g_A c_3 \left(M_\pi^4-M_\pi^2 m_N^2\right)}{3 F_\pi^2
    m_N^3}+\frac{2 g_A c_4 \left(M_\pi^4-M_\pi^2 m_N^2\right)}{3
    F_\pi^2 m_N^3}\right.\\
&\qquad\qquad\left.-\frac{g_A c_2 \left(-8 M_\pi^6+5
      M_\pi^4 m_N^2+48 M_\pi^2 m_N^4\right)}{24 F_\pi^2 m_N^5}\right)
A_0\left(M_\pi^2\right)\\
&+\left(\frac{4 g_A c_1 M_\pi^2}{F_\pi^2 m_N}-\frac{2 g_A c_4 \left(M_\pi^4-10 m_N^4\right)}{3 F_\pi^2 m_N^3}-\frac{2 g_A c_3 \left(M_\pi^4+2 m_N^4\right)}{3 F_\pi^2 m_N^3}\right.\\
&\qquad\qquad\left.-\frac{g_A c_2 \left(2 M_\pi^6-3 M_\pi^4
      m_N^2+M_\pi^2 m_N^4+3 m_N^6\right)}{6 F_\pi^2 m_N^5}\right)
A_0\left(m_N^2\right)\\
&+\left(\frac{2 g_A c_4 \left(-M_\pi^6+2 M_\pi^4 m_N^2+8 M_\pi^2 m_N^4\right)}{3 F_\pi^2 m_N^3}-\frac{g_A c_2 \left(2 M_\pi^8-7 M_\pi^6 m_N^2+8 M_\pi^4 m_N^4\right)}{6 F_\pi^2 m_N^5}\right.\\
&\qquad\qquad\left.
+\frac{4 g_A c_1 M_\pi^4}{F_\pi^2 m_N}-\frac{2 g_A c_3 \left(M_\pi^6-2 M_\pi^4 m_N^2+4 M_\pi^2 m_N^4\right)}{3 F_\pi^2 m_N^3}\right) B_0\left(m_N^2,M_\pi^2,m_N^2\right)\punkt
  \end{aligned}
\end{equation*}

\subsubsection{Heavy-baryon chiral perturbation theory}
\label{sec:heavy-baryon}

The HB expression for the nucleon mass reads
\begin{equation}
\begin{aligned}
  \label{eq:2}
  m &= m_N + \delta m^{(2)}+ \delta m^{(3)}+ \delta m^{(4)} \komma\\
  \delta m^{(2)} &=4 c_1 M_\pi^2 \komma\\
  \delta m^{(3)} &= -\frac{3 g_A^2 M_\pi^2 J_0(0)}{4 F_\pi^2}\komma\\
  \delta m^{(4)} &= M_\pi^4 \left(2 e_{115}+2 e_{116}+16
    e_{38}-\frac{8 c_1 l_3}{F_\pi^2}-\frac{3 c_2}{128 F_\pi^2 \pi
      ^2}+\frac{3 g_A^2}{64 F_\pi^2 m_N \pi ^2}\right)\\
  &-M_\pi^2 \left(\frac{-32 c_1+3 (c_2+4 c_3)}{4 F_\pi^2}+\frac{3 g_A^2}{4 F_\pi^2 m_N}\right) A_0(M_\pi^2) \komma\\
\end{aligned}
\end{equation}

whereas the Z-Factor is given by
\begin{equation}
  \begin{aligned}
    \label{eq:3}
    Z_N &= 1+ \delta Z_N^{(3)}+\delta Z_N^{(4)} \komma\\
    \delta Z_N^{(3)} &= -\frac{3 g_A^2 M_\pi^2}{32 F_\pi^2 \pi ^2}+\frac{9 g_A^2 A_0(M_\pi^2) }{4 F_\pi^2}\komma\\
    \delta Z_N^{(4)} &= -\frac{9 g_A^2 M_\pi^2 J_0(0)}{8 F_\pi^2 m_N}\punkt
  \end{aligned}
\end{equation}

The effective axial coupling constant in the HB approach is given by
\begin{equation}
  \begin{aligned}
    \label{eq:4}
    g &= g_A + \delta g^{(3)}+ \delta g^{(4)} \komma\\
    \delta g^{(3)} &= M_\pi^2 \left(-4 d_{16}+2 d_{18}+\frac{g_A^3}{16 F_\pi^2 \pi ^2}\right)-\frac{\left(g_A+2 g_A^3\right) A_0(M_\pi^2) }{F_\pi^2}\komma\\
    \delta g^{(4)} &= M_\pi^2 \left(-\frac{4 g_A (c_3-2 c_4)}{3
        F_\pi^2}+\frac{g_A+g_A^3}{F_\pi^2 m_N}\right)
    J_0(0)\punkt
  \end{aligned}
\end{equation}

\section{Renormalization of LECs}
\label{sec:renormalization-lecs}

\subsection{Mesonic Sector}
\label{sec:mesonic-sector}
The $\beta$-functions in the mesonic sector read
\begin{equation}
  \label{eq:32}
    \beta_{l_1} = \frac{1}{3}\komma\quad
    \beta_{l_2} = \frac{2}{3}\komma\quad
    \beta_{l_3} = -\frac{1}{2}\komma\quad
    \beta_{l_4} = 2\punkt
\end{equation}

\subsection{Baryonic Sector}
\label{sec:baryonic-sector}
In the baryonic sector we have to differentiate between
the EOMS and HB renormalization rules.
\subsubsection{Covariant chiral perturbation theory}
\label{sec:covar-chir-pert}
In EOMS scheme, the $\beta$-functions of the $c_i$ read at order $Q^3$ 
\begin{equation}
  \label{eq:33}
  \begin{aligned}
    \beta^{(3)}_{c_1}&=-\frac{3 g_A^2 m_N}{4}\komma\\
    \beta^{(3)}_{c_2}&=\left(-1+g_A^2\right)^2 m_N\komma\\
    \beta^{(3)}_{c_3}&=\frac{1}{2} \left(1-6 g_A^2+g_A^4\right) m_N\komma\\
    \beta^{(3)}_{c_4}&=\frac{1}{2} \left(-1-2 g_A^2+3 g_A^4\right) m_N\komma
  \end{aligned}
\end{equation}
while the contributions at order $Q^4$ have the form  
\begin{equation}
  \label{eq:35}
  \begin{aligned}
    \beta^{(4)}_{c_1}&=9 g_A^2 c_1 m_N^2\komma\\
    \beta^{(4)}_{c_2}&=\frac{1}{3} \left(-4 c_4+g_A^2 (3 c_2+8 c_3+4 c_4)\right) m_N^2\komma\\
    \beta^{(4)}_{c_3}&=\frac{1}{6} \left(g_A^2 (21 c_2+54 c_3-52 c_4)+20 c_4\right) m_N^2\komma\\
    \beta^{(4)}_{c_4}&=\frac{1}{6} \left(3 c_2+8 c_3-20 c_4-g_A^2 (15 c_2+24 c_3+2 c_4)\right) m_N^2\,.
  \end{aligned}
\end{equation}
The corresponding finite shifts are given by
\begin{equation}
  \label{eq:36}
  \begin{aligned}
    \delta\bar c_{1,f}^{(3)} &=\frac{3 g_A^2 m_N}{128 \pi ^2}\komma\\
    \delta\bar c_{2,f}^{(3)} &=-\frac{\left(2+g_A^4\right) m_N}{32 \pi ^2}\komma\\
    \delta\bar c_{3,f}^{(3)} &=\frac{9 g_A^4 m_N}{64 \pi ^2}\komma\\
    \delta\bar c_{4,f}^{(3)} &=-\frac{g_A^2 \left(5+g_A^2\right) m_N}{64 \pi ^2}
  \end{aligned}
\end{equation}
at order $Q^3$ and 
\begin{equation}
  \label{eq:37}
  \begin{aligned}
    \delta\bar c_{1,f}^{(4)} &=\frac{3 g_A^2 c_1 m_N^2}{16 \pi ^2}\komma\\
    \delta\bar c_{2,f}^{(4)} &=\frac{\left(-2 c_4+g_A^2 (9 c_2+16 c_3+14 c_4)\right) m_N^2}{144 \pi ^2}\komma\\
    \delta\bar c_{3,f}^{(4)} &=\frac{\left(g_A^2 (-9 c_2+216 c_3-272 c_4)+16 c_4\right) m_N^2}{1152 \pi ^2}\komma\\
    \delta\bar c_{4,f}^{(4)} &=\frac{\left(9 \left(-1+g_A^2\right) c_2-8 \left(4 c_3+\left(2+11 g_A^2\right) c_4\right)\right) m_N^2}{1152 \pi ^2}
  \end{aligned}
\end{equation}
at order $Q^4$, respectively. 
Similarly, for the LECs $d_i$, we obtain 
\begin{equation}
  \label{eq:38}
  \begin{aligned}
    \beta^{(3)}_{d_1}+  \beta^{(3)}_{d_2}&=\frac{1}{24} \left(1-4 g_A^2+3 g_A^4\right)\komma\\
    \beta^{(3)}_{d_3}&=0\komma\\
    \beta^{(3)}_{d_4}&=\frac{1}{8} g_A \left(-1+g_A^2\right)^2\komma\\
    \beta^{(3)}_{d_5}&=\frac{1}{24} \left(1-g_A^2\right)\komma\\
    \beta^{(3)}_{d_{10}}&=\frac{1}{4} g_A \left(-1+g_A^4\right)\komma\\
    \beta^{(3)}_{d_{11}}&=-\frac{1}{4} g_A \left(3-4 g_A^2+g_A^4\right)\komma\\
    \beta^{(3)}_{d_{12}}&=\frac{1}{2} g_A \left(-1+g_A^2\right)^2\komma\\
    \beta^{(3)}_{d_{13}}&=-\frac{1}{2} g_A \left(-1+g_A^2\right)^2\komma\\
    \beta^{(3)}_{d_{14}}-\beta^{(3)}_{d_{15}}&=\frac{1}{2} \left(-1+g_A^2\right)^2\komma\\
    \beta^{(3)}_{d_{16}}&=\frac{1}{2} g_A \left(-1+g_A^2\right)\komma\\
    \beta^{(3)}_{d_{18}}&=0
  \end{aligned}
\end{equation}
and
\begin{equation}
  \label{eq:41}
  \begin{aligned}
      \beta^{(4)}_{d_1}+\beta^{(4)}_{d_2}&=\frac{\left(\left(7+11 g_A^2\right) c_2-16 \left(-1+g_A^2\right) c_3+2 \left(-5+g_A^2\right) c_4\right) m_N}{12 }\komma\\
    \beta^{(4)}_{d_3}&=-\frac{5 \left(-1+g_A^2\right) c_2 m_N}{3}\komma\\
    \beta^{(4)}_{d_4}&=\frac{g_A \left(\left(9-15 g_A^2\right) c_2+4 (3 c_3+5 c_4)-4 g_A^2 (3 c_3+14 c_4)\right) m_N}{24 }\komma\\
    \beta^{(4)}_{d_5}&=\frac{\left(24 \left(-3+2 g_A^2\right) c_1-3
        c_2+8 c_3+4 c_4+2 g_A^2 (-2 c_3+c_4)\right) m_N}{24 }\komma\\
    \beta^{(4)}_{d_{10}}&=-\frac{g_A \left(\left(-71+63 g_A^2\right)
        c_2+64 \left(-3+g_A^2\right) c_3+24 \left(3+g_A^2\right)
        c_4\right) m_N}{24 }\komma
  \end{aligned}
\end{equation}
\begin{equation*}
  \begin{aligned}
    \beta^{(4)}_{d_{11}}&=\frac{g_A \left(\left(-33+13 g_A^2\right) c_2-96 c_3+8 \left(-17+11 g_A^2\right) c_4\right) m_N}{24 }\komma\\
    \beta^{(4)}_{d_{12}}&=\frac{g_A \left(c_2-13 g_A^2 c_2-16 \left(-3+g_A^2\right) c_3+112 \left(-1+g_A^2\right) c_4\right) m_N}{24 }\komma\\
    \beta^{(4)}_{d_{13}}&=\frac{g_A \left(\left(11-15 g_A^2\right) c_2+16 \left(\left(-3+g_A^2\right) c_3-7 \left(-1+g_A^2\right) c_4\right)\right) m_N}{24 }\komma\\
    \beta^{(4)}_{d_{14}}-\beta^{(4)}_{d_{15}}&=\frac{\left(g_A^2 (-13 c_2+8 c_3-12 c_4)+12 c_4\right) m_N}{6 }\komma\\
    \beta^{(4)}_{d_{16}}&=\frac{g_A \left(-24 \left(-2+g_A^2\right) c_1+c_2+6 (c_3-3 c_4)\right) m_N}{6 }\komma\\
    \beta^{(4)}_{d_{18}}&=\frac{g_A (24 c_1+c_2-4 (c_3+c_4)) m_N}{6 }\komma 
  \end{aligned}
\end{equation*}
while the finite shifts have the form
\begin{equation}
  \label{eq:39}
  \begin{aligned}
    \delta\bar d^{(3)}_{1,f}+ \delta\bar d^{(3)}_{2,f}&=-\frac{12+39 g_A^2+11 g_A^4}{768
       \pi ^2}\komma\\
    \delta\bar d^{(3)}_{3,f}&=\frac{3+3 g_A^2+g_A^4}{96 \pi ^2}\komma\\
    \delta\bar d^{(3)}_{4,f}&=\frac{g_A \left(7+2 g_A^2+g_A^4\right)}{256 \pi ^2}\komma\\
    \delta\bar d^{(3)}_{5,f}&=\frac{g_A^2 \left(9+g_A^2\right)}{512 \pi ^2}\komma\\
    \delta\bar d^{(3)}_{10,f}&=\frac{g_A \left(18+81 g_A^2+31 g_A^4\right)}{384 \pi ^2}\komma\\
    \delta\bar d^{(3)}_{11,f}&=-\frac{g_A^3 \left(33+19 g_A^2\right)}{384 \pi ^2}\komma\\
    \delta\bar d^{(3)}_{12,f}&=-\frac{g_A \left(36+46 g_A^2+29 g_A^4\right)}{192 \pi ^2}\komma\\
    \delta\bar d^{(3)}_{13,f}&=\frac{g_A \left(12+22 g_A^2+13 g_A^4\right)}{192 \pi ^2}\komma\\
    \delta\bar d^{(3)}_{14,f}-\delta\bar d^{(3)}_{15,f}&=\frac{g_A^4}{192 \pi ^2}\komma\\
    \delta\bar d^{(3)}_{16,f}&=\frac{g_A+g_A^3}{32 \pi ^2}\komma\\
    \delta\bar d^{(3)}_{18,f}&=\frac{g_A^3}{192 \pi ^2}
  \end{aligned}
\end{equation}
and
\begin{equation}
  \label{eq:40}
  \begin{aligned}
    \delta\bar d^{(4)}_{1,f}+ \delta\bar d^{(4)}_{2,f}&=-\frac{\left(\left(4+8 g_A^2\right) c_2+\left(10-22 g_A^2\right) c_3+\left(5+38 g_A^2\right) c_4\right) m_N}{576  \pi ^2}\komma\\
    \delta\bar d^{(4)}_{3,f}&=\frac{\left(\left(-34+4 g_A^2\right) c_2+3 \left(-10 c_3+c_4+5 g_A^2 c_4\right)\right) m_N}{288  \pi ^2}\komma\\
    \delta\bar d^{(4)}_{4,f}&=\frac{g_A \left(6 c_3+184 c_4+g_A^2 (-9 (c_2+2 c_3)+2 c_4)\right) m_N}{1152  \pi ^2}\komma\\
    \delta\bar d^{(4)}_{5,f}&=\frac{\left(72 \left(2+g_A^2\right)
        c_1-2 \left(1+19 g_A^2\right) c_3+\left(-1+10 g_A^2\right)
        c_4\right) m_N}{1152  \pi ^2}\komma
  \end{aligned}
\end{equation}
\begin{equation*}
  \begin{aligned}
    \delta\bar d^{(4)}_{10,f}&=-\frac{g_A \left(\left(41+195
          g_A^2\right) c_2+704 g_A^2 (c_3-3 c_4)+48 (10 c_3-3
        c_4)\right) m_N}{4608  \pi ^2}\komma\\
    \delta\bar d^{(4)}_{11,f}&=\frac{g_A \left(87 c_2+5 g_A^2 c_2+1056 c_3-16 \left(5+64 g_A^2\right) c_4\right) m_N}{4608  \pi ^2}\komma\\
    \delta\bar d^{(4)}_{12,f}&=-\frac{g_A \left(\left(-329+341 g_A^2\right) c_2+32 \left(-33+4 g_A^2\right) c_3+64 \left(5+7 g_A^2\right) c_4\right) m_N}{4608  \pi ^2}\komma\\
    \delta\bar d^{(4)}_{13,f}&=\frac{g_A \left(\left(-485+33 g_A^2\right) c_2+32 (-33 c_3+4 c_4)+64 g_A^2 (5 c_3+7 c_4)\right) m_N}{4608  \pi ^2}\komma\\
    \delta\bar d^{(4)}_{14,f}-\delta\bar d^{(4)}_{15,f}&=\frac{\left(24 c_4+g_A^2 (67 c_2-56 c_3+96 c_4)\right) m_N}{1152 \pi ^2}\komma\\
    \delta\bar d^{(4)}_{16,f}&=-\frac{g_A \left(72 \left(-1+g_A^2\right) c_1+c_2+18 (c_3-c_4)\right) m_N}{288 \pi ^2}\komma\\
    \delta\bar d^{(4)}_{18,f}&=\frac{g_A (c_2-c_3-c_4) m_N}{144 \pi ^2}\punkt
  \end{aligned}
\end{equation*}
Finally, we also list the $\beta$-functions and the finite shifts for
the LECs $e_i$:
\begin{equation}
  \label{eq:42}
  \begin{aligned}
          \beta^{(4)}_{e_{10}}&=\frac{1}{192} g_A \left(\left(101-41
              g_A^2\right) c_2+16 \left(\left(-3+g_A^2\right) c_3-7
              \left(-1+g_A^2\right) c_4\right)\right)\\
          &\qquad-\frac{g_A \left(-1+g_A^2\right)^2}{16 m_N}\komma\\
          \beta^{(4)}_{e_{11}}&=-\frac{1}{24} g_A \left(\left(-35+39 g_A^2\right) c_2+22 \left(-1+g_A^2\right) c_3+\left(-29+83 g_A^2\right) c_4\right) \\
          &\qquad+\frac{g_A \left(19-40 g_A^2+21 g_A^4\right)}{96 m_N}\komma\\
          \beta^{(4)}_{e_{12}}&=\frac{1}{24} g_A \left(\left(-25+29 g_A^2\right) c_2+22 \left(-1+g_A^2\right) c_3+\left(-29+83 g_A^2\right) c_4\right) \\
          &\qquad-\frac{g_A \left(19-40 g_A^2+21 g_A^4\right)}{96 m_N}\komma\\
          \beta^{(4)}_{e_{13}}&=\frac{1}{12} g_A \left(-1+g_A^2\right) c_2\komma\\
          \beta^{(4)}_{e_{14}}&=\frac{1}{96} \left(-\left(8+25 g_A^2\right) c_2+4 \left(2 \left(-6+g_A^2\right) c_3+5 \left(-1+g_A^2\right) c_4\right)\right) \\
          &\qquad+\frac{\left(-1+g_A^2\right)^2}{32 m_N}\komma\\
          \beta^{(4)}_{e_{15}}&=\frac{g_A^2 c_2}{6}\komma\\
          \beta^{(4)}_{e_{16}}&=0\komma\\
          \beta^{(4)}_{e_{17}}&=\frac{1}{48} \left(\left(-1+15 g_A^2\right) c_2+2 \left(1-7 g_A^2\right) c_4\right)+\frac{1-4 g_A^2+3 g_A^4}{96 m_N}\komma\\
          \beta^{(4)}_{e_{18}}&=\frac{1}{4} \left(c_2-g_A^2
            c_2\right)\komma
  \end{aligned}
\end{equation}
\begin{equation*}
  \begin{aligned}
          2\beta^{(4)}_{e_{19}}-\beta^{(4)}_{e_{22}}-\beta^{(4)}_{e_{36}}&=2 c_1+\frac{1}{24} \left(-5+36 g_A^2\right) c_2+\frac{3 c_3}{4}+\frac{5 g_A^2 c_3}{12}+c_4-g_A^2 c_4 \\
          &\qquad-\frac{\left(-1+g_A^2\right)^2}{8 m_N}\komma\\
          \beta^{(4)}_{e_{20}}+\beta^{(4)}_{e_{35}}&=-\frac{1}{12}
          \left(-6+g_A^2\right) c_2\komma\\
          2\beta^{(4)}_{e_{21}}-\beta^{(4)}_{e_{37}}&=\frac{1}{24} \left(-24 c_1+c_2-4 g_A^2 c_2+4 \left(4 c_3-3 g_A^2 c_3+7 g_A^2 c_4\right)\right) \\
          &\qquad+\frac{1-g_A^2}{24 m_N}\komma\\ 
          \beta^{(4)}_{e_{22}}-4\beta^{(4)}_{e_{38}}&=\frac{1}{48} \left(-72 \left(2+g_A^2\right) c_1+12 c_2-39 g_A^2 c_2+36 c_3+8 g_A^2 c_3-4 c_4+4 g_A^2 c_4\right) \\
          &\qquad+\frac{\left(-1+g_A^2\right)^2}{16 m_N}\komma\\ 
          \beta^{(4)}_{e_{34}}&=\frac{1}{48} g_A \left(-48 c_1+\left(-25+49 g_A^2\right) c_2+4 \left(-7 c_3+9 g_A^2 c_3-17 c_4+43 g_A^2 c_4\right)\right) \\
          &\qquad-\frac{g_A \left(7-16 g_A^2+9 g_A^4\right)}{48 m_N}
  \end{aligned}
\end{equation*}
and the finite pieces
\begin{equation}
  \label{eq:43}
  \begin{aligned}
        \delta\bar e^{(4)}_{10,f}&=-\frac{g_A \left(\left(503+493 g_A^2\right) c_2-32 \left(\left(9+2 g_A^2\right) c_3-8 \left(13+6 g_A^2\right) c_4\right)\right)}{36864 \pi ^2}\\
          &\qquad-\frac{g_A \left(30+38 g_A^2+17 g_A^4\right)}{1536 m_N \pi ^2}\komma\\
    \delta\bar e^{(4)}_{11,f}&=-\frac{g_A \left(\left(69+70 g_A^2\right) c_2-4 \left(-5+8 g_A^2\right) c_3+\left(-244+233 g_A^2\right) c_4\right)}{2304 \pi ^2}\\
          &\qquad+\frac{g_A \left(46-47 g_A^2+87 g_A^4\right)}{3072 m_N \pi ^2}\komma\\
    \delta\bar e^{(4)}_{12,f}&=\frac{g_A \left(\left(29+50 g_A^2\right) c_2+8 \left(1+g_A^2\right) c_3+\left(38+375 g_A^2\right) c_4\right)}{2304 \pi ^2}\\
          &\qquad+\frac{g_A \left(162+435 g_A^2+13 g_A^4\right)}{3072 m_N \pi ^2}\komma\\
    \delta\bar e^{(4)}_{13,f}&=\frac{g_A \left(\left(40-7 g_A^2\right) c_2+\left(65+2 g_A^2\right) c_3-19 \left(c_4+4 g_A^2 c_4\right)\right)}{576 \pi ^2}\\
          &\qquad-\frac{g_A \left(48+139 g_A^2+35 g_A^4\right)}{768 m_N \pi ^2}\komma\\
    \delta\bar e^{(4)}_{14,f}&=\frac{-40 c_4+g_A^2 (-53 c_2+184 c_3+160 c_4)}{18432 \pi ^2}-\frac{6-12 g_A^2+11 g_A^4}{1536 m_N \pi ^2}\komma\\
    \delta\bar e^{(4)}_{15,f}&=\frac{12 c_4+g_A^2 (7 c_2-2 (4 c_3+c_4))}{576 \pi ^2}+\frac{21+22 g_A^2+7 g_A^4}{768 m_N \pi ^2}\komma\\
    \delta\bar e^{(4)}_{16,f}&=\frac{3 c_4+g_A^2 (-3 c_2-2
      c_3+c_4)}{288 \pi ^2}-\frac{6+12 g_A^2+5 g_A^4}{256 m_N \pi
      ^2}\komma
  \end{aligned}
\end{equation}
\begin{equation*}
  \begin{aligned}
    \delta\bar e^{(4)}_{17,f}&=\frac{\left(4+6 g_A^2\right) c_2+6 \left(1+g_A^2\right) c_3+\left(3+20 g_A^2\right) c_4}{2304 \pi ^2}\\
          &\qquad+\frac{g_A^2 \left(11+13 g_A^2\right)}{3072 m_N \pi ^2}\komma\\
    \delta\bar e^{(4)}_{18,f}&=-\frac{6 \left(1+g_A^2\right) c_2+\left(-6+4 g_A^2\right) c_3+\left(15+19 g_A^2\right) c_4}{1152 \pi ^2}\\
          &\qquad-\frac{3+12 g_A^2+8 g_A^4}{768 m_N \pi ^2}\komma\\
    2 \delta\bar e^{(4)}_{19,f}-\delta\bar e^{(4)}_{22,f}-\delta\bar
    e^{(4)}_{36,f}&=-\frac{72 c_4+g_A^2 (144 c_1-69 c_2+28 c_3+96
      c_4)}{4608 \pi ^2}+\frac{3-7 g_A^2+27 g_A^4}{768 m_N \pi
      ^2}\komma\\
  \delta\bar e^{(4)}_{20,f}+\delta\bar e^{(4)}_{35,f}&=\frac{g_A^2 (72 c_1-5 c_2+72 c_3-24 c_4)-48 c_4}{2304 \pi ^2}+\frac{-12+40 g_A^2+19 g_A^4}{1536 m_N \pi ^2}\komma\\
  2 \delta\bar e^{(4)}_{21,f}-\delta\bar e^{(4)}_{37,f}&=\frac{-4 c_2-34 c_3+19
    c_4+g_A^2 (72 c_1+12 c_2-6 c_3+35 c_4)}{1152 \pi ^2}\\
          &\qquad+\frac{6+25
    g_A^2+3 g_A^4}{1536 m_N \pi ^2}\komma\\
  \delta\bar e^{(4)}_{22,f}-4 \delta\bar e^{(4)}_{38,f}&=-\frac{8 c_4+g_A^2 (720 c_1+153 c_2-136 c_3+16 c_4)}{9216 \pi ^2}\\
          &\qquad-\frac{2+12 g_A^2+3 g_A^4}{512 m_N \pi ^2}\komma\\
  \delta\bar e^{(4)}_{34,f}&=\frac{g_A \left(576 c_1+3 \left(4+11 g_A^2\right) c_2-110 c_3+42 g_A^2 c_3-101 c_4+137 g_A^2 c_4\right)}{2304 \pi ^2}\\
          &\qquad-\frac{g_A \left(17-11 g_A^2+42 g_A^4\right)}{1536 m_N \pi ^2}\punkt
  \end{aligned}
\end{equation*}

\subsubsection{Heavy-baryon chiral perturbation theory}
\label{sec:heavy-baryon-chiral}
In the HB formulation, the employed $\beta$-functions at order $Q^3$ read
\begin{equation}
  \label{eq:44}
  \begin{aligned}
    \beta_{d_1}&= -\frac{g_A^4}{6}\komma \quad
    &\beta_{d_2}&= \frac{1}{12} \left(-1-5 g_A^2\right)\komma \\
    \beta_{d_3}&= \frac{1}{6} \left(3+g_A^4\right)\komma \quad
    &\beta_{d_4}&= 0\komma \quad\\
    \beta_{d_5}&= \frac{1}{24} \left(1+5 g_A^2\right)\komma \quad
    &\beta_{d_{10}}&= \frac{1}{2} \left(g_A+5 g_A^3+4 g_A^5\right)\komma \\
    \beta_{d_{11}}&= \frac{1}{6} \left(3 g_A-9 g_A^3-4 g_A^5\right)\komma 
    &\beta_{d_{12}}&= -g_A \left(2+g_A^2+2 g_A^4\right)\komma \\
    \beta_{d_{13}}&= g_A^3+\frac{2 g_A^5}{3}\komma \quad
    &\beta_{d_{14}}&= \frac{g_A^4}{3}\komma \\
    \beta_{d_{15}}&= 0\komma \quad
    &\beta_{d_{16}}&= \frac{g_A}{2}+g_A^3\komma \\
    \beta_{d_{18}}&= 0 \,. \quad
  \end{aligned}
\end{equation}
For the $\beta$-functions at order $Q^4$ the following results are obtained:
\begin{equation}
  \label{eq:45}
  \begin{aligned}
    \beta_{e_{10}}&=-\frac{1}{6} g_A \left(3+8 g_A^2\right) c_4-\frac{g_A \left(3+19 g_A^2+13 g_A^4\right)}{24 m_N}\komma\\
    \beta_{e_{11}}&=-\frac{g_A c_4}{3}+\frac{g_A \left(-7+35 g_A^2+12 g_A^4\right)}{48 m_N}\komma\\
    \beta_{e_{12}}&=\frac{4}{3} g_A \left(1+g_A^2\right) c_4+\frac{g_A \left(61+57 g_A^2+26 g_A^4\right)}{48 m_N}\komma\\
    \beta_{e_{13}}&=-\frac{2}{3} \left(g_A+2 g_A^3\right) c_4-\frac{g_A \left(73+54 g_A^2+21 g_A^4\right)}{24 m_N}\komma\\
    \beta_{e_{14}}&= \frac{1}{12} (-c_2-6 c_3)-\frac{g_A^2 \left(3+g_A^2\right)}{12 m_N}\komma\\
    \beta_{e_{15}}&= \frac{9+2 g_A^2+11 g_A^4}{24 m_N}\komma\\
    \beta_{e_{16}}&= \frac{-3-2 g_A^2-2 g_A^4}{4 m_N}\komma\\
    \beta_{e_{17}}&= -\frac{c_4}{12}+\frac{-1+7 g_A^2+4 g_A^4}{48 m_N}\komma\\
    \beta_{e_{18}}&= -\frac{2 g_A^2 c_4}{3}-\frac{g_A^2 \left(3+4 g_A^2\right)}{12 m_N}\komma\\
    2\beta_{e_{19}}-\beta_{e_{22}}-\beta_{e_{36}}&= 2 c_1-\frac{5
      c_2}{24}+\frac{3 c_3}{4}+\frac{-1+g_A^2-6 g_A^4}{8 m_N}\komma\\
    \beta_{e_{20}}+\beta_{e_{35}}&= \frac{c_2}{2}+\frac{6+16 g_A^2+15 g_A^4}{24 m_N}\komma\\
    2\beta_{e_{21}}-\beta_{e_{37}}&= \frac{1}{3} \left(2+9 g_A^2\right) c_4+\frac{2+16 g_A^2+9 g_A^4}{12 m_N}\komma\\
    \beta_{e_{22}}-4\beta_{e_{38}}&= \frac{1}{4} (-12 c_1+c_2+3
    c_3)\komma\\
    \beta_{e_{34}}&= \frac{2 g_A c_4}{3}+\frac{g_A-7 g_A^3-6 g_A^5}{24 m_N}\punkt
  \end{aligned}
\end{equation}

%%%%%%%%%%%%%%%%%%%%%%%%%%%%%%%%%%%%%%%%%%%%%%%%%%%%%%%%%%%%%%%%%%%%%%%%%

\clearpage
\section{Tables}

\begin{table}[ht]
  \centering
  \begin{tabular*}{0.5\textwidth}{@{\extracolsep{\fill}} c| r r r}\hline\hline
    $Q^2$ &HB-NN &HB-$\pi$N &Cov\phantom{---} \\\hline\hline
$c_1$&-1.69(4)&-1.60(5)&-2.19(5)\\ $c_2$&3.18(8)&3.63(9)&2.52(7)\\ $c_3$&-6.08(5)&-6.24(5)&-6.25(6)\\ $c_4$&4.61(2)&5.22(3)&4.32(2)\\ 
\hline\hline
$\chi^2_{\pi N}$/dof &0.72 &0.69 &0.67\\\hline\hline
$\bar\chi^2_{\pi N}$/dof &116 &98 &413\\\hline\hline
  \end{tabular*}
\vskip 10pt
  \begin{tabular*}{0.5\textwidth}{@{\extracolsep{\fill}} c| r r r}\hline\hline
    $Q^3$ &\multicolumn{1}{c}{HB-NN} &\multicolumn{1}{c}{HB-$\pi$N} &\multicolumn{1}{c}{Cov}\\\hline\hline
$c_1$&-1.24(2)&-1.64(2)&-1.55(2)\\ $c_2$&4.89(5)&3.51(3)&3.60(4)\\ $c_3$&-7.25(2)&-6.63(2)&-6.54(2)\\ $c_4$&4.74(4)&4.01(4)&3.86(3)\\ $d_{1+2}$&3.39(4)&4.37(4)&4.09(4)\\ $d_3$&-3.47(7)&-3.34(7)&-2.50(4)\\ $d_5$&    0.00(4)&-0.56(4)&-0.86(4)\\ $d_{14-15}$&-7.39(13)&-7.49(13)&-6.05(10)\\ 
\hline\hline
$\chi^2_{\pi N}$/dof &1.04 &1.03 &0.97\\\hline\hline
$\bar\chi^2_{\pi N}$/dof &14.6 &13.0 &13.5\\\hline\hline
  \end{tabular*}
\vskip 10pt
  \begin{tabular*}{0.5\textwidth}{@{\extracolsep{\fill}} c| r r r}\hline\hline
    $Q^4$ &\multicolumn{1}{c}{HB-NN} &\multicolumn{1}{c}{HB-$\pi$N} &\multicolumn{1}{c}{Cov}\\\hline\hline
$c_1$&-1.31(8)&-1.15(8)&-0.82(7)\\ $c_2$&1.88(23)&2.39(22)&3.56(16)\\ $c_3$&-4.43(9)&-4.44(9)&-4.59(9)\\ $c_4$&3.24(17)&3.45(17)&3.44(13)\\ $d_{1+2}$&5.95(9)&5.60(9)&5.43(5)\\ $d_3$&-5.64(6)&-3.84(4)&-4.58(8)\\ $d_5$&-0.11(4)&-0.89(4)&-0.40(4)\\ $d_{14-15}$&-11.61(9)&-9.45(8)&-9.94(7)\\ $e_{14}$&0.86(29)&1.28(32)&-0.63(24)\\ $e_{15}$&-11.36(81)&-13.26(79)&-7.33(45)\\ $e_{16}$&10.73(95)&8.29(95)&1.86(37)\\ $e_{17}$&-0.66(46)&-0.73(47)&-0.90(32)\\ $e_{18}$&4.47(87)&4.17(90)&3.17(45)\\ 
\hline\hline
$\chi^2_{\pi N}$/dof &1.90 &1.83 &1.94\\\hline\hline
$\bar\chi^2_{\pi N}$/dof &4.5 &4.1 &4.9\\\hline\hline
  \end{tabular*}
  \caption{LECs determined from fits at order $Q^2$, $Q^3$, $Q^4$ with
    $T_\pi<100$ MeV.}
\label{tab:Fit}
\end{table}

\newcolumntype{C}{>{\centering\arraybackslash}p{2.5em}}
\newcolumntype{D}{>{\centering\arraybackslash}p{3.5em}}

\begin{table}[ht]
  \centering
  \begin{tabular*}{0.7\textwidth}{@{\extracolsep{\fill}} D|| C C C C
   C C C C}\hline\hline
HBNN&$c_1$&$c_2$&$c_3$&$c_4$&$d_{1+2}$&$d_3$&$d_5$&$d_{14-15}$\\ \hline\hline$c_1$&6&91&-39&23&-15&1&6&4\\ \cline{3-3}$c_2$&10&21&-73&28&7&-6&-3&0\\ \cline{4-4}$c_3$&-2&-7&4&-17&-43&17&15&7\\ \cline{5-5}$c_4$&2&5&-1&16&-22&15&-4&50\\ \cline{6-6}$d_{1+2}$&-2&1&-4&-4&18&-57&-4&-17\\ \cline{7-7}$d_3$&0&-2&2&4&-16&44&-78&2\\ \cline{8-8}$d_5$&1&-1&1&-1&-1&-21&17&17\\ \cline{9-9}$d_{14-15}$&1&0&2&25&-9&1&9&163\\ 
\hline\hline
  \end{tabular*}
\vskip 2pt
  \begin{tabular*}{0.7\textwidth}{@{\extracolsep{\fill}} D|| C C C C
   C C C C}\hline\hline
HB$\pi$N&$c_1$&$c_2$&$c_3$&$c_4$&$d_{1+2}$&$d_3$&$d_5$&$d_{14-15}$\\ \hline\hline$c_1$&4&86&-1&22&-15&6&2&14\\ \cline{3-3}$c_2$&6&11&-52&29&16&-9&-3&4\\ \cline{4-4}$c_3$&0&-3&3&-9&-58&31&6&17\\ \cline{5-5}$c_4$&2&3&-1&12&2&14&-15&52\\ \cline{6-6}$d_{1+2}$&-1&2&-4&0&15&-50&-6&1\\ \cline{7-7}$d_3$&1&-2&4&3&-13&46&-82&1\\ \cline{8-8}$d_5$&0&0&0&-2&-1&-24&19&7\\ \cline{9-9}$d_{14-15}$&4&2&4&24&0&1&4&168\\ 
\hline\hline
  \end{tabular*}
\vskip 2pt
  \begin{tabular*}{0.7\textwidth}{@{\extracolsep{\fill}} D|| C C C C
   C C C C}\hline\hline
Cov&$c_1$&$c_2$&$c_3$&$c_4$&$d_{1+2}$&$d_3$&$d_5$&$d_{14-15}$\\ \hline\hline$c_1$&6&83&20&27&7&14&-16&21\\ \cline{3-3}$c_2$&7&12&-38&31&10&1&-12&0\\ \cline{4-4}$c_3$&1&-3&5&-2&-2&22&-7&38\\ \cline{5-5}$c_4$&2&4&0&12&10&9&-14&46\\ \cline{6-6}$d_{1+2}$&1&1&0&1&17&-7&-56&19\\ \cline{7-7}$d_3$&1&0&2&1&-1&19&-77&-9\\ \cline{8-8}$d_5$&-2&-2&-1&-2&-10&-14&18&3\\ \cline{9-9}$d_{14-15}$&5&0&9&16&8&-4&1&102\\ 
\hline\hline
  \end{tabular*}
  \caption{The upper and lower triangle correspond to 
    the correlation and the covariance matrices for the fits at
    $Q^3$. The correlation and covariance values are given in units of
  $10^{-2}$ and $10^{-4}$, respectively.}
\label{tab:Q3piNCorrCov}
\end{table}

\newcolumntype{C}{>{\centering\arraybackslash}p{2.5em}}

\begin{table}[ht]
  \centering
  \begin{tabular*}{1.0\textwidth}{@{\extracolsep{\fill}} D|| C C C C
   C C C C C C C C C}\hline\hline
HB-NN&$c_1$&$c_2$&$c_3$&$c_4$&$d_{1+2}$&$d_3$&$d_5$&$d_{14-15}$&$e_{14}$&$e_{15}$&$e_{16}$&$e_{17}$&$e_{18}$\\ \hline\hline$c_1$&61&90&12&39&35&-20&-28&-26&-30&38&-78&9&-35\\ \cline{3-3}$c_2$&162&531&-31&38&41&-23&-35&-43&-24&58&-94&10&-35\\ \cline{4-4}$c_3$&8&-64&82&3&-14&7&16&39&-1&-56&46&-5&1\\ \cline{5-5}$c_4$&52&147&5&288&94&-61&-65&-55&-29&29&-38&15&-86\\ \cline{6-6}$d_{1+2}$&25&87&-12&148&85&-68&-66&-56&-26&36&-43&11&-80\\ \cline{7-7}$d_3$&-9&-31&4&-62&-38&36&-9&42&23&-24&25&-30&63\\ \cline{8-8}$d_5$&-9&-34&6&-46&-26&-2&18&37&15&-27&36&13&45\\ \cline{9-9}$d_{14-15}$&-19&-93&34&-89&-49&24&15&90&25&-48&50&-42&66\\ \cline{10-10}$e_{14}$&-67&-163&-3&-144&-70&39&18&69&835&-78&48&-20&31\\ \cline{11-11}$e_{15}$&240&1077&-412&395&268&-116&-93&-367&-1832&6534&-81&21&-32\\ \cline{12-12}$e_{16}$&-579&-2064&400&-613&-381&142&145&454&1309&-6269&9065&-16&38\\ \cline{13-13}$e_{17}$&33&107&-19&121&47&-82&25&-183&-268&774&-689&2140&-62\\ \cline{14-14}$e_{18}$&-236&-696&5&-1271&-639&326&164&539&767&-2220&3096&-2502&7518\\ 
\hline\hline
  \end{tabular*}
\vskip 2pt
  \begin{tabular*}{1.0\textwidth}{@{\extracolsep{\fill}} D|| C C C C
   C C C C C C C C C}\hline\hline
HB$-\pi$N&$c_1$&$c_2$&$c_3$&$c_4$&$d_{1+2}$&$d_3$&$d_5$&$d_{14-15}$&$e_{14}$&$e_{15}$&$e_{16}$&$e_{17}$&$e_{18}$\\ \hline\hline$c_1$&60&93&8&39&37&-20&-33&-19&-15&32&-80&12&-36\\ \cline{3-3}$c_2$&159&494&-29&39&43&-22&-40&-34&-8&47&-93&13&-37\\ \cline{4-4}$c_3$&5&-58&80&2&-16&6&19&39&-8&-51&46&-4&1\\ \cline{5-5}$c_4$&53&149&3&300&94&-62&-72&-51&-18&23&-38&19&-87\\ \cline{6-6}$d_{1+2}$&25&85&-13&144&78&-62&-80&-49&-9&26&-43&11&-79\\ \cline{7-7}$d_3$&-6&-20&2&-45&-23&17&4&42&20&-23&25&-33&65\\ \cline{8-8}$d_5$&-11&-38&7&-53&-30&1&18&35&-1&-20&38&8&53\\ \cline{9-9}$d_{14-15}$&-11&-58&26&-68&-33&13&11&58&25&-47&46&-45&63\\ \cline{10-10}$e_{14}$&-37&-59&-23&-101&-25&26&-2&61&1007&-78&35&-20&21\\ \cline{11-11}$e_{15}$&196&825&-359&318&182&-74&-66&-281&-1957&6273&-75&22&-27\\ \cline{12-12}$e_{16}$&-583&-1959&388&-618&-358&96&155&328&1040&-5623&8954&-19&38\\ \cline{13-13}$e_{17}$&42&133&-18&153&47&-65&16&-161&-301&833&-830&2191&-64\\ \cline{14-14}$e_{18}$&-250&-732&11&-1356&-629&240&202&433&611&-1949&3261&-2681&8066\\ 
\hline\hline
  \end{tabular*}
\vskip 2pt
  \begin{tabular*}{1.0\textwidth}{@{\extracolsep{\fill}} D|| C C C C
   C C C C C C C C C}\hline\hline
Cov&$c_1$&$c_2$&$c_3$&$c_4$&$d_{1+2}$&$d_3$&$d_5$&$d_{14-15}$&$e_{14}$&$e_{15}$&$e_{16}$&$e_{17}$&$e_{18}$\\ \hline\hline$c_1$&49&92&48&35&-11&80&-78&3&-37&1&-63&11&-36\\ \cline{3-3}$c_2$&104&266&9&38&5&62&-70&-19&-18&15&-82&11&-40\\ \cline{4-4}$c_3$&30&14&82&10&-34&62&-40&45&-45&-37&24&0&-9\\ \cline{5-5}$c_4$&32&81&12&173&70&-6&-41&-43&-25&18&-39&-14&-80\\ \cline{6-6}$d_{1+2}$&-3&4&-15&43&22&-51&-14&-47&-1&31&-24&-1&-61\\ \cline{7-7}$d_3$&43&78&43&-6&-18&59&-77&31&-29&-19&-30&1&5\\ \cline{8-8}$d_5$&-23&-49&-15&-23&-3&-25&18&2&32&-3&50&1&35\\ \cline{9-9}$d_{14-15}$&2&-21&28&-39&-15&16&1&48&-3&-33&38&-21&52\\ \cline{10-10}$e_{14}$&-61&-71&-98&-79&-2&-53&32&-5&565&-65&41&-14&27\\ \cline{11-11}$e_{15}$&4&108&-151&105&64&-64&-6&-101&-695&2006&-66&17&-22\\ \cline{12-12}$e_{16}$&-162&-491&80&-187&-42&-84&78&97&359&-1094&1353&-17&43\\ \cline{13-13}$e_{17}$&24&58&0&-57&-1&3&1&-45&-103&235&-196&1010&-46\\ \cline{14-14}$e_{18}$&-114&-290&-35&-472&-128&15&65&159&284&-442&703&-659&1996\\ 
\hline\hline
  \end{tabular*}
  \caption{The upper and lower triangle correspond to 
    the correlation and the covariance matrices for the fits at
    $Q^4$. The correlation and covariance values are given in units of
  $10^{-2}$ and $10^{-4}$, respectively.}
\label{tab:Q4piNCorrCov}
\end{table}

\begin{table}[ht]
  \centering
  \begin{tabular*}{0.5\textwidth}{@{\extracolsep{\fill}} c| r r r}\hline\hline
    $Q^2+\delta^1$ &HB-NN &HB-$\pi$N &Cov\phantom{---} \\\hline\hline
$c_1$&-1.02(3)&-0.84(4)&-0.88(3)\\ $c_2$&0.26(6)&0.85(6)&0.64(4)\\ $c_3$&-0.98(3)&-1.13(3)&-1.00(3)\\ $c_4$&0.48(4)&1.09(3)&1.00(3)\\ 
\hline\hline
$\chi^2_{\pi N}$/dof &0.51 &0.50 &0.53 \\\hline\hline
$\bar\chi^2_{\pi N}$/dof &11 &3.5 &3.3\\\hline\hline
  \end{tabular*}
\vskip 10pt
  \begin{tabular*}{0.5\textwidth}{@{\extracolsep{\fill}} c| r r r}\hline\hline
    $Q^3+\delta^1$ &\multicolumn{1}{c}{HB-NN} &\multicolumn{1}{c}{HB-$\pi$N} &\multicolumn{1}{c}{Cov}\\\hline\hline
$c_1$&-1.35(2)&-1.45(1)&-1.13(1)\\ $c_2$&1.27(3)&0.89(2)&1.24(2)\\ $c_3$&-2.71(1)&-2.52(1)&-2.29(1)\\ $c_4$&2.06(2)&1.77(2)&1.73(2)\\ $d_{1+2}$&-0.47(3)&-0.08(3)&0.24(2)\\ $d_3$&-0.72(6)&-0.59(5)&-0.68(3)\\ $d_5$&0.71(4)&0.43(3)&0.29(3)\\ $d_{14-15}$&-0.16(6)&-0.40(6)&-0.37(4)\\ 
\hline\hline
$\chi^2_{\pi N}$/dof &0.98 &1.09 &1.08\\\hline\hline
$\bar\chi^2_{\pi N}$/dof &2.2 &2.2 &2.2\\\hline\hline
  \end{tabular*}
\vskip 10pt
  \begin{tabular*}{0.5\textwidth}{@{\extracolsep{\fill}} c| r r r}\hline\hline
    $Q^4+\delta^1$ &\multicolumn{1}{c}{HB-NN} &\multicolumn{1}{c}{HB-$\pi$N} &\multicolumn{1}{c}{Cov}\\\hline\hline
$c_1$&-1.34(6)&-1.19(6)&-1.15(5)\\ $c_2$&0.94(17)&1.34(15)&1.57(10)\\ $c_3$&-2.35(5)&-2.33(5)&-2.54(5)\\ $c_4$&2.39(13)&2.45(12)&2.61(10)\\ $d_{1+2}$&1.24(7)&1.41(6)&1.29(3)\\ $d_3$&-1.79(5)&-1.16(3)&-1.83(5)\\ $d_5$&0.38(3)&-0.07(3)&0.37(3)\\ $d_{14-15}$&-1.92(7)&-1.67(5)&-2.22(5)\\ $e_{14}$&1.20(20)&1.00(18)&0.49(13)\\ $e_{15}$&-2.74(54)&-2.72(51)&-1.07(29)\\ $e_{16}$&1.30(62)&-0.91(63)&-1.54(22)\\ $e_{17}$&-0.83(30)&-0.49(29)&-0.94(19)\\ $e_{18}$&-1.64(61)&-1.50(58)&-1.22(29)\\ 
\hline\hline
$\chi^2_{\pi N}$/dof &1.64 &1.72 &1.71\\\hline\hline
$\bar\chi^2_{\pi N}$/dof &2.0 &2.0 &2.0\\\hline\hline
  \end{tabular*}
  \caption{LECs determined from fits at order $Q^2+\delta^1$,
    $Q^3+\delta^1$, $Q^4+\delta^1$ with $T_\pi<100$ MeV.}
\label{tab:FitD}
\end{table}

\newcolumntype{C}{>{\centering\arraybackslash}p{2.5em}}

\begin{table}[ht]
  \centering
  \begin{tabular*}{0.7\textwidth}{@{\extracolsep{\fill}} D|| C C C C
   C C C C}\hline\hline
HB-NN&$c_1$&$c_2$&$c_3$&$c_4$&$d_{1+2}$&$d_3$&$d_5$&$d_{14-15}$\\ \hline\hline$c_1$&3&88&-26&34&17&-36&34&-1\\ \cline{3-3}$c_2$&5&11&-68&42&32&-46&34&-8\\ \cline{4-4}$c_3$&-1&-3&2&-23&-38&36&-19&11\\ \cline{5-5}$c_4$&1&3&-1&5&13&-25&17&-39\\ \cline{6-6}$d_{1+2}$&1&3&-2&1&9&-64&16&-8\\ \cline{7-7}$d_3$&-4&-9&3&-3&-12&39&-85&25\\ \cline{8-8}$d_5$&2&4&-1&1&2&-19&13&-19\\ \cline{9-9}$d_{14-15}$&0&-2&1&-5&-2&10&-4&40\\ 
\hline\hline
  \end{tabular*}
\vskip 2pt
  \begin{tabular*}{0.7\textwidth}{@{\extracolsep{\fill}} D|| C C C C
   C C C C}\hline\hline
HB$-\pi$N&$c_1$&$c_2$&$c_3$&$c_4$&$d_{1+2}$&$d_3$&$d_5$&$d_{14-15}$\\ \hline\hline$c_1$&2&81&12&23&5&-19&19&18\\ \cline{3-3}$c_2$&2&4&-46&30&24&-31&21&11\\ \cline{4-4}$c_3$&0&-1&1&-5&-32&26&-9&9\\ \cline{5-5}$c_4$&0&1&0&2&13&-2&-5&-18\\ \cline{6-6}$d_{1+2}$&0&1&-1&1&7&-49&-1&11\\ \cline{7-7}$d_3$&-1&-3&2&0&-7&26&-85&5\\ \cline{8-8}$d_5$&1&1&0&0&0&-15&11&-7\\ \cline{9-9}$d_{14-15}$&1&1&1&-2&2&2&-1&31\\ 
\hline\hline
  \end{tabular*}
\vskip 2pt
  \begin{tabular*}{0.7\textwidth}{@{\extracolsep{\fill}} D|| C C C C
   C C C C}\hline\hline
Cov&$c_1$&$c_2$&$c_3$&$c_4$&$d_{1+2}$&$d_3$&$d_5$&$d_{14-15}$\\ \hline\hline$c_1$&2&80&17&18&-1&-15&16&30\\ \cline{3-3}$c_2$&2&4&-43&26&13&-24&14&19\\ \cline{4-4}$c_3$&0&-1&2&-6&-22&21&-4&18\\ \cline{5-5}$c_4$&0&1&0&2&11&4&-10&-14\\ \cline{6-6}$d_{1+2}$&0&1&-1&0&5&-21&-38&16\\ \cline{7-7}$d_3$&-1&-2&1&0&-1&10&-81&1\\ \cline{8-8}$d_5$&1&1&0&0&-2&-7&8&-5\\ \cline{9-9}$d_{14-15}$&2&2&1&-1&2&0&-1&19\\ 
\hline\hline
  \end{tabular*}
  \caption{The upper and lower triangle correspond to 
    the correlation and the covariance matrices for the fits at
    $Q^3+\delta^1$. The correlation and covariance values are given in units of
  $10^{-2}$ and $10^{-4}$, respectively.}
\label{tab:Q3piNCorrCovD}
\end{table}

\newcolumntype{C}{>{\centering\arraybackslash}p{2.5em}}

\begin{table}[ht]
  \centering
  \begin{tabular*}{1.0\textwidth}{@{\extracolsep{\fill}} D|| C C C C
   C C C C C C C C C}\hline\hline
HB-NN&$c_1$&$c_2$&$c_3$&$c_4$&$d_{1+2}$&$d_3$&$d_5$&$d_{14-15}$&$e_{14}$&$e_{15}$&$e_{16}$&$e_{17}$&$e_{18}$\\ \hline\hline$c_1$&38&94&23&44&39&-17&-31&-20&-7&20&-78&11&-40\\ \cline{3-3}$c_2$&97&279&-11&42&42&-20&-33&-33&-9&38&-92&13&-41\\ \cline{4-4}$c_3$&8&-10&30&10&-4&10&-4&32&18&-59&35&-5&-6\\ \cline{5-5}$c_4$&35&91&7&166&95&-53&-61&-57&-25&26&-43&14&-88\\ \cline{6-6}$d_{1+2}$&17&48&-1&84&47&-66&-55&-58&-28&33&-45&12&-83\\ \cline{7-7}$d_3$&-5&-17&3&-36&-24&27&-26&54&21&-23&23&-37&60\\ \cline{8-8}$d_5$&-6&-18&-1&-26&-13&-4&11&18&15&-19&34&22&39\\ \cline{9-9}$d_{14-15}$&-9&-40&13&-52&-28&20&4&51&24&-41&43&-50&69\\ \cline{10-10}$e_{14}$&-8&-30&20&-63&-38&21&10&34&382&-86&41&-9&22\\ \cline{11-11}$e_{15}$&66&341&-175&178&122&-65&-34&-156&-902&2911&-71&14&-25\\ \cline{12-12}$e_{16}$&-300&-948&118&-339&-191&73&70&189&498&-2350&3815&-17&42\\ \cline{13-13}$e_{17}$&21&67&-8&54&26&-58&23&-108&-52&224&-310&920&-59\\ \cline{14-14}$e_{18}$&-154&-417&-19&-695&-350&192&80&303&260&-836&1588&-1095&3772\\ 
\hline\hline
  \end{tabular*}
\vskip 2pt
  \begin{tabular*}{1.0\textwidth}{@{\extracolsep{\fill}} D|| C C C C
   C C C C C C C C C}\hline\hline
HB$-\pi$N&$c_1$&$c_2$&$c_3$&$c_4$&$d_{1+2}$&$d_3$&$d_5$&$d_{14-15}$&$e_{14}$&$e_{15}$&$e_{16}$&$e_{17}$&$e_{18}$\\ \hline\hline$c_1$&37&95&14&43&41&-19&-35&-16&-8&25&-81&10&-40\\ \cline{3-3}$c_2$&89&236&-17&41&43&-22&-35&-29&-10&42&-93&12&-39\\ \cline{4-4}$c_3$&5&-14&28&12&-1&8&-3&34&16&-60&39&-6&-7\\ \cline{5-5}$c_4$&33&79&8&156&95&-53&-71&-54&-24&23&-40&12&-88\\ \cline{6-6}$d_{1+2}$&15&40&0&72&37&-58&-74&-51&-27&31&-44&8&-83\\ \cline{7-7}$d_3$&-4&-11&1&-22&-11&11&-11&48&17&-20&23&-35&61\\ \cline{8-8}$d_5$&-7&-17&-1&-27&-14&-1&10&26&21&-24&37&17&50\\ \cline{9-9}$d_{14-15}$&-5&-23&9&-35&-16&8&4&28&21&-37&38&-49&67\\ \cline{10-10}$e_{14}$&-9&-28&16&-55&-30&10&12&21&331&-84&39&-3&19\\ \cline{11-11}$e_{15}$&78&324&-160&147&95&-33&-37&-100&-779&2587&-71&9&-22\\ \cline{12-12}$e_{16}$&-312&-899&130&-317&-170&48&71&127&450&-2290&3970&-14&39\\ \cline{13-13}$e_{17}$&17&54&-9&43&14&-32&15&-73&-16&135&-253&814&-57\\ \cline{14-14}$e_{18}$&-142&-354&-20&-642&-293&115&91&206&202&-643&1439&-949&3421\\ 
\hline\hline
  \end{tabular*}
\vskip 2pt
  \begin{tabular*}{1.0\textwidth}{@{\extracolsep{\fill}} D|| C C C C
   C C C C C C C C C}\hline\hline
Cov&$c_1$&$c_2$&$c_3$&$c_4$&$d_{1+2}$&$d_3$&$d_5$&$d_{14-15}$&$e_{14}$&$e_{15}$&$e_{16}$&$e_{17}$&$e_{18}$\\ \hline\hline$c_1$&27&94&55&41&1&72&-71&3&-14&-19&-65&5&-41\\ \cline{3-3}$c_2$&47&95&24&41&8&59&-64&-14&-5&-8&-79&9&-44\\ \cline{4-4}$c_3$&14&11&24&25&-8&58&-51&34&-22&-43&10&-9&-19\\ \cline{5-5}$c_4$&20&39&12&93&74&-5&-41&-50&-28&9&-42&-22&-84\\ \cline{6-6}$d_{1+2}$&0&2&-1&21&9&-43&-19&-49&-22&24&-23&-5&-68\\ \cline{7-7}$d_3$&18&27&13&-2&-6&23&-79&38&-2&-31&-29&-7&9\\ \cline{8-8}$d_5$&-11&-19&-7&-12&-2&-11&9&-6&18&13&45&12&31\\ \cline{9-9}$d_{14-15}$&1&-7&8&-24&-7&9&-1&25&7&-27&33&-23&59\\ \cline{10-10}$e_{14}$&-10&-6&-14&-36&-9&-1&7&5&181&-78&46&11&17\\ \cline{11-11}$e_{15}$&-28&-23&-60&26&20&-43&11&-38&-301&820&-52&-2&-4\\ \cline{12-12}$e_{16}$&-72&-167&11&-86&-14&-30&29&35&133&-323&465&-7&42\\ \cline{13-13}$e_{17}$&5&16&-8&-39&-3&-6&7&-21&27&-11&-30&344&-33\\ \cline{14-14}$e_{18}$&-61&-123&-27&-234&-59&13&27&86&67&-36&263&-178&840\\ 
\hline\hline
  \end{tabular*}
  \caption{The upper and lower triangle correspond to 
    the correlation and the covariance matrices for the fits at
    $Q^4+\delta^1$. The correlation and covariance values are given in units of
  $10^{-2}$ and $10^{-4}$, respectively.}
\label{tab:Q4piNCorrCovD}
\end{table}

\newcolumntype{C}{>{\centering\arraybackslash}p{6.5em}}
\begin{table}[h]
  \centering
  \begin{tabular*}{0.8\textwidth}{@{\extracolsep{\fill}} c| C | C|
    C| c}\hline\hline
     $Q^3$&\multicolumn{1}{c}{HB-NN} &\multicolumn{1}{c}{HB-$\pi$N} &\multicolumn{1}{c}{Cov} &\multicolumn{1}{c}{RS} \\\hline\hline
$d_{00}^+[M_\pi^{-1}]$&-2.34(4)(1.97)&-1.44(3)(95)&-1.72(3)(50)&-1.36(3)\\
    $d_{10}^+[M_\pi^{-3}]$&2.20(3)(3.06)&1.32(2)(1.98)&1.83(2)(1.08)&1.16(2)\\
    $d_{01}^+[M_\pi^{-3}]$&1.75(1)(96)&1.55(1)(85)&1.68(1)(71)&1.16(2)\\
    $d_{20}^+[M_\pi^{-5}]$&0.22(0)(1.07)&0.22(0)(1.07)&0.06(0)(48)&0.196(3)\\
    $d_{11}^+[M_\pi^{-5}]$&0.07(0)(57)&0.07(0)(73)&0.02(0)(41)&0.185(3)\\
    $d_{02}^+[M_\pi^{-5}]$&0.035(0)(8)&0.035(0)(18)&0.017(0)(21)&0.0336(6)\\
    $b_{00}^+[M_\pi^{-3}]$&-10.1(2)(4.9)&-10.2(2)(8.8)&-8.0(1)(1.9)&-3.45(7)\\
    $d_{00}^-[M_\pi^{-2}]$&1.78(2)(63)&1.76(2)(1.04)&1.53(1)(14)&1.41(1)\\
    $d_{10}^-[M_\pi^{-4}]$&-0.70(1)(99)&-0.67(1)(1.44)&-0.40(1)(20)&-0.159(4)\\
    $d_{01}^-[M_\pi^{-4}]$&-0.35(0)(14)&-0.44(0)(38)&-0.35(0)(10)&-0.141(5)\\
    $b_{00}^-[M_\pi^{-2}]$&15.3(2)(8.9)&12.2(2)(5.5)&13.8(1)(5.8)&10.49(11)\\
    $b_{10}^-[M_\pi^{-4}]$&0.97(0)(4.79)&0.97(0)(5.32)&0.34(0)(3.79)&1.00(3)\\
    $b_{01}^-[M_\pi^{-4}]$&0.19(0)(19)&0.19(0)(28)&0.06(0)(32)&0.21(2)\\ 
$a_{0+}^+[M_\pi^{-1}10^{-3}]$&80.7(4)(2.1)&81.2(4)(1.0)&81.4(4)(2.0)&85.4(9)\\ $a_{0+}^-[M_\pi^{-3}10^{-3}]$&4.6(6)(3.8)&6.4(6)(3.5)&7.1(7)(7.1)&-0.9(1.4)\\ 
\hline\hline
  \end{tabular*}\vskip 10pt
  \begin{tabular*}{0.8\textwidth}{@{\extracolsep{\fill}} c| C |  C|
    C| c}\hline\hline
   $Q^4$  &\multicolumn{1}{c}{HB-NN} &\multicolumn{1}{c}{HB-$\pi$N} &\multicolumn{1}{c}{Cov} &\multicolumn{1}{c}{RS} \\\hline\hline
$d_{00}^+[M_\pi^{-1}]$&-0.37(12)(46)&-0.48(12)(22)&-1.22(9)(12)&-1.36(3)\\
    $d_{10}^+[M_\pi^{-3}]$&-0.86(20)(71)&-0.67(20)(46)&0.75(11)(25)&1.16(2)\\
    $d_{01}^+[M_\pi^{-3}]$&0.79(4)(22)&0.70(4)(20)&0.97(3)(16)&1.16(2)\\
    $d_{20}^+[M_\pi^{-5}]$&1.29(9)(25)&1.30(9)(25)&0.54(4)(11)&0.196(3)\\
    $d_{11}^+[M_\pi^{-5}]$&0.64(4)(13)&0.80(4)(17)&0.43(2)(9)&0.185(3)\\
    $d_{02}^+[M_\pi^{-5}]$&0.033(7)(2)&0.052(8)(4)&-0.004(6)(5)&0.0336(6)\\
    $b_{00}^+[M_\pi^{-3}]$&-5.2(2)(1.1)&-1.44(21)(2.04)&-6.05(10)(45)&-3.45(7)\\
    $d_{00}^-[M_\pi^{-2}]$&1.15(2)(15)&0.71(2)(24)&1.40(1)(3)&1.41(1)\\
    $d_{10}^-[M_\pi^{-4}]$&0.30(3)(23)&0.77(3)(34)&-0.21(1)(5)&-0.159(4)\\
    $d_{01}^-[M_\pi^{-4}]$&-0.210(4)(33)&-0.060(4)(89)&-0.247(3)(23)&-0.141(5)\\
    $b_{00}^-[M_\pi^{-2}]$&6.4(7)(2.1)&6.7(8)(1.3)&8.0(5)(1.3)&10.49(11)\\
    $b_{10}^-[M_\pi^{-4}]$&5.8(5)(1.1)&6.3(5)(1.2)&4.13(27)(88)&1.00(3)\\
    $b_{01}^-[M_\pi^{-4}]$&0.38(16)(4)&0.47(16)(6)&0.38(11)(7)&0.21(2)\\ 
$a_{0+}^+[M_\pi^{-1}10^{-3}]$&82.8(3)(5)&82.2(3)(2)&83.3(3)(5)&85.4(9)\\ $a_{0+}^-[M_\pi^{-3}10^{-3}]$&3.1(9)(1.0)&2.9(9)(8)&-0.01(88)(1.66)&-0.9(1.4)\\ 
\hline\hline
  \end{tabular*}
  \caption{Subtreshold and threshold parameters predicted at order
    $Q^3$ and $Q^4$ in comparison with RS values. The statistical and
    theoretical uncertainties are given in the first and second
    bracket, respectively.}
\label{tab:SubThrPara}
\end{table}

\begin{table}[h]
  \centering
  \begin{tabular*}{0.8\textwidth}{@{\extracolsep{\fill}} c| C |  C|
    C| c}\hline\hline
     $Q^3+\delta^1$&\multicolumn{1}{c}{HB-NN} &\multicolumn{1}{c}{HB-$\pi$N} &\multicolumn{1}{c}{Cov} &\multicolumn{1}{c}{RS} \\\hline\hline
$d_{00}^+[M_\pi^{-1}]$&-1.09(3)(34)&-0.85(2)(14)&-1.33(2)(18)&-1.36(3)\\
    $d_{10}^+[M_\pi^{-3}]$&0.72(2)(48)&0.48(1)(21)&1.15(1)(31)&1.16(2)\\
    $d_{01}^+[M_\pi^{-3}]$&1.23(0)(23)&1.17(0)(21)&1.25(0)(15)&1.16(2)\\
    $d_{20}^+[M_\pi^{-5}]$&0.40(0)(13)&0.40(0)(16)&0.24(0)(14)&0.196(3)\\
    $d_{11}^+[M_\pi^{-5}]$&0.24(0)(14)&0.24(0)(21)&0.19(0)(10)&0.185(3)\\
    $d_{02}^+[M_\pi^{-5}]$&0.021(0)(19)&0.021(0)(25)&0.005(0)(19)&0.0336(6)\\
    $b_{00}^+[M_\pi^{-3}]$&-6.0(1)(4.1)&-6.3(1)(5.5)&-5.8(1)(2.0)&-3.45(7)\\
    $d_{00}^-[M_\pi^{-2}]$&1.63(1)(58)&1.60(1)(77)&1.54(1)(20)&1.41(1)\\
    $d_{10}^-[M_\pi^{-4}]$&-0.42(1)(70)&-0.39(1)(94)&-0.29(1)(20)&-0.159(4)\\
    $d_{01}^-[M_\pi^{-4}]$&-0.22(0)(8)&-0.26(0)(21)&-0.22(0)(5)&-0.141(5)\\
    $b_{00}^-[M_\pi^{-2}]$&9.90(9)(60)&8.67(7)(77)&10.81(7)(63)&10.49(11)\\
    $b_{10}^-[M_\pi^{-4}]$&1.91(0)(44)&1.91(0)(97)&1.28(0)(68)&1.00(3)\\
    $b_{01}^-[M_\pi^{-4}]$&0.07(0)(25)&0.07(0)(21)&-0.07(0)(36)&0.21(2)\\ 
$a_{0+}^+[M_\pi^{-1}10^{-3}]$&86.0(3)(1.0)&86.5(3)(2.6)&86.2(3)(1.0)&85.4(9)\\ $a_{0+}^-[M_\pi^{-3}10^{-3}]$&4.0(5)(3.2)&4.1(5)(3.0)&2.0(5)(3.4)&-0.9(1.4)\\ 
\hline\hline
  \end{tabular*}\vskip 10pt
  \begin{tabular*}{0.8\textwidth}{@{\extracolsep{\fill}} c| C | C|
    C| r}\hline\hline
   $Q^4+\delta^1$  &\multicolumn{1}{c}{HB-NN} &\multicolumn{1}{c}{HB-$\pi$N} &\multicolumn{1}{c}{Cov} &\multicolumn{1}{c}{RS} \\\hline\hline
$d_{00}^+[M_\pi^{-1}]$&-0.75(8)(8)&-0.88(9)(3)&-1.15(6)(4)&-1.36(3)\\
    $d_{10}^+[M_\pi^{-3}]$&0.23(14)(11)&0.43(14)(5)&0.84(7)(7)&1.16(2)\\
    $d_{01}^+[M_\pi^{-3}]$&1.00(3)(5)&0.96(3)(5)&1.10(2)(3)&1.16(2)\\
    $d_{20}^+[M_\pi^{-5}]$&0.53(6)(3)&0.56(6)(4)&0.37(2)(3)&0.196(3)\\
    $d_{11}^+[M_\pi^{-5}]$&0.37(3)(3)&0.44(3)(5)&0.29(1)(2)&0.185(3)\\
    $d_{02}^+[M_\pi^{-5}]$&0.040(5)(4)&0.046(5)(6)&0.025(3)(4)&0.0336(6)\\
    $b_{00}^+[M_\pi^{-3}]$&-1.95(13)(94)&-0.74(14)(1.29)&-3.75(6)(47)&-3.45(7)\\
    $d_{00}^-[M_\pi^{-2}]$&1.04(2)(14)&0.83(2)(18)&1.33(1)(5)&1.41(1)\\
    $d_{10}^-[M_\pi^{-4}]$&0.29(3)(16)&0.54(2)(22)&-0.09(1)(5)&-0.159(4)\\
    $d_{01}^-[M_\pi^{-4}]$&-0.148(3)(18)&-0.054(2)(47)&-0.178(2)(12)&-0.141(5)\\
    $b_{00}^-[M_\pi^{-2}]$&9.30(54)(14)&9.27(52)(18)&10.82(38)(15)&10.49(11)\\
    $b_{10}^-[M_\pi^{-4}]$&2.13(35)(10)&2.88(33)(23)&1.95(17)(16)&1.00(3)\\
    $b_{01}^-[M_\pi^{-4}]$&0.32(10)(6)&0.27(10)(5)&0.29(6)(8)&0.21(2)\\ 
$a_{0+}^+[M_\pi^{-1}10^{-3}]$&85.9(2)(3)&85.3(2)(6)&86.8(2)(3)&85.4(9)\\ $a_{0+}^-[M_\pi^{-3}10^{-3}]$&3.0(8)(7)&2.7(8)(7)&2.0(7)(8)&-0.9(1.4)\\ 
\hline\hline
  \end{tabular*}
  \caption{Subtreshold and threshold parameters predicted at order
    $Q^3+\delta^1$ and $Q^4+\delta^1$ in comparison with RS values. The statistical and
    theoretical uncertainties are given in the first and second
    bracket, respectively.}
\label{tab:SubThrParaD}
\end{table}

\clearpage
\section{Figures}

\vspace{1.5cm}

\clearpage

\pagebreak
\begin{figure}[ht]
\vspace{-1cm}
  \centering
  \includegraphics[width=0.65\textwidth]{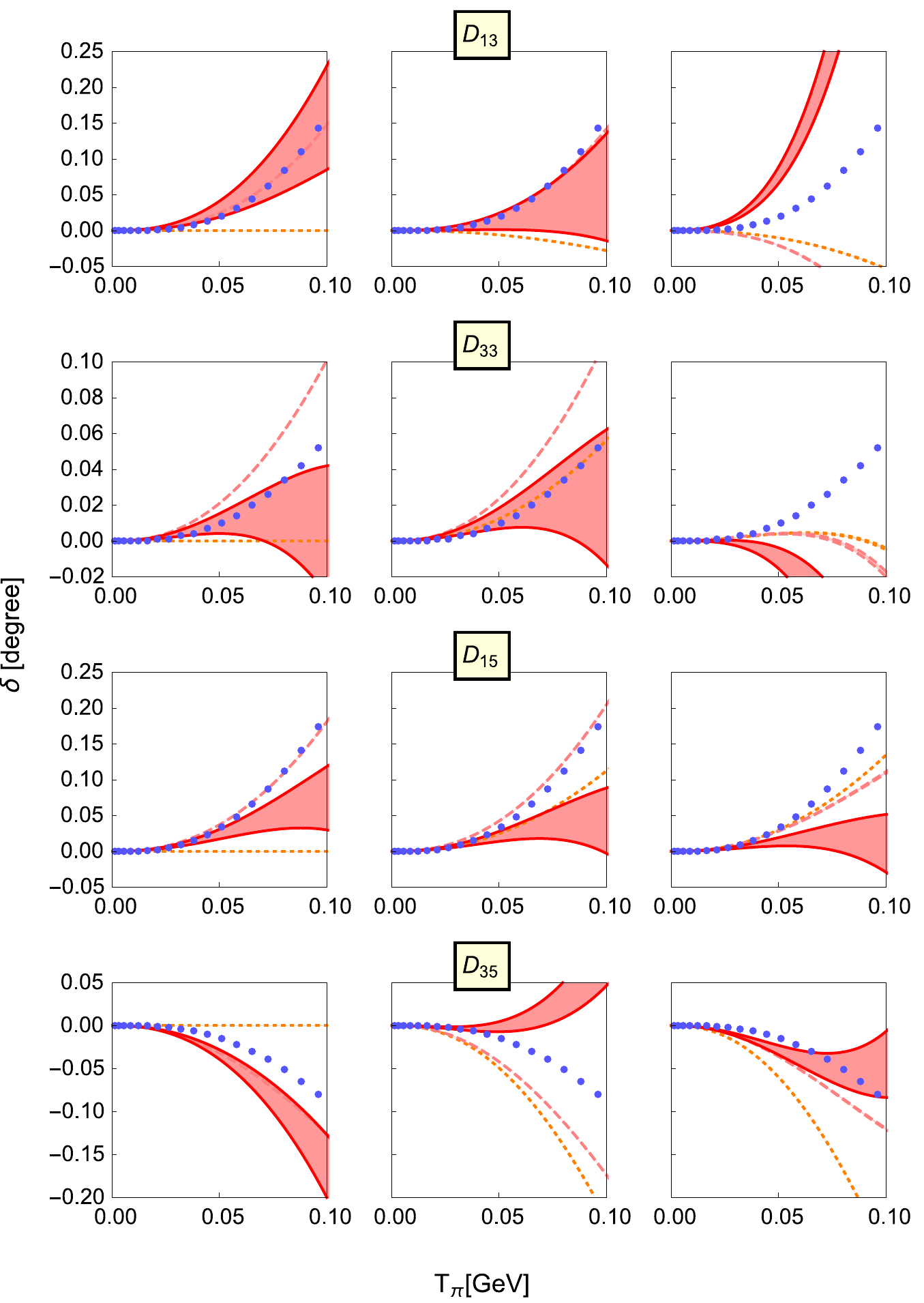}
\caption{(Color online) Predictions including statistical uncertainties for $D$ waves up to
  $T_\pi=100$~MeV. For remaining notation see Fig.~\ref{fig:SnPwavesStat}.}
\label{fig:DwavesStat}
\end{figure}

\begin{figure}[ht]
\vspace{-1cm}
  \centering
  \includegraphics[width=0.65\textwidth]{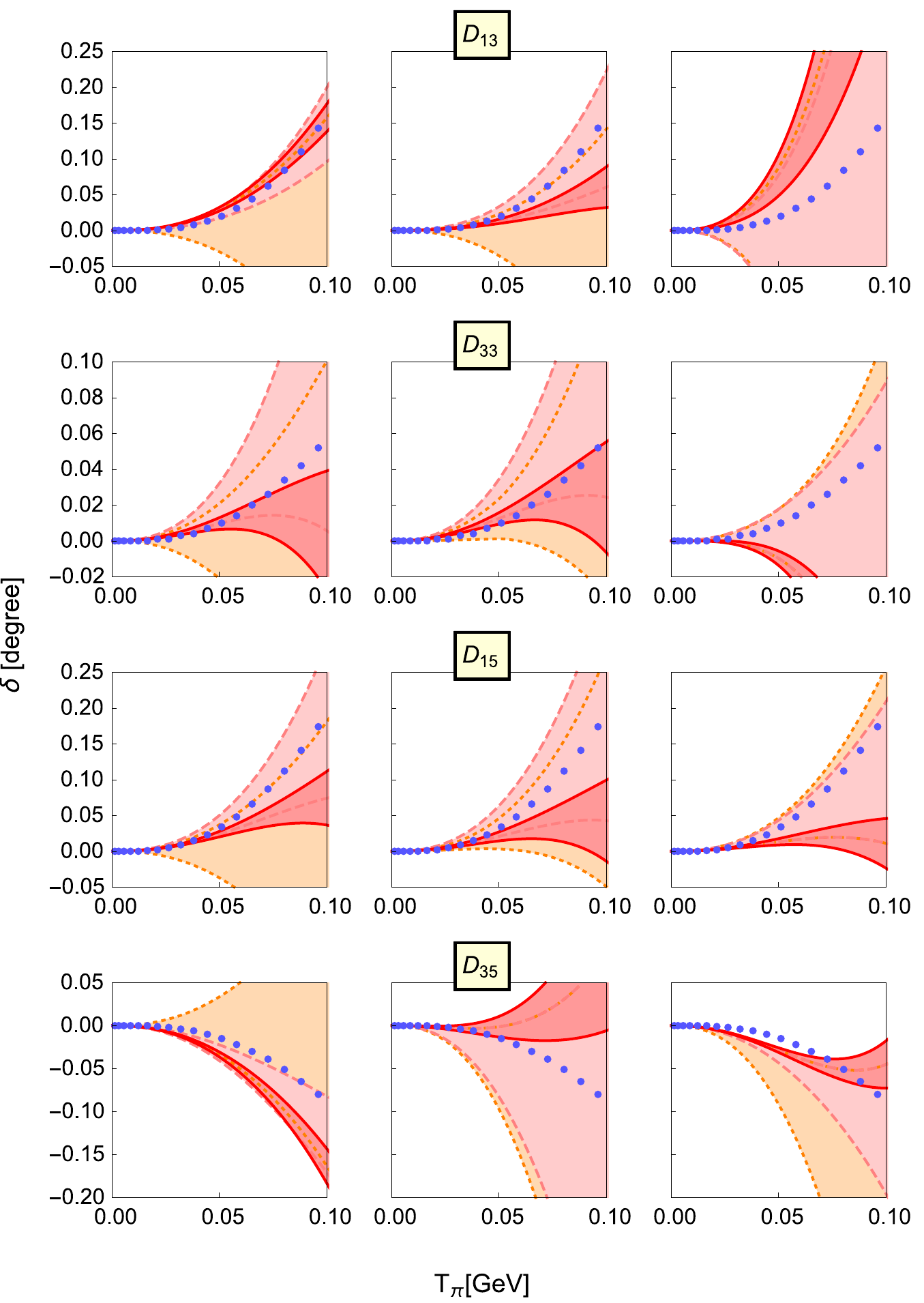}
\caption{(Color online) Predictions including theoretical uncertainties for $D$ waves up to
  $T_\pi=100$~MeV. For remaining notation see Fig.~\ref{fig:SnPwaves}.}
\label{fig:Dwaves}
\end{figure}

\begin{figure}[ht]
\vspace{-1cm}
  \centering
  \includegraphics[width=0.65\textwidth]{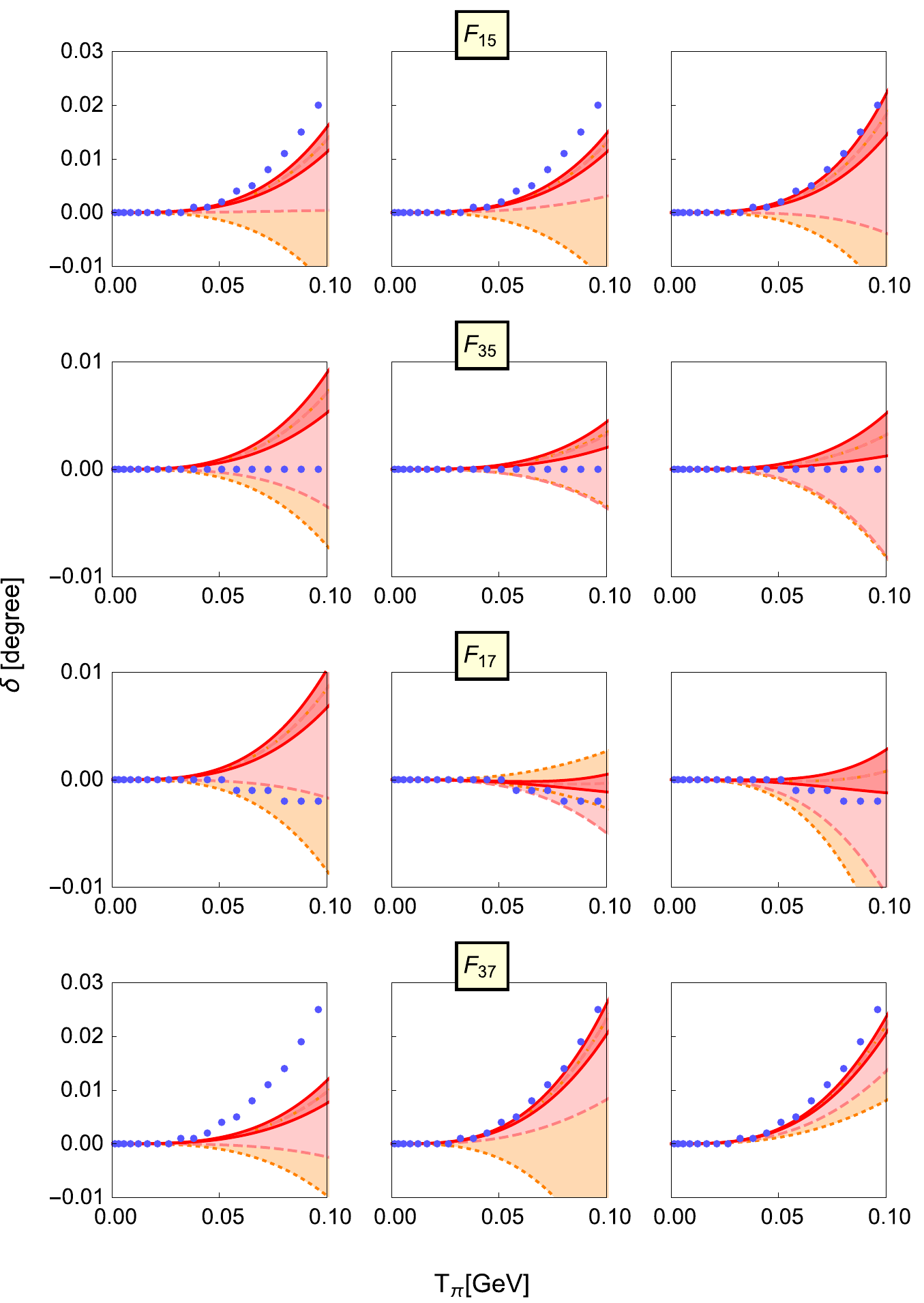}
\caption{(Color online) Predictions including theoretical uncertainties for $D$ waves up to
  $T_\pi=100$~MeV. For remaining notation see Fig.~\ref{fig:SnPwaves}.}
\label{fig:Fwaves}
\end{figure}

\begin{figure}[ht]
  \centering
\includegraphics[width=0.7\textwidth]{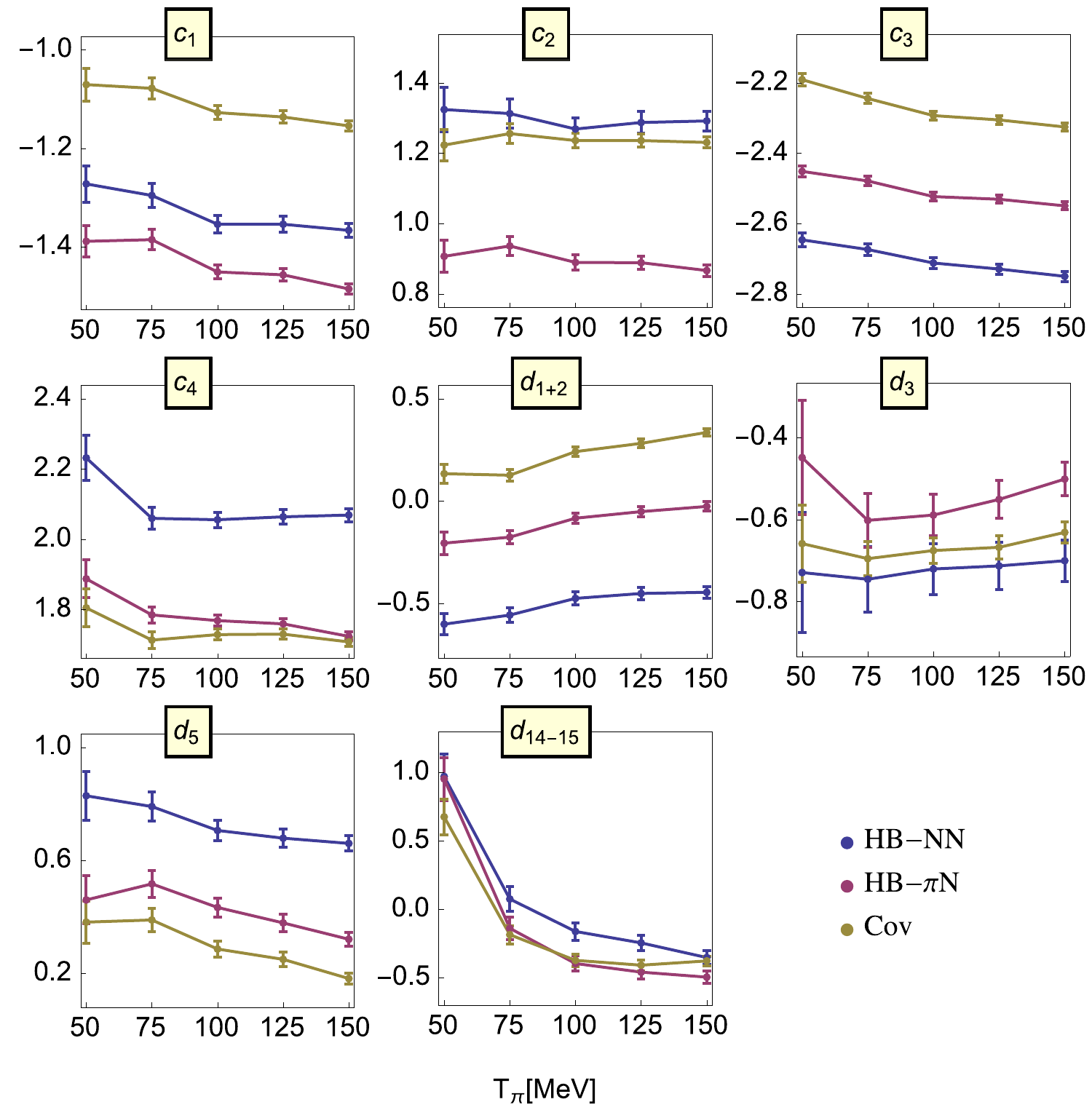}
  \caption{Change of LECs at $Q^3+\delta^1$ over maximum fit energy $T_\pi$.}
  \label{fig:LECsQ3D}
\end{figure}

\begin{figure}[ht]
  \centering
\includegraphics[width=0.7\textwidth]{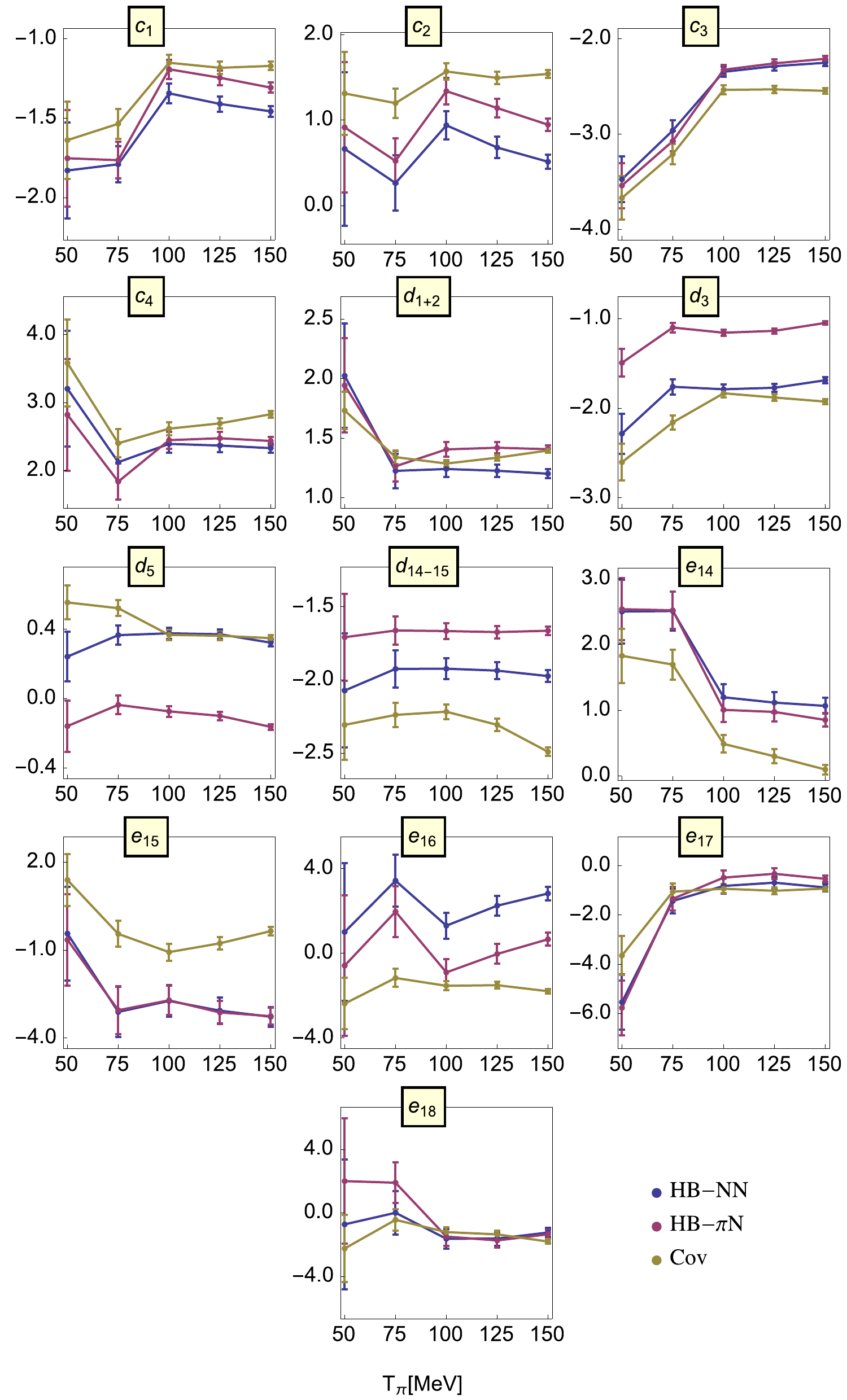}
  \caption{Change of LECs at $Q^4+\delta^1$ over maximum fit energy $T_\pi$.}
  \label{fig:LECsQ4D}
\end{figure}

\begin{figure}[ht]
\vspace{-1cm}
  \centering
  \includegraphics[width=0.65\textwidth]{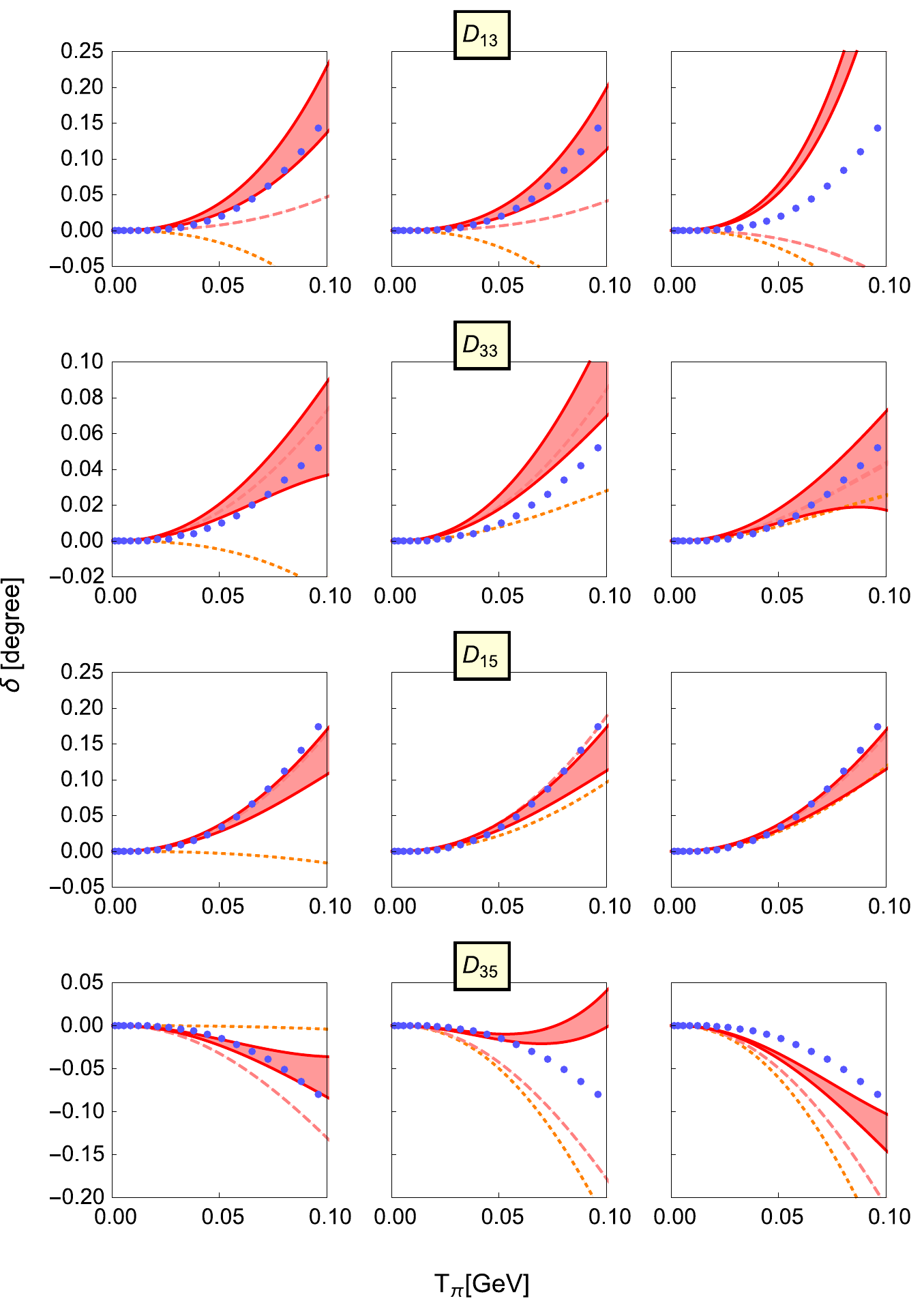}
\caption{(Color online) Predictions including statistical uncertainties for $D$ waves up to
  $T_\pi=100$~MeV. For remaining notation see Fig.~\ref{fig:SnPwavesStatD}.}
\label{fig:DwavesStatD}
\end{figure}

\begin{figure}[ht]
\vspace{-1cm}
  \centering
  \includegraphics[width=0.65\textwidth]{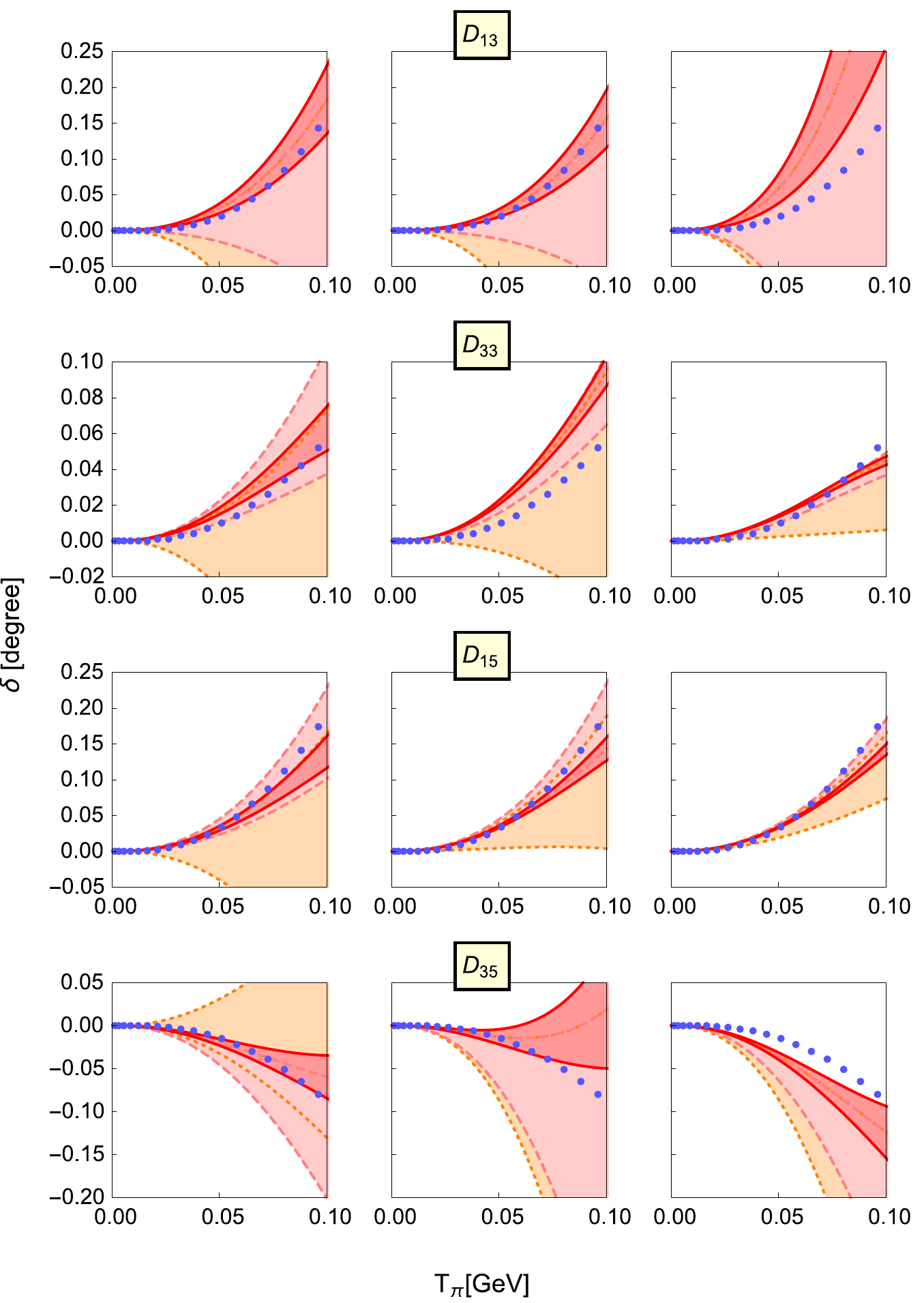}
\caption{(Color online) Predictions including theoretical uncertainties for $D$ waves up to
  $T_\pi=100$~MeV. For remaining notation see Fig.~\ref{fig:SnPwavesD}.}
\label{fig:DwavesD}
\end{figure}

\begin{figure}[ht]
\vspace{-1cm}
  \centering
  \includegraphics[width=0.65\textwidth]{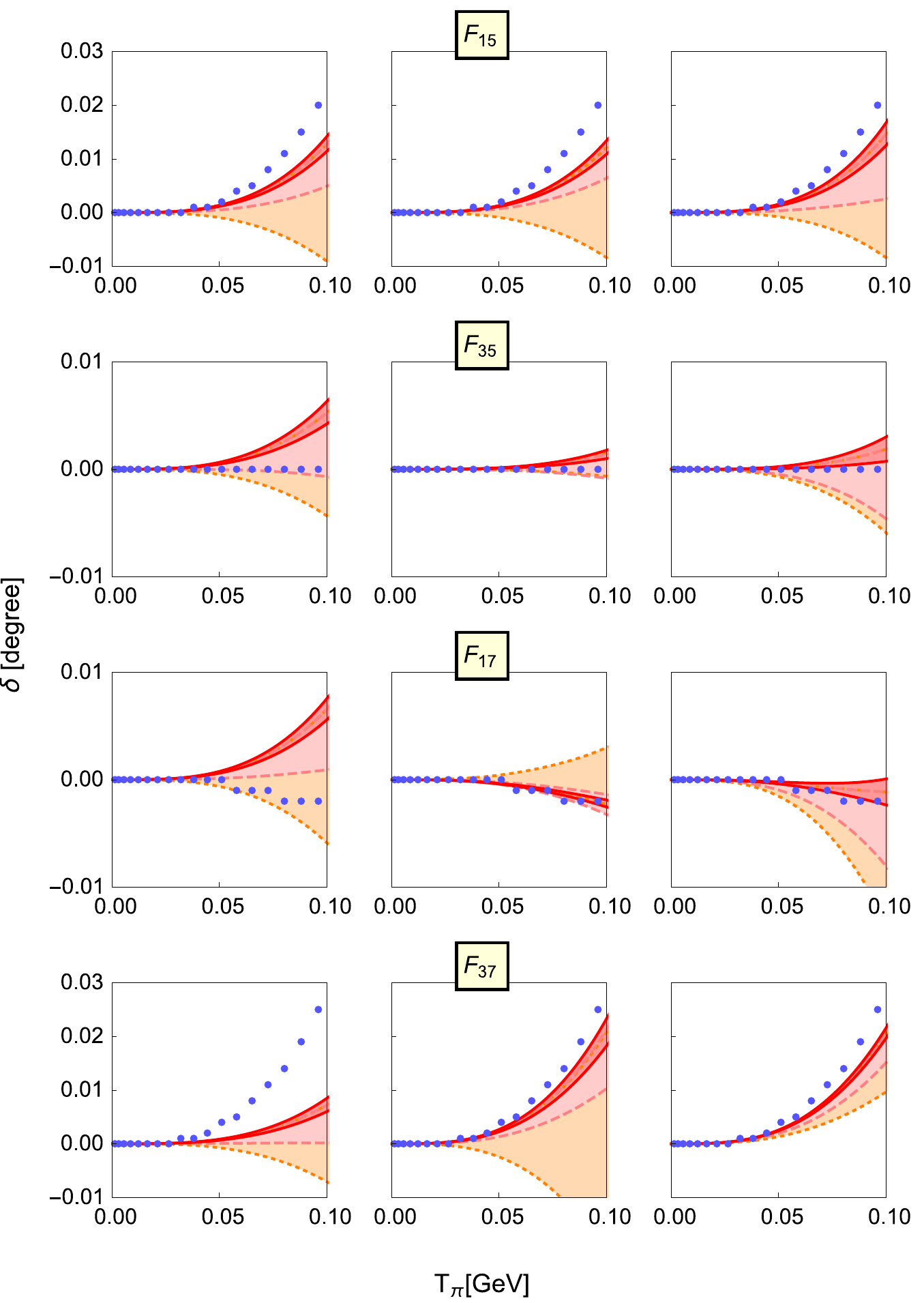}
\caption{(Color online) Predictions including theoretical uncertainties for $D$ waves up to
  $T_\pi=100$~MeV. For remaining notation see Fig.~\ref{fig:SnPwavesD}.}
\label{fig:FwavesD}
\end{figure}

\clearpage

\end{document}